НИКУЛЬЧЕВ Е. В.

# ГЕОМЕТРИЧЕСКИЙ ПОДХОД К МОДЕЛИРОВАНИЮ НЕЛИНЕЙНЫХ СИСТЕМ ПО ЭКСПЕРИМЕНТАЛЬНЫМ ДАННЫМ

Монография

Р е ц е н з е н т ы :

зав. каф. компьютерных технологий и систем Санкт-Петербургского гос. ун-та,
д-р физ.-мат. наук, профессор *Е. И. Веремей,*
зав. каф. прикладной математики и информатики Астраханского гос. ун-та,
д-р физ.-мат. наук *Ю. Ю. Тарасевич*

В монографии изложен геометрический метод моделирования нелинейных динамических систем по экспериментальным данным. Основой метода является качественный подход к анализу нелинейных моделей и построение групп симметрий аттракторов динамических систем с управлениями.

Приведено теоретическое обоснование, включая теоремы о центральном многообразии, определяющие условия существования рассматриваемого класса моделей в локальной области с учетом групповых свойств; алгоритмы оценки инвариантных характеристик, методы построения идентифицируемых моделей и описание результатов, полученных при использовании метода для моделирования управляемых технических процессов. Включены два приложения, являющиеся развитием предложенного подхода: выявление групп симметрий по фазовым портретам динамических систем и методика построения нейросетевых прогнозирующих моделей.

Монография рассчитана на аспирантов и специалистов в области прикладной математики и математического моделирования систем.

# ОГЛАВЛЕНИЕ

# Предисловие

Современное развитие техники и технологий определяет высокие требования к системам управления. Методы совершенствования качества и надежности функционирования управляемых систем в значительной степени определяются используемыми математическими методами и моделями. Это особенно проявляется в работающих промышленных объектах, динамическое поведение контролируемых параметров которых, несмотря на управление, носит нелинейный характер. Построение моделей, адекватных динамическому поведению, определяет проектирование качественных и надежных систем автоматического управления. Ниже проводится краткий аналитический обзор подходов к построению математических моделей нелинейных явлений, использованию качественной теории нелинейных систем [117, 143], объектом исследования которой является фазовые портреты. Проведен также анализ современных методов нелинейной динамики, связанный с задачами исследования хаотических систем.

<u>Подходы к моделированию управляемых технических систем</u>. При решении задач управления техническими системами одной из важных задач является формализация эволюционного поведения объекта управления, процесс моделирования также включает математическое описание целей управления, требований к качеству и надежности функционирования. На основании построенных моделей требуется сформировать алгоритмы управления. К сожалению, современное состояние промышленности (несмотря на усовершенствование технологий) таково, что проектировщики оперируют только с базовыми линейными моделями, возлагая решение задач регулирования нелинейными явлениями на промышленные контроллеры. В результате имеют место реально функционирующие управляемые системы, работающие не только не оптимально, но, даже и в режиме ручной подрегулировки управляющих процессов. Таким образом, для реальных, уже функционирующих систем повышение требований к качеству и надежности не может быть в полной мере обеспечено классическими моделями и средствами управления. Сложность стоящих задач определяется конструкциями промышленных устройств и протекающими нелинейными процессами такими как, процессы теплообмена, течения вязких жидкостей, волновые явления, химические реакции и др.

Можно выделить три классических подхода к моделированию нелинейных явлений в технических системах. Первый заключается в использовании стохастических моделей, при этом считается, что нелинейные колебания являются реализацией случайного процесса, характеристики которого требуется найти. В этом случае, очевидно, не могут быть строго обоснованы предположения и допущения, которые не носят явно вероятностный характер, а наблюдаемое поведение присуще целому классу нелинейных объектов. Оценка адекватности при построении стохастических моделей заключается в проверке соответствия гипотез первоначальным предположениям. Второй подход — построение моделей исходя из физических свойств всех протекающих в технических системах явлений и вывод эволюционных уравнений. Несмотря на кажущуюся очевидность такого решения, оно является практически нереальным для технических систем, т. к. закрытость и структурная сложность конструкций определяет громоздкость и большую наукоемкость такого физического моделирования, тем более что получение необходимых для управления эволюционных уравнений из полученных моделей является отдельной сложной задачей. Например, уравнение неравномерного нагрева в заданных условиях технологической конструкции явно нетривиальная задача, при этом из решения требуется получить зависимость температуры нагрева от изменения параметров какого-либо управляющего механизма.

Перечисленные сложности моделирования реальных функционирующих систем определяют третий, наиболее ориентированный на широкое использование подход к моделированию, заключающийся в выборе вида математической модели в виде эволюционного уравнения и последующей идентификации параметров, либо непараметрической идентификации модели. Модель считается адекватной, если оценка заданного критерия адекватности, вычисленная как зависимость невязки модели от экспериментальных данных находится в допустимых пределах.

В работе рассматриваются модели с управлением в виде системы дифференциальных уравнений с сосредоточенными параметрами с независимыми функциями (управлениями) или их дискретные аналоги, описываемые конечно-разностными уравнениями. Необходимость рассмотрения непрерывных объектов в дискретном времени определяется, с одной стороны, широким применением в практике автоматического управления цифровых управляющих устройств, с другой — конечно-разностные аналоги дифференциальных систем дают возможность получения вычислительно-надежных и эффективных методов управления.

Выбор класса объектов для моделирования обусловлен практической значимостью и разработанностью методов и средств управления для этих моделей. Например, моделирование распределенных систем с использованием разработанного аппарата представляется возможным, но, сколько-нибудь общие, инструментальные и методические средства синтеза управления для этого класса систем отсутствуют.

Таким образом, метод моделирования, изложенный в настоящей монографии, состоит в построении, на основании исследования экспериментальных данных, уравнений, решение которых имеет адекватное динамическое поведение, моделируя при этом качественную динамическую сложность изучаемого процесса. Важным свойством функционирующих систем является устойчивость. Следовательно, требуется рассмотрение моделей на бесконечном интервале времени, т. е. динамических систем.

Моделирование динамической системы состоит из выделения трех компонентов: 1) определение фазового пространства в условиях ограничений; 2) выбор дискретности или непрерывности времени и 3) закона эволюции, т. е. отображение любой заданной точки в фазовом пространстве и любого значения времени в однозначно определенное состояние системы. При этом для дискретного времени ищется закон эволюции в виде
$$x(t+1) = \psi(x,t,u),$$
$$y = h(x(t)), \qquad t \in \mathbb{Z},$$
для непрерывного времени:
$$\dot{x} = f(x,u,t),$$
$$y = \tilde{h}(x(t)), \qquad t \in \mathbb{R}^1,$$
где $x$ — вектор состояний; $y$ — измеряемые процессы (далее, если это не определено особенностями задачи будем предполагать, что наблюдению доступны все состояния); $t$ — время; $f, \psi, h, \tilde{h}$ — в определенном смысле непрерывные и гладкие вектор-функции.

Использование качественной теории нелинейных систем в теории управления традиционно ограничено исследованиями устойчивости по Ляпунову. Применению методов функций Ляпунова для синтеза управлений посвящены работы Н. Н. Красовского, В. И. Зубова, их последователей и учеников. Н. Н. Красовский установил связь метода функций Ляпунова с методом динамического программирования Беллмана и показал, что принципу оптимальности удовлетворяют только те оптимизирующие функции, которые являются функциями Ляпунова для замкнутой системы. Найденные по этим законам управления оптимальны и обеспечивают устойчивость движения. В соответствии с теорией В. И. Зубова, управления строятся из условия реализации наибольшей скорости убывания функции Ляпунова.

В настоящее время происходит развитие геометрической теории управления, обусловленное переходом методов моделирования систем от построения пространства состояний к многообразиям. Развитию этого направления посвящены геометрические исследования А. Г. Бутковского [33], Г. В. Кондратьева [74], А. П. Крищенко [78], В. И. Елкина [50], А. А. Аграчеева [1], Ю. Л. Сачкова [1], В. И. Краснощекова [77], К. Г. Гареева [38], а также в работах по развитию аппарата групп симметрий управляемых систем Ю. Н. Павловского [109], Г. Н. Яковенко [150] и др.

Однако про геометрическую теорию управления сложно сказать как о сформировавшемся научном направлении со своей терминологией и системой обозначений. Характерно, что первый учебник на русском языке [1] вышел в 2006 году.

<u>Методы качественной теории нелинейных систем</u>. Традиционно, исследование управляемых систем происходит во временной области.

Развитие геометрических принципов управления, позволивших получение решения важных задач в терминах симметрий, а также необходимость исследования для нелинейных систем качественного поведения определяют переход к многообразию и методам нелинейной динамики.

В основе нелинейной динамики гладких систем лежат работы А. Пуанкаре, А. М. Ляпунова, Ж. Адамара. На раннем этапе вклад в развитие внесли Д. Биркхгоф, Е. Хопф, С. Катуни. Уже к тридцатым годам прошлого века сформировалась математическая теория колебаний двумерных систем [3, 4, 5, 23, 28, 113]. А. А. Андроновым и Л. С. Понтрягиным определено понятие грубых, структурно-устойчивых систем, А. А. Андроновым, Е. А. Леонтович рассмотрены основные бифуркации предельных циклов, А. А. Андроновым, Л. С. Понтрягиным исследованы полные топологические инварианты для грубых систем, результаты обобщены Е. А. Леонтович и А. Г. Майером. В качественной теории выделяют основные типы фазовых портретов: состояние равновесия (которое соответствует стационарному состоянию во временной области), предельный цикл (соответствует автоколебаниям), инвариантный тор с квазипериодической траекторией (соответствует модуляциям) [3, 53, 63]. Важным классом траекторий являются траектории, устойчивые по Пуассону. При этом движение стремится к своему начальному положению, однако для малой фиксированной окрестности исходного положения последовательность соответствующих времён возвращения может быть неограниченна, то есть движение оказывается непредсказуемым. Все типы движения, соответствующие непереходному поведению, по классификации Биркгофа [28], следующие — стационарные, периодические, квазипериодические, почти периодические и устойчивые по Пуассону траектории. Причем, Марковым установлено, что устойчивая по Пуассону траектория является равномерно устойчивой по Ляпунову, то она должна быть почти периодической. Динамика гладких систем тесно связана с эргодической теорией, основанной на геометрической теории и комбинаторном подходе А. Н. Колмогорова. Следует упомянуть работы Д. В. Аносова, А. Б. Катка для построения гладких систем. Для сложных траекторий создана гиперболическая теория [129].

С. Смейлом [128] и Л. П. Шильниковым [144] в 70-м году одновременно были опубликованы статьи, в которых показано, что системы со сложным поведением орбит могут быть структурно-устойчивыми. Таким образом, портреты с гомоклинической траекторией Пуанкаре обладают бесконечным множеством сосуществующих периодических траекторий и континуумом устойчивых по Пуассону траекторий. В большинстве случаев пространство параметров может быть разбито на две области — с простым или сложным поведением. Одним из основных признаков сложного поведения является наличие в системе гомоклинической траектории Пуанкаре.

Развитие нелинейной динамики определилось открытием хаотических систем, модели которых имеют простой вид. В середине семидесятых начались исследования модели [85]:

$$\dot{x} = -\sigma(x - y),$$
$$\dot{y} = rx - y - xz,$$
$$\dot{z} = -bz + xy,$$

хаотическое поведение решений которой, численно определил Э. Лоренц в 1962 году [206]. Система Лоренца построена как галеркинское приближение задачи о плоском слое жидкости. Модель Лоренца явилась фактическим доказательством существования хаоса. В системе существует так называемый странный аттрактор: поведение траекторий неустойчиво при малых гладких возмущениях системы. Для аттракторов лоренцева типа (с единственным положением равновесия — седло) характерно, что при помощи конечного числа бифуркаций к ним можно перейти от систем с тривиальной динамикой [26, 27, 40].

С появлением многочисленных и тщательных исследований системы Лоренца [17, 26, 62, 65, 139, 225, 230, 238, 242 и др.], динамический хаос стал общепризнанным. Однако позже было строго доказано, что более близкая к физике является математическая модель Чуа [166, 167], в которой также имеет место динамический хаос [20].

Объектом анализа теории динамических систем является гамильтонова динамика, которая в части анализа интегрируемых систем привела к КАМ-теории (А. Н. Колмогоров, В. И. Арнольд, Ю. Мозер [72, 11, 12]). Согласно теории многие качественные особенности сохраняются под действием возмущений, а также возникают в типичных ситуациях, например, в окрестности эллиптической точки.

В общем, можно сделать вывод, что имеются теоретические предпосылки для использования моделей нелинейной динамики для задач моделирования технических объектов и имеется возможность использования методов управления для хаотических объектов.

<u>Управление системами с нелинейным динамическим поведением.</u> Динамическая природа хаотических режимов и их чувствительности по отношению к малым возмущениям определяют возможности по эффективному управлению. Целью внешнего воздействия может быть реализация в системе периодического режима вместо хаоса или попадание в заданную область фазового пространства. Этот подход, предложенный группой американских исследователей из университета штата Мериленд [213], стал основой для решения прикладных задач.

Успешные примеры управления хаосом реализованы в механических системах, электронных устройствах, лазерах. В обзоре [80] приведен пример расчета управления космического аппарата при полете на Луну. Предлагается с помощью малых контролируемых воздействий решить задачу с существенной экономией топлива.

Другое направление применения идей и методов нелинейной динамики связано с проблемой обработки сигналов [9]. Предложены методики, позволяющие выяснить, произведен ли сигнал динамической системой, а также получить информацию о свойствах и характеристиках этой системы. Таким образом, аппарат нелинейной динамики превращается в инструмент исследования, позволяющий сделать заключение или предположение о структуре объекта, сконструировать его динамическую модель и т. д. Разработку методов и алгоритмов анализа сигналов можно считать важным направлением нелинейной динамики, непосредственно связанным с возможными приложениями.

Существуют примеры успешного использования методов нелинейной динамики для анализа и обработки сигналов, конструирования моделей, а также методик управления хаосом применительно к проблемам медицины и биологии [7, 92].

В радиотехнике и электронике разработаны генераторы шумоподобных колебаний, функционирующие в режиме динамического хаоса [8, 15, 19, 79, 202, 211]. Например, генераторы с запаздывающей связью, на лампе бегущей волны и на лампе обратной волны, предложенные академиком Гинзбургом. Одно из возможных приложений хаоса состоит в использовании генерируемых динамическими системами хаотических сигналов в целях коммуникации.

Основанные на хаотической природе сигналов создаются и широко используются новые методы кодирования информации, например, с целью сделать ее труднодоступной для перехвата.

Результаты, полученные в нелинейной динамике, используются при сжатии и хранении информации. Интересным примером такого рода может служить предложенная в Институте радиотехники и электроники РАН схема кодирования с использованием одномерных отображений.

Существуют работы по аналитическому проектированию автоматических регуляторов. Например, работы А. А. Колесникова [70], Н. А. Магницкого и С. В. Сидорова [87, 88] и др. [14, 87, 159, 173, 193]. Созданию и классификации методов управления посвящены исследования представителей школы Саратовского государственного университета (В. С. Анищенко и др.) [8, 15 и др.].

Широкое распространение получили методы по проектированию систем управления, основанные на нечеткой логике и принципах адаптивности (Л. М. Пекора, Дж. Х. Пенг, К. Танака и др.) [216, 217, 218, 224, 234].

Для класса задач управления хаотическими системами имеются результаты по оценке времени управления хаосом [137].

<u>Построение инвариантных характеристик по наблюдаемым данным.</u> Значительный интерес для практического применения представляют методы вычисления инвариантных характеристик и реконструкции аттрактора по временному ряду, полученному в ходе экспериментального исследования нелинейных систем (Ф. Такенс, Д. Рюэль, Н. Паккард, А. Вольф, Г. Г. Малинецкий, В. С. Анищенко, и др.) [6, 26, 29, 90, 91, 214, 233, 239]. Практическое применение и классификации методов реконструкции инвариантных характеристик изложены в четвертой главе.

Анализ публикаций по использованию инвариантных характеристик показывает, что доведено до практического применения в основном моделирование на основе нейросетевых и нечетких моделей [76]. В области управляемых систем, такие модели ограничивают методы управления адаптивным подходом, что не позволяет в полной мере использовать нелинейную теорию динамических систем.

<u>Реконструкция систем по экспериментальным данным</u>. Проблема определения вида динамической системы по ее одномерной реализации относится к классу некорректных задач. В отличие от задачи анализа данная проблема неоднозначна, т. к. существует бесконечное множество динамических систем различного вида и различной сложности, способных воспроизвести имеющийся сигнал с заданной степенью точности. К настоящему моменту разработаны лишь общие рекомендации для случаев, когда исходная система не является слишком сложной [7].

Метод глобальной реконструкции уравнений динамической системы по ее одномерной реализации был предложен в [169, 170]. Алгоритм состоит в следующем. По одномерной реализации процесса в некоторой системе, которая считается «черным ящиком», восстанавливается фазовый портрет, по теореме Такенса, топологически эквивалентный аттрактору исходной системы [232]. По априорно заданным уравнениям, находится методом наименьших квадратов набор неизвестных коэффициентов.

В настоящее время имеется значительное количество работ, развивающих и совершенствующих предложенный метод [9, 90, 152, 153, 154, 161, 180, 221, 226 и др.]. Например, в работах Р. Брауна и др. [159] для воссоздания динамических уравнений по экспериментальному временному ряду с широкополосным сплошным спектром использовалась дополнительная информация о динамических и статистических свойствах исходной системы, содержащаяся в реализации. При получении уравнений учитывались значения показателей Ляпунова [108] и плотности вероятности, рассчитанные по исходному временному ряду. Однако результирующие эволюционные уравнения имели очень громоздкий вид, неудобный для применения. В работе [160] для записи модельных уравнений использовались скрытые переменные. В [159] описывается метод синхронизации модели с исходными данными. В ряде работ О. Л. Аносова предложен алгоритм восстановления скалярного дифференциального уравнения для систем с задержкой.

Однако особенностью многих работ является то, что предлагаемые методы проиллюстрированы на примерах простых маломерных модельных систем, когда заранее известно, каким должен быть результат глобальной реконструкции. При этом не показаны существенные преимущества, даваемые каким-либо усовершенствованным методом по сравнению с [169].

Описываемые в доступных публикациях алгоритмы тестируются на ряде известных модельных систем, имеющих малую размерность и достаточно простой вид правых частей [169]. Например, в обзоре [7] авторами аргументация в пользу новых сложных алгоритмов представляется неубедительной, т. к. работоспособность методов не продемонстрирована на примере сложных временных рядов, генерируемых реальными «черными ящиками».

К настоящему времени существует незначительное количество публикаций, в которых описывается применение данных методик к сигналам, порожденным реальными системами, об операторе эволюции которых ничего не известно. Все это, с одной стороны, создает предпосылки для разработки методологических основ моделирования нелинейных явлений по экспериментальным данным, с другой — нерешенной остается задача создания математических методов и моделей, позволяющих описывать качественное динамическое поведение реальных технических систем с заранее неизвестными структурами моделей.

Сложность задачи состоит в необходимости работать с зашумленными данными при обработке экспериментальных временных рядов. С одной стороны, более желательным является использование метода последовательного дифференцирования для восстановления фазовой траектории, поскольку при этом можно получить модель, содержащую в общем случае приблизительно в $n$ раз меньше коэффициентов при различных нелинейностях, чем при использовании метода задержки. Но дифференцирование неизбежно будет приводить к усилению шумовой компоненты в производных высокого порядка. Без предварительной фильтрации зависимость от времени уже второй производной может оказаться шумоподобным процессом. Кроме того, традиционные методы вложения [91, 123] имеют очевидные недостатки при анализе существенно неоднородных реализаций, т. е. сигналов, в которых участки с быстрым движением чередуются с участками медленных движений.

Произвольный выбор нелинейностей, как правило, не позволяет осуществить удачную реконструкцию динамических уравнений для реальных систем. В частности, в работе [181] указывается на наличие трех типичных случаев:

1. Восстановленные уравнения локально описывают фазовую траекторию исходной системы. При этом реконструированная модель неустойчива в том смысле, что решение полученных уравнений воспроизводит исследуемый сигнал только в течение короткого промежутка времени.

2. Имеет место плохая локальная предсказуемость фазовой траектории, однако наблюдается визуальное сходство фазовых портретов. Решение восстановленных уравнений устойчиво по Пуассону. В этом случае аттрактор реконструированной модели имеет метрические характеристики, близкие к характеристикам исходного аттрактора.

3. Имеет место хорошая локальная предсказуемость фазовой траектории с любой ее точки при значениях времени, превышающих характерное время корреляции. Фазовый портрет реконструированной модели идентичен исходному, а сама система является устойчивой по Пуассону.

Алгоритм глобальной реконструкции в последнее время стал использоваться не только для получения математической модели, но и для классификации динамических режимов [200]. Метод классификации предполагает переход из фазового пространства исходной динамической системы в пространство коэффициентов восстановленных уравнений. Особенность данного подхода состоит в том, что применение к экспериментальным данным алгоритма реконструкции не нацелено на получение модели, способной воспроизводить исходный режим. Коэффициенты в получаемых уравнениях являются количественными характеристиками, несущими информацию о линейных и нелинейных корреляциях в исходном сигнале. В результате аппроксимации коэффициентов по участкам реализации динамической системы исследователь получает множество точек в пространстве коэффициентов. Разным классам динамических систем отвечают непересекающиеся в данном пространстве области, что и дало основание использовать описанную методику в целях классификации.

Монография имеет следующую структуру по главам:

глава 1 — определение вида и класса динамических моделей; описание системы определений и обозначений, используемых в работе, некоторые необходимые для последовательного изложения положения из качественной теории нелинейных систем;

глава 2 — изложение основ теории групп симметрий систем с управлениями и разработка теоретических основ группового анализа дискретных систем;

глава 3 — построение теории о центральном многообразии, определяющей условия существования выбранного класса моделей в локальной области с учетом групповых свойств;

глава 4 — алгоритмы оценки инвариантных характеристик, метода построения моделей и методики идентификации их параметров;

глава 5 — разработка технологии повышения качества функционирования управляемых технических систем и описание результатов, полученных при внедрении в промышленность.

# ГЛАВА 1. ТЕОРЕТИЧЕСКИЕ ОСНОВЫ ГЕОМЕТРИЧЕСКОГО ПОДХОДА К МОДЕЛИРОВАНИЮ СИСТЕМ

В главе определен объект исследования, применительно к которому сформулированы понятия качественной теории систем; проведен анализ и обобщены положения геометрического подхода к моделированию фазовых траекторий систем с управлением; сформулированы классы решаемых задач и ограничения применимости метода.

## *1.1. Математические модели нелинейных систем*

Проведем выбор системы допущений к формальному описанию нелинейных моделей, реконструируемых по экспериментальным данным с учетом требуемых свойств и ограничений, связанный с классом управляемых систем.

Предположим, что данные, полученные в результате функционирования управляемых систем, являются выходными процессами нелинейной аффинной динамической системы, определяемой преобразованием $D \times U \to T_x X$:

$$\dot{x} = f(x(t), t) + h(x(t))u(t), \qquad (1.1)$$

где $x$ — вектор состояний системы из $\mathbb{R}^n$; $u(t)$ — $p$-мерный вектор управлений из множества допустимых управлений $U \subseteq \mathbb{R}^p$ ($p \leq n$); многообразие $X$, обычно отождествляемое с $\mathbb{R}^n$, либо с $\mathbb{R}^{n+1}$, в последнем случае многообразие включает независимую переменную $t$ (время); $\dot{x} \in T_x X$; $T_x X$ — касательное пространство к $X$ в точке $x$, определяемое допустимыми управлениями; $f(\cdot) = (f_1, ...., f_n)$, $h(\cdot) = (h_1, ...., h_n)$ — $\mathbb{C}^r$-гладкие ($r \geq 1$) вектор-функции, определенные в некоторой области $D \subseteq \mathbb{R}^n$, которая рассматривается как фазовое пространство (ограниченное, неограниченное или совпадающее с евклидовым пространством $\mathbb{R}^n$).

На правую часть уравнения управляемой системы наложены ограничения:
1. $x \mapsto f(x)$ — гладкое векторное поле на $X$ при любом фиксированном $u \in U$;
2. $(x, u) \mapsto f(x) + h(u)$ — непрерывное отображение $x \in X$, $u \in U$;
3. $(x, u) \mapsto \partial[f(x) + h(u)]/\partial x$ — непрерывное отображение в любых локальных координатах на $X$ при $x \in X$, $u \in U$.

Допустимые управления — измеримые локально ограниченные отображения

$$u: t \mapsto u(t) \in U.$$

Пусть допущением является то, что в зависимости от выбора управляющих сигналов, система (1.1) может быть преобразована к эквивалентной автономной, т. е. при наличии управления параметризует автономную систему. Таким образом, вспомогательным объектом являются автономные системы обыкновенных дифференциальных уравнений, записанные в виде

$$\dot{x} = f(x). \qquad (1.2)$$

Дифференцируемое отображение $\varphi : \tau \to D$, где $\tau$ — интервал на оси $t$, называется решением $x = \varphi(t)$ системы (1.1), если

$$\dot{\varphi}(t) = f(\varphi(t)) \quad \text{для любого } t \in \tau. \qquad (1.3)$$

По предположению, условия теоремы Коши выполняются, следовательно; для любых значений $x_0 \in D$ и $t_0 \in \mathbb{R}^1$ существует единственное решение $\varphi$, удовлетворяющее начальному условию

$$x_0 = \varphi(t_0). \qquad (1.4)$$

Решение определено на некотором интервале $(t_+, t_-)$, содержащем значение $t = t_0$. Вообще, граничные точки $t_+$ и $t_-$ могут принимать как конечные, так и бесконечные значения. Решения системы (1.2) обладают следующими свойствами:

1. Если $x = \varphi(t)$ — решение системы (1.2), то очевидно, что $x. = \varphi(t + C)$ также является решением, определенным на интервале $(t_+ - C, t_- - C)$.

2. Решения $x = \varphi(t)$ и $x = \varphi(t + C)$ можно рассматривать как решения, соответствующие одной начальной точке $x_0$, но различным начальным моментам времени $t_0$.

3. Решение, удовлетворяющее условию (1.4), можно записать в виде $x. = \varphi(t - t_0, x_0)$, где $\varphi(0, x_0) = x_0$.

4. Если $x_1 = \varphi(t_1 - t_0, x_0)$, то $\varphi(t - t_0, x_0) = \varphi(t - t_1, x_1)$. Обозначая $t_1 - t_0$ через новое $t_1$, а $t - t_2$ — через $t_2$, получаем так называемое групповое свойство решений:

$$\varphi(t_2, \varphi(t_1, x_0)) = \varphi(t_1 + t_2, x_0). \qquad (1.5)$$

Известно, что решение $x_1 = \varphi(t_1 - t_0, x_0)$ задачи Коши (1.3) для $\mathbb{C}^r$-гладкой системы (1.1) является гладким ($\mathbb{C}^r$) относительно времени и начальных данных $x_0$. Первая производная $\eta(t - t_0, x_0) \equiv \dfrac{\partial \varphi}{\partial x_0}$ — удовлетворяет так называемому уравнению в вариациях $\dot{\eta} = f'(\varphi(t - t_0, x_0))\eta$ с начальным условием: $\eta(0, x_0) = I$ ($I$ — единичная матрица). Уравнение в вариациях — это линейная неавтономная система, полученная формальным дифференцированием выражения (1.2).

В связи с тем, что теория управления, с одной стороны, изучает динамические модели, с другой — многие вопросы и результаты качественной теории не являются общепринятыми и однозначно определенными. Приведем основные результаты качественной теории нелинейных систем в соответствии с [4, 28], а также с их современными эквивалентными представлениями в [143].

Рассмотрим траекторию $\Lambda$, отличную от состояния равновесия, соответствует такому решению $\varphi(t)$ системы (1.2), что $\varphi(t_1) = \varphi(t_2)$ при $t_1 \neq t_2$. Тогда $\varphi(t)$ определено для всех значений $t$ и периодично, а $\Lambda$ — гладкая замкнутая кривая. Если $\tau$ — наименьший период $\varphi(t)$, то параметрическое уравнение траектории $\Lambda$ принимает вид $x = \varphi(t)$, где $t_0 \leq t \leq t_0 + \tau$, причем в данном интервале различные значения $t$ соответствуют различным точкам траектории $\Lambda$. Траекторию $\Lambda$, соответствующую периодическому решению $\varphi(t)$, называют периодической.

Любая другая траектория, не являющаяся ни положением равновесия, ни периодической траекторией, есть незамкнутая кривая [4]. Отсюда следует, что незамкнутая траектория не имеет точек самопересечения.

Любые два решения, которые отличаются лишь выбором начального момента времени $t_0$, соответствуют одной и той же траектории. И наоборот: два различных решения, отвечающие одной и той же траектории, совпадают с точностью до сдвига по времени $t \to t + C$. Таким образом, решения, отвечающие одной и той же периодической траектории, периодичны и имеют равные периоды [5]. Как известно, любая траектория, лежащая в ограниченной области, является целой, т. е. для нее определено решение для $t \in (-\infty, +\infty)$.

Для траекторий, отличных от положений равновесия, можно задать положительное направление движения, совпадающее с направлением возрастания $t$. В каждой точке такой траектории направление определятся при помощи соответствующего касательного вектора. Вместе с системой (1.1) в нелинейной динамике принято рассматривать соответствующую «обращенную во времени» систему

$$\dot{x} = -f(x). \qquad (1.5)$$

Здесь векторное поле системы получается из векторного поля системы (1.1) при изменении направления каждого касательного вектора на противоположное. При этом решение $x = \varphi(t)$ системы (1.1) соответствует решению $x = \varphi(-t)$ системы (1.5) и наоборот. Очевидно, что системы (1.1) и (1.5) имеют одни и те же фазовые кривые с точностью до замены времени $t \to -t$.

Важным свойством является возможность при рассмотрении фазовых портретов перемасштабирование времени. Для системы (1.1), траектории вида $x = \varphi(t - t_0, x_0)$, проходящей через точку $x_0$ при $t = t_0$, при параметризации в соответствии с правилом

$$d\tilde{t} = \frac{dt}{f(\varphi(t - t_0, x_0))}$$

или

$$d\tilde{t} = t_0 + \int_{t_0}^{t} \frac{d\theta}{f(\varphi(\theta - t_0, x_0))}$$

имеется траектория системы

$$\dot{x} = f(x)F(x). \quad (1.6)$$

Здесь $\mathbb{C}^r$-гладкая функция $F(x): D \mapsto \mathbb{R}^1$ не обращается в нуль в $D$. Следовательно, системы (1.1) и (1.6) имеют одни и те же фазовые кривые с точностью до замены времени. Траектории обеих систем имеют одинаковое направление при $F(x) > 0$ и противоположное при $F(x) < 0$.

С точки зрения динамики, целые траектории или траектории, которые можно определить, по крайней мере, для всех положительных $t$ на бесконечном промежутке времени, представляют особый интерес. Системы, решения которых могут быть определены на бесконечном промежутке времени, были названы Биркгофом [28] динамическими.

Общепринято определение динамической системы состоит из ее трех компонентов:

1) фазового пространства $D$;

2) типа времени $t$: непрерывного ($t \in \mathbb{R}^1$) или дискретного ($t \in \mathbb{Z}$);

3) закона эволюции, т. е. отображение заданной точки $x \in D$ и любого $t$ в однозначно определенное состояние $\varphi(t, x) \in D$, удовлетворяющее теоретико-групповым свойствам:

1. $\varphi(0, x) = x.$
2. $\varphi(t_1, \varphi(t_2, x)) = \varphi(t_1 + t_2, x)).$
3. $\varphi(t, x)$ непрерывно по $(x, t)$.

Если переменная $t$ непрерывна, приведенные условия определяют непрерывную динамическую систему, являющейся с точки зрения геометрии потоком, т. е. однопараметрической группой гомеоморфизмов фазового пространства $D$. Фиксируя $x$ и изменяя $t$ от $-\infty$ до $+\infty$, получаем ориентируемую кривую, называемую фазовой траекторией. Принята следующая классификация фазовых траекторий [4]:

− состояния равновесия;
− периодические траектории;
− незамкнутые траектории.

Заметим, что в качестве фазового пространства $D$, как правило, используют область в $\mathbb{R}^n$, либо на $n$-мерным торе

$$\underbrace{\mathbb{S}^1 \times \mathbb{S}^1 \times ... \times \mathbb{S}^1}_{k\ раз},$$

гладкой поверхности или многообразии. При этом соответствие между гладким потоком и векторным полем путем определения поля скоростей определяется соотношением

$$f(x) = \left.\frac{d\varphi(t,x)}{dt}\right|_{t=0}.$$

Для дискретной динамической системы
$$x(k+1) = \psi(x(k)), \qquad (1.7)$$

Последовательность $\{x(k)\}_{k=-\infty}^{+\infty}$ называется траекторией точки $x_0$. Существует три типа траекторий:

1. Точка $x(0)$. Эта точка является неподвижной точкой гомеоморфизма $\psi(x)$, т. е. отображается при помощи $\psi(x)$ в себя.

2. Цикл $(x(0), ..., x(k-1))$, где $x(i) = \psi(x(0))$, $i = \overline{0, k-1}$; $x_0 = \psi^k(x_0)$; $x_i \neq x_j$ при $i \neq j$. Здесь число $k$ является периодом; каждая точка $x_i$ — периодической точкой с периодом $k$.

3. Бесконечная в обе стороны траектория, т. е. последовательность $\{x_k\}_{k=-\infty}^{+\infty}$, где $x_i \neq x_j$ при $i \neq j$. Как и в случае потоков, такую траекторию называют незамкнутой.

В соответствии с качественной теорией нелинейных систем [3–5, 143] введем некоторые понятия.

Множество $A$ называется инвариантным относительно динамической системы, если $A = \varphi(t, A)$ для любого $t$. В этом выражении $\varphi(t, A)$ обозначает множество $\bigcup_{\mathbf{x} \in A} \varphi(t, A)$. Из данного определения следует, что если $x \in A$, то траектория $\varphi(t, A)$ лежит в множестве $A$.

Точка $x_0$ называется блуждающей, если существует открытая окрестность $U(x_0)$ и такое положительное значение $T$, что
$$U(x_0) \bigcap \varphi(t, U(x_0)) = \varnothing \text{ при } t > T. \qquad (1.8)$$

Применяя к (1.8) преобразование $\varphi(-t, \cdot)$, получаем, что
$$\varphi(-t, U(x_0)) \bigcap U(x_0) = \varnothing \text{ при } t > T. \qquad (1.9)$$

Следовательно, определение блуждающей точки является симметричным относительно обращения времени [47].

Множество блуждающих является открытым и инвариантным. Открытость множества следует из того, что вместе с $x_0$ любая точка в окрестности $U(x_0)$ является блуждающей. Инвариантность множества следует из того, что если точка $x_0$ — блуждающая, то точка $\varphi(t_0, x_0)$ также будет блуждающей при любом значении $t_0$.

Например, для динамической системы
$$\dot{x} = f(x, \theta),$$
$$\theta = 1,$$

заданной в $\mathbb{R}^{n+1}$ имеет место (1.8), поскольку $\theta(t) = \theta_0 + t$ монотонно возрастает с увеличением $t$. Следовательно, каждая точка фазового пространства является блуждающей. Состояния равновесия, и точки, принадлежащие периодическим траекториям, являются неблуждающими. Все точки двояко-асимптотических траекторий, которые при $t \to \pm\infty$ стремятся к состояниям равновесия и периодическим траекториям, также неблуждающие. Такая двояко-асимптотическая траектория незамкнута и называется гомоклинической. Точки, принадлежащие устойчивым по Пуассону траекториям, также неблуждающие.

Точка $x_0$ называется положительно устойчивой по Пуассону, если для произвольной окрестности $U(x_0)$ и любого $T > 0$ существует такое значение $t > T$, что

$$\varphi(t, x_0) \subset U(x_0).$$

В этих условиях если для любого $T > 0$ существует такое значение $t$, что $t < -T$, то $x_0$ — отрицательно устойчивая по Пуассону точка. Если точка одновременно и положительно, и отрицательно устойчива по Пуассону, то она называется устойчивой по Пуассону.

Заметим, что если точка $x_0$ положительно (отрицательно) устойчива по Пуассону, то любая точка на траектории $\varphi(t, x_0)$ также является положительно (отрицательно) устойчивой по Пуассону.

Таким образом, вводятся [143] $P^+$-траектории (положительно устойчивая по Пуассону), $P^-$-траектории (отрицательно устойчивая по Пуассону) и просто $P$-траектории (устойчивая по Пуассону). Непосредственно из (1.8) следует, что $P^+$, $P^-$ и $P$-траектории состоят из неблуждающих точек.

Очевидно, что состояния равновесия и периодические траектории являются замкнутыми $P$-траекториями. Если $P^+$ ($P^-$, $P$)-траектория незамкнута, то ее замыкание $\Sigma$ содержит континуум незамкнутых P-траекторий [28]). $P$-траектория последовательно пересекает любую $\varepsilon$-окрестность $U_\varepsilon(x_0)$ бесконечное число раз. Для потоков множество значений, для которых $P$-траектория пересекает окрестность $U_\varepsilon(x_0)$, состоит из бесконечного множества временных интервалов $I_n(\varepsilon)$ и $t_n(\varepsilon)$ — одно из значений из $I_n(\varepsilon)$.

Величина

$$\tau_n(\varepsilon) = t_{n+1}(\varepsilon) - t_n(\varepsilon)$$

называется временем возвращения Пуанкаре. Здесь последовательность $\{t_n(\varepsilon)\}_{-\infty}^{+\infty}$ такая, что $t_n(\varepsilon) < t_{n+1}(\varepsilon)$ и $\varphi(t_n(\varepsilon), x_0) \subset U_\varepsilon(x_0)$. В литературе по обыкновенным дифференциальным уравнениям нетривиальным возвращением часто называют наличие непериодических рекуррентных точек, что является примером сложного асимптотического поведения.

Таким образом для незамкнутой P-траектории возможны 2 случая.

1. Рекуррентная последовательность, т. е. $\{\tau_n(\varepsilon)\}$ ограничена для любого конечного $\varepsilon$, т. е. существует число $K(\varepsilon)$, что $\tau_n(\varepsilon) < K(\varepsilon)$ при любом $n$.

2. Последовательность $\{\tau_n(\varepsilon)\}$ неограничена при любом достаточно малом $\varepsilon$.

В первом случае, все траектории, лежащие в ее замыкании $\Sigma$, также рекуррентные, а само замыкание — минимальное множество, т. е. множество непусто, инвариантно, замкнуто и не содержит собственных подмножеств, обладающих указанными тремя свойствами. Основное свойство такой траектории состоит в том, что она возвращается в $\varepsilon$-окрестность точки $x_0$ в промежуток времени, не превышающий $K(\varepsilon)$. Однако в отличие от периодических траекторий, время возвращения которых фиксировано, времена возвращения рекуррентной траектории не ограничены.

Во втором случае замыкание $\Sigma$ незамкнутой $P$-траектории называется квазиминимальным множеством. В данном случае в $\Sigma$ всегда существуют другие инвариантные замкнутые подмножества, которые могут быть состояниями равновесия, периодическими траекториями, инвариантными торами и т. д. В этом случае времена возвращения Пуанкаре могут быть сколь угодно велики.

По определению, аттрактор $Atr$ — это замкнутое инвариантное множество, имеющее такую окрестность (поглощающую область) $U(Atr)$, что траектория $\varphi(t,x)$ произвольной точки $x$, принадлежащей $U(Atr)$, удовлетворяет условию

$$\rho(\varphi(t,x), Atr) \to 0 \text{ при } t \to +\infty, \qquad (1.10)$$

где

$$\rho(x, Atr) = \inf_{x_0 \in Atr} \| x - x_0 \|.$$

Примерами аттракторов являются состояния равновесия, устойчивые периодические траектории и устойчивые инвариантные торы, содержащие квазипериодические траектории.

Для странных аттракторов хаотических систем, являющиеся инвариантными замкнутыми множествами, состоящими только из неустойчивых траекторий выполняется условие квазиминимальности.

## 1.2. Качественное исследование динамических систем

В данном разделе проведем анализ и обоснованный выбор методов качественной теории нелинейной динамики.

Исследование моделей систем, представленных в виде уравнений (1.1) включает поиск решение, т. е. интегрирование системы. Эта цель достижима только для линейных систем с постоянными коэффициентами и для некоторых очень специальных уравнений, которые можно проинтегрировать в квадратурах. Поэтому для многих задач уместно использование нелинейной динамики, которая исследует качественные свойства: устойчивость, количество состояний равновесия, существование периодических траекторий и т. д.

Качественное исследование включает два этапа:
- определение возможных типов траекторий, имеющих различное поведение и «формы»;
- описание для каждой группы топологически схожих траекторий.

Первый этап заключается в определении траектории движения при $(t \to +\infty)$ и при $(t \to -\infty)$. При этом делается положение, что траектория $L$, задаваемая уравнением $x = \varphi(t)$, остаётся в некоторой ограниченной области фазового пространства при $t \geq t_0 (t \leq t_0)$.

Для полноты изложения имеет смысл введение следующих понятий.

Точку $x^*$ называют $\omega$-предельной точкой траектории $L$, если для последовательности $\{t_k\}$, где $t_k \to +\infty$,
$$\lim_{k \to \infty} \varphi(t_k) = x^*.$$

Точку $x^*$ называют $\alpha$-предельной, если $t_k \to -\infty$ при $k \to \infty$.

Обозначим множество всех $\omega$-предельных точек, принадлежащих траектории $L$, через $\Omega_L$, а множество $\alpha$-предельных точек — через $A_L$. Состояние равновесия является единственной предельной точкой самого себя. В случае если траектория $L$ — периодическая, все её точки являются $\alpha$- и $\omega$-предельными: $L = \Omega_L = A_L$. Если $L$ — незамкнутая устойчивая по Пуассону траектория, то множества $\Omega_L$ и $A_L$ совпадают с её замыканием $\bar{L}$. Множество $\bar{L}$ является либо минимальным (если $L$ — рекуррентная траектория), либо квазиминимальным множеством, если времена возвращения Пуанкаре траектории $L$ не ограничены. Все состояния равновесия, а также периодические и устойчивые по Пуассону траектории самопредельны [66].

Структура множеств $\Omega_L$ и $A_L$ подробно исследована для двумерных систем [4, 5] на плоскости. Пуанкаре и Бендиксон [13] установили, что множество $\Omega_L$ может быть только одного из трёх приведённых ниже топологических типов:
(1) состояние равновесия;
(2) периодическая траектория;
(3) контур, образованный состояниями равновесия и траекториями, стремящимся к данным состояниям равновесия при $t \to \pm\infty$.

Используя указанную общую классификацию, можно перечислить все типы положительных полутраекторий систем на плоскости:
- состояния равновесия;
- периодические траектории;
- полутраектории, стремящиеся к состоянию равновесия;
- полутраектории, стремящиеся к периодической траектории;
- полутраектории, стремящиеся к предельному множеству типа (3).

Для многомерных систем классификация значительно сложнее, т. к. кроме состояний равновесия и периодических траекторий, предельные множества могут быть минимальными или квазиминимальными множествами различных топологических типов — таких, например, как странные аттракторы, которые могут быть гладкими или негладкими многообразиями или фрактальными множествами с локальной структурой прямого произведения диска на канторово множество, и даже еще более экзотическими множествами [143].

В качественной теории динамических систем [10] для построения фазовых портретов используется понятие топологической эквивалентности.

Как известно [18], две системы называются топологически эквивалентными, если существует гомеоморфизм соответствующих фазовых пространств, отображающий траектории одной системы в траектории второй. Следовательно, состояния равновесия, а также периодические и незамкнутые траектории одной системы соответственно отображаются в состояния равновесия, периодические и незамкнутые траектории другой системы.

Понятие топологической эквивалентности двух динамических систем определяет методы разбиения фазового пространства на области существования траекторий различных топологических типов. Такие структуры должны быть инвариантны относительно всех возможных гомеоморфизмов фазового пространства.

Пусть $N$ — ограниченная область фазового пространства, а $H = \{h_i\}$ — множество гомеоморфизмов в $N$. Вводя метрику
$$\mathrm{dist}(h_1, h_2) = \sup_{x \in N} \| h_1 x - h_2 x \|,$$
получим, что траекторию $L\,(L \subset N)$ можно назвать особой, если для достаточно малого значения $\varepsilon < 0$ для всех гомеоморфизмов $h_i$, переводящих траектории в траектории и удовлетворяющих условию $dist(h_i, I) < \varepsilon$, где $I$ — тождественное отображение, выполняется условие

$$h_i L = L.$$

Все изолированные состояния равновесия и периодически траектории являются особыми траекториями. Незамкнутые траектории также могут быть особыми. Например, все траектории двумерной системы, стремящиеся к седловым состояниям равновесия при $t \to +\infty$ при $t \to -\infty$, являются особыми. Такие траектории называют сепаратрисами [99].

Будем использовать следующее понятие топологической эквивалентности [143]. Две траектории $L_1$ и $L_2$ эквивалентны, если любого $\varepsilon > 0$ существуют такие переводящие траектории в траектории гомеоморфизмы $h_1, h_2, ..., h_{m(\varepsilon)}$, что
$$L_2 = h_{m(\varepsilon)} \cdots h_1 L_1,$$
где $\mathrm{dist}(h_k, I) < \varepsilon\ (k = \overline{1, m(\varepsilon)})$.

Заметим, что метод исследования, изложенный в классической монографии [5] не применим для общего случая многомерных систем. Множество особых траекторий в трехмерной системе может быть бесконечным или континуальным. Таким образом, решение задач нахождения полного топологического инварианта не является реальным.

## 1.3. Топологическая классификация грубых состояний равновесия

Для построения методов моделирования приведем сведения о построении топологически эквивалентных линейных системах и классификации грубых состояний нелинейных систем. Приведенные исследования соответствуют [143] и приводятся для обоснованности разработанных методов моделирования.

Поскольку разработанный геометрический метод моделирования использует модели вида
$$\dot{x} = Ax + g(x), \qquad (1.11)$$
рассмотрим соответствующую линеаризованную систему
$$\dot{x} = Ax. \qquad (1.12)$$
В соответствии с качественной нелинейной теорией опишем технику оценки топологической эквивалентности системы поведения системы (1.11) поведению линеаризованной системы (1.12).

Как известно, две $n$-мерные системы
$$\dot{x} = F_1(x) \text{ и } \dot{x} = F_2(x),$$
определенные в областях $D_1$ и $D_2$, соответственно, топологически эквивалентны в подобластях $U_1 \subseteq D_1$ и $U_2 \subseteq D_2$, если существует гомеоморфизм
$$\eta : U_1 \to U_2,$$
под действием которого, траектория (полутраектория, отрезок траектории) первой системы отображается в траекторию (полутраекторию, отрезок траектории) второй системы, с сохранении ориентации (направления движения).

Заметим, что эквивалентности нелинейной системы (1.11) и линеаризованной (1.12) в состоянии равновесия не имеет смысла, если есть хотя бы один характеристический показатель на мнимой оси. Таким образом, вопрос о топологической эквивалентности в окрестности негрубого состояния равновесия не имеет смысла.

Приведем два важных примера из [143], которые иллюстрируют это утверждение и демонстрируют основные типы траекторий для случая систем на плоскости.

Рассмотрим нелинейную систему
$$\begin{aligned}\dot{x} &= \omega y + g_1(x, y),\\ \dot{y} &= \omega x + g_2(x, y),\end{aligned} \qquad (1.13)$$

где функции $g_1$ и $g_2$, а также их первые производные обращаются в нуль в начале координат. Состояние равновесия имеет пару чисто мнимых показателей $\lambda_{1,2} = \pm i\omega$ ($\omega > 0$).

Общее решение соответствующей линеаризованной системы имеет вид:
$$x = x_0 \cos(\omega t) - y_0 \sin(\omega t),$$
$$y = y_0 \cos(\omega t) - x_0 \sin(\omega t),$$
фазовые траектории которой являются замкнутыми кривыми, в центре которых лежит начало координат (рис. 1.1). Такое состояние равновесия называется центром.

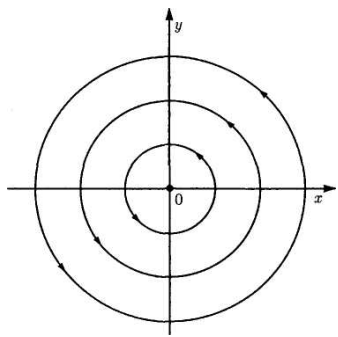

Рис. 1.1. Состояние типа «центр»

Для нелинейной системы фазовый портрет может значительно отличаться от заданного. Например, если $g_1 = -x(x^2 + y_2)$, $g_2 = -y(x^2 + y_2)$, то общее решение уравнения (1.13) в полярных координатах имеет вид:
$$r^2 = \frac{1}{2t + r_0^{-2}}, \quad \varphi = \omega t + \varphi_0.$$

В данном случае все фазовые траектории принадлежат к типу «седло» (см. рис. 1.2).

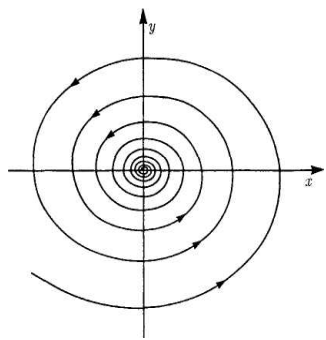

Рис. 1.2. Пример фазового портрета нелинейной системы.

Таким образом, в любой малой окрестности такого состояния равновесия невозможно отыскать гомеоморфизм, при помощи которого траектории данной системы отображаются в траектории линеаризованной системы, в связи с тем, что гомеоморфизм отображает замкнутые кривые в замкнутые кривые. Заметим, что в случае систем с управлением, рассматривая семейство систем с независимой управляющей функцией, преобразование может быть найдено в заданном классе.

Рассмотрим систему вида

$$\dot{x} = g_1(x, y),$$
$$\dot{y} = \lambda y + g_2(x, y). \qquad (1.14)$$

При условии, что один характеристический показатель $\lambda_1$ равен нулю, а второй отрицательный. В (1.14) предполагается, что функции $g_1$ и $g_2$ вместе со своими первыми производными обращаются в нуль в начале координат.

По решению линеаризованной

$$x = x_0, \quad y = e^{-\lambda t} y_0.$$

Построим фазовый портрет, показанный на рис. 1.3. Ось $Ox$ целиком состоит из состояний равновесия линеаризованной системы, каждое из которых притягивает только пару траекторий. Очевидно, что нелинейная система может содержать континуум состояний равновесия только при очень специальном выборе функций $g_1$ и $g_2$. Следовательно, между исходной и линеаризованной системами топологической эквивалентности не существует.

На рис. 1.4 показан фазовый портрет для случая, когда $g_1 = x^2$ и $g_2 = 0$. Из рисунков видно, что два локальных фазовых портрета (так называемый «седло-узел») не могут быть топологически эквивалентными.

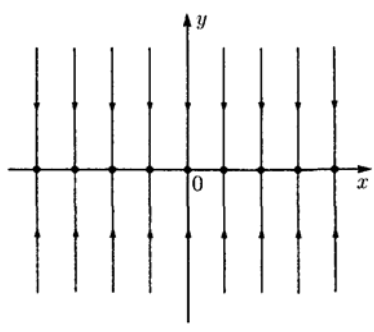

Рис. 1.3. Пример фазовых траекторий линеаризованной системы.

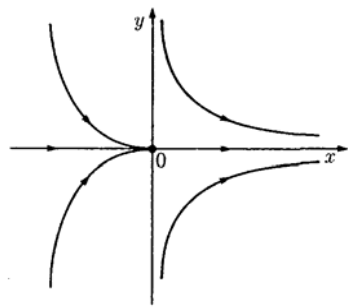

Рис. 1.4. Негрубая точка типа «седло-узел».

Задача исследования локальной топологической эквивалентности грубых состояний равновесия сформулирована в теореме Гробмана-Хартмана [191], которую можно изобразить следующей коммутативной диаграммой:

$$\begin{array}{ccc} f: & U_1 \to U_2 \\ & {}^h\downarrow \quad \downarrow^h \\ Df_0: & V_1 \to V_2 \end{array}.$$

Здесь $f$ непрерывное дифференцируемое отображение открытого множества $U \subset \mathbb{R}^n$ в $\mathbb{R}^n$; $U_1$, $U_2$, $V_1$, $V_2$ — окрестности гиперболической неподвижной точки $O \in U$.

Таким образом, существуют окрестности $U_1$ и $U_2$, в которых исходная и линеаризованная системы топологически эквивалентны.

Изложим результаты о топологической эквивалентности линейных систем. Топологический тип грубого состояния равновесия определяется двумя числами $(k, n-k)$, где $k$ — количество характеристических показателей, лежащих слева от мнимой оси, а $(n-k)$ — справа от нее.

Известно [18], что линейные системы с состояниями равновесия одно типа топологически эквивалентны. Использование этого результата конструктивно в том смысле, что гомеоморфизм $\eta: \mathbb{R}^n \mapsto \mathbb{R}^n$ можно построить в явном виде.

Например, для двух линейных систем, первая из которых в начале координат имеет фокус:

$$\begin{aligned} \dot{x} &= -x + y, \\ \dot{y} &= -x - y, \end{aligned} \quad (1.15)$$

а вторая — узел

$$\begin{aligned} \dot{x} &= -x, \\ \dot{y} &= -\frac{1}{3}y. \end{aligned} \quad (1.16)$$

Эти системы топологически эквивалентны, так как гомеоморфизм

$$(x, y) \mapsto (x\cos(\tau) - y^3 \sin(\tau), y^3 \cos(\tau) - x\sin(\tau)),$$

где $\tau(x,y) = -\ln(x^2 + y^6)/2$, отображает траектории системы (1.16) в траектории системы (1.15).

Следовательно, $n$-мерная система может иметь только $(n+1)$ различных топологических типов грубых состояний равновесия. В частности, любая система с грубым состоянием равновесия типа $(k, n-k)$ локально топологически эквивалентна системе

$$\dot{x} = A_k x, \quad (1.17)$$

где

$$A_k = \begin{pmatrix} -I_k & 0 \\ 0 & I_{n-k} \end{pmatrix};$$

$I_i$ — единичная матрица размерности $i$.

Вводя обозначения

$$x = \begin{pmatrix} s \\ z \end{pmatrix},$$

где $s \in \mathbb{R}^k$, $z \in \mathbb{R}^{n-k}$, система (1.17) преобразуется в вид:

$$\dot{s} = -z \qquad (1.18)$$
$$\dot{z} = s,$$

с общим решением

$$s(t) = e^{-I_k t} s_0, \qquad (1.19)$$
$$z(t) = e^{I_{n-k} t} z_0.$$

Рассмотрим $k = n$. При $t \to +\infty$ все траектории системы (1.19) стремятся к состоянию равновесия, т. е. любая траектория из достаточно малой окрестности состояния равновесия типа ($n$, 0) нелинейной системы также стремится к состоянию равновесия. Такое состояние равновесия принято называть устойчивым топологическим узлом, или стоком. В условиях, когда траектория, начинающаяся в малой окрестности состояния равновесия типа (0, $n$), стремится к нему при $t \to -\infty$, и выходящая за окрестность при $t \to +\infty$, состояние равновесия называется неустойчивым топологическим узлом, или источником.

Остальные грубые состояния равновесия являются топологическими седлами. Из теоремы Гробмана-Хартмана следует, что топологическое седло исходной нелинейной системы имеет локально устойчивое и локально неустойчивое многообразия $W_{loc}^*$ и $W_{loc}^0$ размерности $k$ и ($n-k$), соответственно. Таким образом, если $h$ — локальный гомеоморфизм, под действием которого траектории линеаризованной системы отображаются на траектории нелинейной системы, то образы $hE^*$ и $hE^0$ устойчивого и неустойчивого инвариантных подпространств линеаризованной системы как раз являются устойчивыми и неустойчивыми многообразиями. Аналогично линейному случаю, положительная полутраектория, выходящая из любой точки многообразия $W_{loc}^*$, лежит в нем полностью и стремится к состоянию равновесия $O$ при $t \to +\infty$. Подобным образом и отрицательная полутраектория, начинающаяся в любой точке многообразия $W_{loc}^0$, лежит в нем полностью и стремится к состоянию равновесия $O$ при $t \to -\infty$. Траектории точек вне $W_{loc}^* \bigcup W_{loc}^*$ покидают любую окрестность седла при $t \to \pm\infty$. Многообразия $W_{loc}^*$ и $W_{loc}^0$ являются инвариантными, т. е. включают в себя все траектории (до тех пор, пока траектории остаются в некоторой окрестности топологического седла).

Очевидно, что если две системы $X_1$ и $X_2$ топологически эквивалентны, то при помощи гомеоморфизма, осуществляющего топологическую эквивалентность, состояния равновесия системы $X_1$ отображаются на состояния равновесия системы $X_2$. Если $O_1$ — состояние равновесия системы, а $O_2$ — образ $O_1$ относительно гомеоморфизма, то траектория, асимптотическая к $O_1$ при $t \to +\infty$ (или, соответственно, $t \to -\infty$), отображается в траекторию, асимптотическую к $O_1$ при $t \to +\infty$ ($t \to -\infty$). Следовательно, устойчивые (неустойчивые) многообразия локально топологически эквивалентных седел имеют равные размерности. Таким образом, для цели исследования важно, что в соответствии с [143] два грубых состояния равновесия локально топологически эквивалентны тогда и только тогда, когда они принадлежат одному и тому же топологическому типу.

Топологический подход позволил решить задачу классификации грубых состояний равновесия. Отметим, что метод не позволяет определить гладкость инвариантных многообразий, поэтому этот вопрос требует дополнительных исследований.

В заключении приведем результат о экспоненциальной устойчивости узла. В [143] показано, что при малом $\delta > 0$ для любого $x_0$ ($\|x_0\| < \delta$) траектория $x(t)$ системы (1.11) с начальной точкой $x_0$ при любых значениях $t \geq 0$ удовлетворяет неравенству
$$\|x(t)\| \leq C e^{(\max \operatorname{Re} \lambda_i + \varepsilon)t} \|y_0\|,$$
где $\varepsilon(\delta) > 0$ — константа; $C > 0$ — множитель, зависящий от выбора базиса в $\mathbb{R}^n$.

единственности решений, т. е. через начальную точку $(t_0, x_0)$, принадлежащую области, проходит единственное решение

$$x = \varphi(t, t_0, x_0). \qquad (2.2)$$

Пусть для вектор-функции $f$ выполняется тождество:

$$\varphi(t_0, t_0, x_0) = x_0,$$

из которого следует равенство

$$\det \left\| \frac{\partial \varphi^i(t, t_0, x_0)}{\partial x_0^k} \right|_{t=t_0} \right\| = 1,$$

а на некотором интервале $t \in [t_0, t_1]$ в некоторой окрестности начальной точки $(t_0, x_0)$ справедливо условие:

$$\det \left\| \frac{\partial \varphi^i(t, t_0, x_0)}{\partial x_0^k} \right\| \neq 0.$$

Инфинитезимальная образующая однопараметрической группы преобразований, представляющая в данном случае оператор дифференцирования по $t$, имеет вид [57]:

$$X_0 = \frac{\partial}{\partial t} + \sum_{k=1}^{n} f^k(t, x) \frac{\partial}{\partial x^k}. \qquad (2.3)$$

Оператор (2.3) группы удовлетворяет системе уравнений Ли:

$$\begin{cases} \dfrac{dt}{da} = 1, & t(0) = t_0, \\ \dfrac{dx}{da} = f(t, x), & x(0) = x_0. \end{cases}$$

Здесь $a$ — независимая переменная (групповой параметр). [Примеры использования см. 60, 77].

Группу сдвига пространства $\mathbb{R}^{n+1}(t, x)$ вдоль решений системы (2.1) образуют преобразования вида

$$\begin{aligned} t &= t_0 + a, \\ x &= \varphi(t_0 + a, t_0, x_0). \end{aligned} \qquad (2.4)$$

с операцией коммутирования операторов (скобки Ли), причем, единицей группы является тождественное преобразование. Функция $\varphi(\cdot)$ определена общим решением (2.2) системы (2.1). Таким образом, группа (2.4) есть однопараметрическое семейство преобразований, переносящих каждую точку $(t_0, x_0)$ вдоль проходящего через эту точку решения в текущую точку $(t, x)$. Из (2.4) следует, что групповой параметр $a$ совпадает с интервалом $a = t - t_0$ изменения независимой в системе (2.1) переменной $t$.

Группа сдвигов является практической значимой в рамках рассматриваемого в работе класса динамических систем с нелинейными колебаниями (например, группами Ли являются группы изометрий на плоскости). Известно [260–263], что группами являются преобразования трансляции на компактных коммутативных группах и линейные потоки на торе. Как уже отмечалось, колебательные системы, допускают преобразования сдвига, а для хаотических систем со странным аттрактором можно считать, что имеет место так называемое слабое нарушение симметрии [134].

Для автономного случая ($\partial f/\partial t = 0$):

$$\dot{x} = f(x), \quad x \in \mathbb{R}^n, \qquad (2.5)$$

общее решение (2.2) имеет вид [105]:

$$x = \varphi(t - t_0, x_0),$$

соответствующая группа сдвигов (2.4) —

$$t = t_0 + a,$$
$$x = \varphi(a, x_0).$$

В автономном случае последняя группа есть прямое произведение групп, одна из которых преобразует переменные $t$, другая — переменную $x$ независимо друг от друга. Следовательно, множество преобразований $x = \varphi(a, x_0)$ само по себе является группой, порожденной оператором

$$X_0 = \frac{\partial}{\partial t} + \sum_{k=1}^{n} f_k(t, x) \frac{\partial}{\partial x^k}.$$

В частности, для линейных систем

$$\dot{x} = A(t)x, \quad x \in \mathbb{R}^n$$

с общим решением (2.2)

$$x = \Phi(t, t_0) x_0,$$

группа сдвигов определяется следующими преобразованиями вдоль решений:

$$t = t_0 + a,$$
$$x = \Phi(t_0 + a, t_0) x_0.$$

Здесь $\Phi(t, t_0)$ — переходная матрица системы. Для стационарной системы ($A = \text{const}$) общее решение

$$x = e^{A(t-t_0)} x_0$$

определяет группу:

$$t = t_0 + a,$$
$$x = e^{Aa} x_0.$$

Таким образом, определены группы симметрий, определяющие основные преобразования решений колебательных систем.

## 2.2. Методика группового анализа систем с управлением

### 2.2.1. Применение классификации систем, допускающих группы симметрий для анализа решений

Рассматривается техника вычисления первых интегралов и классификация групп, предложенная Г. Н. Яковенко [148–151, 95, 81]. Данная классификация определяет методы построения алгебр Ли и методики группового анализа.

Рассмотрим динамическую систему с управлением:

$$\dot{x} = f(t, x, u), \quad x \in \mathbb{R}^n, u \in U \subset \mathbb{R}^r, \qquad (2.6)$$

в D–области пространства $\mathbb{R}^n(t, x)$, где $x = (x_1, ..., x_n)$ — вектор состояний динамической системы; $u = (u_1, ..., u_r)$ — управление, $u \in U \subset \mathbb{R}^r$; $t$ — время; считаем выполненные допущения по гладкости вектор-функции $f$, сформулированные в п. 1.2.

Процедура группового анализа системы (2.6) в соответствии с классом систем состоит из двух этапов: выделение базисных операторов и пополнение этой системы операторов.

Г. Н. Яковенко [148] выделены следующие классы систем, допускающих группы симметрий.

<u>1. B-система</u>, определяется базисом операторов, получающихся из оператора полного дифференцирования по времени $t$ для системы (2.6):

$$X(u) = \frac{\partial}{\partial t} + \sum_{i=1}^{n} f_i(t, x, u) \frac{\partial}{\partial x^i} \qquad (2.7)$$

при различных допустимых значений управлений $u \in U$,

$$X_j = \frac{\partial}{\partial t} + \sum_{i=1}^{n} f_i^j(t, x, u) \frac{\partial}{\partial x^i}, \quad j = \overline{1, p}, \qquad (2.8)$$

линейно несвязанны [109]. Здесь обозначено где $f_i^j(t, x, u) = f_i(t, x, u_j)$. Подстановка в (2.7) любого другого допустимого управления приводит оператору, который линейно связанно выражается через операторы $X_0, X_1, ..., X_p$:

$$X(u) = \sum_{i=1}^{p} f^i(t, x, u) X_j. \qquad (2.9)$$

<u>2. F-система</u>, определяется полная B-системой операторов (2.8), и операторов, являющихся коммутаторами [·, ·] операторов (2.8), вычисленных в процессе пополнения после выделения базиса (2.8) [105]:

$$X_k = \sum_{i=1}^{n} f_k^i(t, x) \frac{\partial}{\partial x^i}, k = \overline{p+1, m}, \qquad (2.10)$$

при этом матрица коэффициентов операторов $X_0, X_1, ..., X_p, X_{p+1}, ..., X_m$ имеет вид:

$$\Theta = \begin{Vmatrix} 1 & \cdots & 1 & 0 & \cdots & 0 \\ f_0^1 & \cdots & f_p^1 & f_{p+1}^1 & \cdots & f_m^1 \\ \vdots & \ddots & \vdots & \vdots & \ddots & \vdots \\ f_0^n & \cdots & f_p^n & f_{p+1}^{n1} & \cdots & f_m^n \end{Vmatrix}.$$

Отметим, что согласно [105] для полной системы выполняется

$$\operatorname{rank} \Theta = m + 1. \qquad (2.11)$$

Описанная процедура вычисления $F$-системы в точке $t, x$ — корректна и содержит конечное число шагов: из построения и из условия (2.11) следует $p \le m \le n$. При разных же значениях $t, x$ разные управления $u$ могут создавать B-систему (2.8), и к операторам (2.10) могут приводить разные последовательности вычисления коммутаторов, вследствие чего, в частности, возможна потеря непрерывности у элементов $f_k^i(t,x)$ матрицы $\Theta$.

Заметим, что можно выделить класс регулярных в D-области систем [41], определяемый условиями: (1) один и тот же набор постоянных допустимых управлений $u_0, ..., u_p$ выделяет B-систему (2.8); (2) одна и та же последовательность коммутаторов приводит к F-системе $X_0, X_1, ..., X_p, X_{p+1}, ..., X_m$, содержащей одно и то же количество $m$ операторов.

3. <u>Ft-система</u> получается, как результат

$$\frac{\partial}{\partial t}, X_0, X_1, ..., X_m, X_{m+1}, ..., X_M$$

пополнения системы операторов $\partial/\partial t, X_0, X_1, ..., X_m$, где $X_0, X_1, ..., X_m$ — операторы F-системы.

В рамках группового анализа введем для (2.6) понятия решения и первого интеграла [51, 109].

Решением системы (2.6) является пара вектор-функций $x(t), u(t)$ ($u(t)$ — допустимое управление), которая при подстановке ее в систему (2.6) приводит к тождеству.

Как известно, первым интегралом системы с управлением (2.6) называется функция $w(t, \mathrm{x})$, которая на любом решении $x(t)$, $u(t)$ системы (2.6) сохраняет постоянное значение [37]

$$w(t, x(t)) = w(t_0, x_0) = const.$$

При этом стационарным первым интегралом системы с управлением (2.6) называется первый интеграл $w(x)$, не зависящий от времени $t$.

Интегральным базисом первых интегралов (стационарных первых интегралов) является такой набор функционально независимых первых интегралов $w^1(t,x), ..., w^Q(t,x)$ $(w^1(x), ..., w^q(x))$, что для любого другого первого интеграла $w(t,x)$ $(w(x))$ справедливо с некоторой функцией $F(\cdot, ..., \cdot)$ равенство

$$w(t,x) = F(w^1(t,x), ..., w^Q(t,x)) \quad (w(x) = F(w^1(x), ..., w^q(x))).$$

Построение функций $w(t,x)$, являющихся первым интегралом системы (2.6) определяется решением полной системы [41, 105]

$$X_0 w = 0, \ X_1 w = 0, ..., X_m w = 0, \quad (2.12)$$

где $X_0, X_1, ..., X_m$ — F-система (2.8), (2.10), соответствующая регулярной системе (2.6). Интегральный базис первых интегралов состоит из $n - m$ функций.

Отсюда следует, что нетривиальные первые интегралы, т. е. те, для которых выполнено условие

$$\sum_{i=1}^{n}\left(\frac{\partial w(t,x)}{\partial x^i}\right)^2 \neq 0.$$

отсутствуют, если количество уравнений в полной системе (2.12) равно $n+1$, где $n$ — размерность регулярной системы (2.6).

Стационарный первый интеграл системы (2.6) строится как решение полной системы [150]

$$\frac{\partial w}{\partial t} = 0; \ X_0 w = 0, ..., X_p w = 0, X_M = 0, \quad (2.13)$$

где $\partial w/\partial t, X_0, ..., X_M$ — Ft-система: результат пополнения системы $\partial w/\partial t, X_0, ..., X_m$ ($X_0, ..., X_m$ — F-система). Интегральный базис стационарных первых интегралов содержит $n - M$ функций.

Так как система (2.12) входит в систему (2.13), любое решение системы (2.13) является первым интегралом. Первое уравнение $\partial w/\partial t = 0$ в (2.13) гарантирует независимость решений от $t$. Количество $n - M$ функционально независимых первых интегралов следует из полноты системы (2.13) [106].

Известно, что [149], что если для F-системы нетривиальные первые интегралы отсутствуют, то система управляема.

В целях практического анализа рассмотрим систему с периодической траекторией ($x, y, u$ — скаляры) [148]:

$$\dot{x} = \sin u + x \cos u,$$
$$\dot{y} = \cos u, \quad u \in U = [0, 2\pi).$$

Системе соответствует оператор (2.7)

$$X(u) = \frac{\partial}{\partial t} + (\sin u + x \cos u)\frac{\partial}{\partial x} + \cos u \frac{\partial}{\partial y}.$$

В условиях, когда множество $U$ состоит, по крайней мере, из двух разных управлений, возможны два альтернативных варианта.

I. Множеству $U$ принадлежат три разных управления $u_0, u_1, u_2$.

Для определенности с учетом $X(u)$ предполагаем
$$0 \le u_0 < u_1 < u_2 < 2\pi.$$
Подстановка $u_0, u_1, u_2$ приводит к операторам:
$$X_0 = X(u_0) = \frac{\partial}{\partial t} + (\sin u_0 + x \cos u_0)\frac{\partial}{\partial x} + \cos u_0 \frac{\partial}{\partial y},$$
$$X_1 = X(u_1) = \frac{\partial}{\partial t} + (\sin u_1 + x \cos u_1)\frac{\partial}{\partial x} + \cos u_1 \frac{\partial}{\partial y},$$
$$X_2 = X(u_2) = \frac{\partial}{\partial t} + (\sin u_2 + x \cos u_2)\frac{\partial}{\partial x} + \cos u_2 \frac{\partial}{\partial y}.$$

Вычисления для матрицы $\Theta$, составленной из коэффициентов операторов $X_0, X_1, X_2$, дают результат
$$\det \Theta = 4 \sin \frac{u_2 - u_1}{2} \sin \frac{u_1 - u_0}{2} \sin \frac{u_0 - u_2}{2}.$$

При условии $\det \Theta \ne 0$ операторы $X_0, X_1, X_2$ — линейно несвязанны, а так как их количество совпадает с размерностью пространства $\mathbb{R}^3(t, x, y)$, то при любом другом управлении $u \in U$ оператор $X(u)$ линейно связанно выразится через операторы $X_i$, $i = 0, 1, 2$. Через них выразится и каждый коммутатор $[X_i, X_j]$, $i, j = 0, 1, 2$. Таким образом, если множество $U$ содержит не менее трех управлений, то система является регулярной во всем пространстве $\mathbb{R}^3(t, x, y)$, а система операторов (2.60) — B-система и F-система. Первые интегралы в этом случае отсутствуют. В $D$-области возможна локальная управляемость.

II. Пусть множество $U$ допустимых значений состоит из двух управлений $u_0, u_1$, и соответствующие операторы $X_0, X_1$ линейно несвязанны — являются B-системой.

Коммутатор:
$$[X_0, X_1] = \sin(u_0 - u_1)\frac{\partial}{\partial x}.$$

Для матрицы $\Theta_1$, составленной из коэффициентов операторов $X_0, X_1$ и $[X_0, X_1]$, выполняется
$$\det \Theta_1 = 2 \sin^2 \frac{u_0 - u_1}{2} \sin \frac{u_0 + u_1}{2} \cos \frac{u_0 - u_1}{2}.$$

Справедливо неравенство
$$\sin \frac{u_0 - u_1}{2} \ne 0,$$
поэтому возможны только два случая, приводящие к вырожденности матрицы $\Theta_1$ и к тому, B-система $X_0, X_1$ есть F-система.

а) Случай
$$\sin \frac{u_0 + u_1}{2} = 0,$$

реализуемый множеством

$$U = \{u_0 = \alpha \neq 0;\ u_1 = 2\pi - \alpha\}.$$

Системе соответствует B-система, являющаяся F-системой:

$$X_0 = \frac{\partial}{\partial t} + (\sin\alpha + x\cos\alpha)\frac{\partial}{\partial x} + \cos\alpha\frac{\partial}{\partial y},$$

$$X_1 = \frac{\partial}{\partial t} + (-\sin\alpha + x\cos\alpha)\frac{\partial}{\partial x} + \cos\alpha\frac{\partial}{\partial y}.$$

Решив систему $X_0 w = 0, X_1 w = 0$, приходим к гарантированному первому интегралу системы

$$w = y - t\cos\alpha.$$

b) Случай

$$\cos\frac{u_0 - u_1}{2} = 0$$

B-система реализуется множеством

$$U = \left\{u_0 = \beta - \frac{\pi}{2};\ u_1 = \beta + \frac{\pi}{2};\ \frac{\pi}{2} \leq \beta < 3\frac{\pi}{2}\right\}.$$

Системе соответствуют B- и F-системы

$$X_0 = \frac{\partial}{\partial t} - (\cos\beta - x\sin\beta)\frac{\partial}{\partial x} + \sin\beta\frac{\partial}{\partial y},$$

$$X_1 = \frac{\partial}{\partial t} + (\cos\beta - x\sin\beta)\frac{\partial}{\partial x} - \sin\beta\frac{\partial}{\partial y}.$$

Решение полной системы $X_0 w = 0, X_1 w = 0,$ приводит к стационарному первому интегралу

$$w = (\cos\beta - x\sin\beta)e^{-y}.$$

Следовательно, управляемость в любой D-области отсутствует.

### 2.2.2. Методика исследования групповых систем

Для практического использования имеет значение класс систем, восходящий к результатам С. Ли [203]. Для этого класса, несмотря на функциональную мощность множества допустимых управлений $u(t)$, преобразование сдвигов вдоль решений принадлежит конечно-параметрической группе. Такой класс называется групповыми системами и определяется следующим образом [51, 148].

В групповой системе

$$\dot{x}_k = \sum_{l=1}^{r} f_k^l(x) u_l(t), \qquad k = \overline{1, n},\ \ u \in U \subset \mathbb{R}. \qquad (2.14)$$

для функций $f_k^l(x)$ выполнены условия

$$\operatorname{rank}\|f_k^l(x)\| = \min\{n, r\}, \qquad (2.15)$$

$$\left\{\sum_{l=1}^{r} c^l f_k^l(x) = 0, \ c^l = const\right\} \Rightarrow \left\{c^l = 0, \ l = \overline{1, r}\right\},$$

$$[X_i, X_j] = \sum_{l=1}^{r} C_{ij}^k X_k, \ \ C_{ij}^k = const, \ \ i, j, k = l = \overline{1, r}, \qquad (2.16)$$

где обозначено

$$X_l = \sum_{k=1}^{n} f_k^l(x) \frac{\partial}{\partial x^k}, \quad l = \overline{1, r};$$

$[X_i, X_j]$ — коммутатор операторов. Следовательно, операторы $\sum_{k=1}^{r} u_k X_k$, $u^k = const$, есть алгебра Ли с базисом $X_k$ и структурными постоянными $C_{ij}^k$.

Если в (2.14) выполнено $n > r$, то полная система уравнений $X_l w = 0$, $l = \overline{1, r}$, имеет $n-r$ функционально независимых решений $w_1(x), \ldots, w^{n-r}(x)$ — первых интегралов системы (2.14) [30]. При переходе на конкретную инвариантную поверхность $w_1(x) = c^1, \ldots, w^{n-r}(x) = c^{n-r}$ количество уравнений уменьшается до $r$. Далее предполагается $n \le r$.

Каждой групповой системе (2.14) ставится в соответствие $r$–параметрическая группа сдвигов вдоль решений системы

$$x_k = g_k(x_{10}, \ldots, x_{n0}, u_1, \ldots, u_r), \quad k = \overline{1, n}, \qquad (2.17)$$

уравнения, которой по базису соответствующей алгебры Ли можно вычислить, например, следующим способом. Каждому оператору $X_l$ система уравнений $dx_i/dv_j = f_i^j(x)$, $x(0) = x_0$ ставится в соответствие решение $x = g_j(x_{10}, \ldots, xn_0, v_j)$ — однопараметрическую группу преобразований; $r$-параметрическая группа (2.17) — суперпозиция этих групп, уравнения которой вычисляются по (2.14) при $u^l(t) = const$. Группе (2.17) соответствует алгебра Ли с базисом (2.68) и структурными постоянными $C_{ij}^k$, определенными в (2.16).

Сопоставим паре $(u(t), t)$ преобразование пространства $\mathbb{R}^n$ — сдвиги вдоль решений $x(t)$ групповой системы (2.14), в которую подставлены функции $u(t)$: из точек $x(0) = x_0$ в точки $x(t)$. Преобразование $x_0 \leftrightarrow x(t)$ принадлежит группе (2.17), т. е. каждой паре $(u(t), t)$ соответствует такой набор параметров $v_1, \ldots, v_r$, что преобразование (2.17) и преобразование $x_0 \leftrightarrow x(t)$ совпадают.

L-система [148] есть групповая система

$$\dot{x}_k = \sum_{l=1}^{r} f_k^l(x) u_l(t), \qquad k = \overline{1, n}, \ u \in U \subset \mathbb{R} \qquad (2.18)$$

при $r = n$.

Принадлежность к L-системам инвариантна по отношению к неособенному преобразованию $\tilde{x} = \varphi(x)$ переменных состояния. Действительно, в переменных $\tilde{x}$ система (2.18) имеет такую же структуру

$$\dot{\tilde{x}}^k = \sum_{l=1}^{r} \tilde{f}_l^k(\tilde{x}) u^l(t), \qquad k = \overline{1, n}, \qquad (2.19)$$

где обозначено

$$\tilde{f}_l^k(\tilde{x}) = \sum_{i=1}^{n} \frac{\partial f^k(x)}{\partial x^i} f_l^i(x) \bigg|_{x \to \tilde{x}}.$$

Функции $\tilde{f}_l^k(\tilde{x})$ удовлетворяют условиям (2.15)–(2.16), причем с теми же постоянными $C_{ij}^k$ в (2.16), что в переменных $x$. Если системы (2.18), (2.19), связанные неособенным преобразованием $x \leftrightarrow \tilde{x}$, считать эквивалентными, то каждому классу эквивалентности соответствуют постоянные $C_{ij}^k$:

$$\begin{cases} \dot{x}^k = \sum_{l=1}^{r} \tilde{f}_l^k(x) u^l(t), \\ \updownarrow x \leftrightarrow \tilde{x} \\ \dot{\tilde{x}}^k = \sum_{l=1}^{r} \tilde{f}_l^k(\tilde{x}) u^l(t), \end{cases}. \qquad (2.20)$$

Приведенное соответствие является взаимно однозначным: по структурным постоянным $C_{ij}^k$ в некоторых переменных вычисляется базис $X_1, \ldots, X_n$, алгебры Ли, по операторам $X_l$ определяется представитель класса эквивалентности (2.20) с возможностью заменой переменных перейти к другому представителю.

### 2.2.3. Методика исследования симметрий по состоянию

Для исследования моделей нелинейных систем разработаем методику группового анализа систем, допускающих симметрии фазовых траекторий в $D$-области, т. е. однопараметрическую группу преобразований ($\tau$ — групповой параметр):

$$\tilde{x} = \tilde{x}(t, x, \tau), \qquad (2.21)$$

с инфинитезимальным оператором фазовой траектории

$$Y = \sum_{k=1}^{n} \varepsilon_i(t, x) \frac{\partial}{\partial x_k}, \ \varepsilon_j = \frac{\partial \tilde{x}_j(t, x, \tau)}{\partial \tau} \bigg|_{\tau=0}. \qquad (2.22)$$

Здесь переменные $t$ и $u$ преобразуются тождественно; преобразования группы не зависят от управления $u$.

Таким образом, неособенное преобразование $\tilde{x} = \tilde{x}(t, x)$ является преобразованием симметрии D-области фазового портрета (2.21), если замена переменных в (2.21) приводит к системе

$$\frac{d\tilde{x}}{dt} = f(t, \tilde{x}, u), \qquad (2.23)$$

с такими же функциями *f* в правой части, что и у исходной системы (2.21). То есть группу симметрий, допускаемая системой (2.6), образуют однопараметрические преобразования вида (2.21), при этом преобразуются только состояния *x*, принадлежащие D-области фазовой траектории.

Следуя определению, группу симметрий фазовых траекторий можно характеризовать следующим переводом решений в решения:

$$\text{graph } \{x(t), u(t)\} \to \text{graph } \{x(t), u(\text{t})\},$$

т. е. группа (2.21) является группой симметрий для каждой системы обыкновенных дифференциальных уравнений, получающейся подстановкой в (2.23) конкретного значения допустимого управления *u*(*t*).

Для группы симметрий системы обыкновенных дифференциальных уравнений определяются из условий коммутирования

$$\forall u(t), \left[ X(u(t)), Y \right] = 0. \qquad (2.24)$$

Заметим, что условие (2.24) можно получить независимо, если продифференцировать (2.22) при замене переменных (2.21) по групповому параметру τ, положить затем τ = 0 и раскрыть полную производную

$$\frac{d\varepsilon(t, x)}{dt} = X(u(t))\varepsilon_i(t, x).$$

В связи с тем, что в операторе (2.24) отсутствует дифференцирование по времени *t*, поэтому формула, полученная после раскрытия коммутатора в (2.24), будет содержать только значения вектор-функции *u*(*t*) и не содержать значений производных от управляющей функции. То есть условие (2.24) имеет вид:

$$\forall u, \left[ X(u), Y \right] = 0. \qquad (2.25)$$

Здесь использован оператор (2.7) с постоянными управлениями *u*.

Выполнение условий (2.24), (2.25) имеет место для регулярных систем [148], а также при выполнении условий

$$[X_k, Y] = 0, \quad k = \overline{0, m}, \qquad (2.26)$$

$$Yf^j = 0, \qquad j = \overline{0, p}, \qquad (2.27)$$

где $X_k$ — операторы (2.8), (2.10) F-системы, $f^j$ — функции в линейной связи (2.9), *Y* — оператор (2.22) группы (2.21).

Совокупность операторов симметрий по состоянию (2.22) является алгеброй Ли [83]: если операторы $Y_1$ и $Y_2$ удовлетворяют условию (2.24) (или (2.25)), то операторы $aY_1 + bY_2$; (*a*, *b* = const) и $[Y_1, Y_2]$ также удовлетворяют (2.24) ((2.25)).

Обозначим алгебру Ли $A_0$, при этом множество операторов $Y$ симметрий фазовых траекторий, удовлетворяет одному из эквивалентных условий (2.24), (2.25).

В отличие от систем обыкновенных дифференциальных уравнений в нормальном виде, у которых множество операторов симметрий всегда имеет функциональную мощность [38], у системы с управлением (2.23) возможны ситуации:

1) Алгебра $A_0$ состоит из единственного оператора $Y = 0$, а группа симметрий — из тождественного преобразования.

2) Множество операторов симметрий не вмещается ни в одну конечномерную алгебру Ли.

Особое положение занимают системы с пустым интегральным базисом первых интегралов. Предположение об отсутствии у регулярной системы (2.23) нетривиальных первых интегралов можно записать в виде [148]:

$$\{\forall u \in U, X(u(t)) = 0\} \Rightarrow \{w \equiv \text{const}\},$$

где $X(u(t))$ — семейство операторов.

Отметим, что если регулярная система (2.6) не имеет нетривиальных первых интегралов, тогда алгебра $A_0$ операторов $Y$ симметрий по состоянию конечномерна, и ее размерность $q$ не превосходит размерности $n$ пространства состояний системы (2.23): $0 \leq q \leq n$ [148].

Конечномерность алгебры Ли с базисом $Y_k$, $k = \overline{1, q}$, и принадлежность коммутаторов $[Y_i, Y_j]$ базисных операторов к этой алгебре, в частности, означают, что для $[Y_i, Y_j]$ справедливы равенства

$$\left[Y_i, Y_j\right] = \sum_{k=1}^{q} C_{ij}^k Y_k, \ i, j = \overline{1, q},$$

где числа $C_{ij}^k$ — структурные постоянные $q$-мерной алгебры Ли: алгебры $A_0$. По базисным операторам $Y_k$, $k = \overline{1, q}$, алгебры Ли вычисляется $q$-параметрическая группа [105]:

$$\tilde{x} = \tilde{x}(t, x, \tau), \qquad \tau \in \mathbb{R}^q. \tag{2.28}$$

которой принадлежат все однопараметрические группы симметрий по состоянию системы (2.23). Числа $C_{ij}^k$ являются структурными постоянными и для группы (2.28).

Размерность $q$ алгебры $A_0$ и соответствующих базисных операторов $Y_k$, $k = \overline{1, q}$, определяются характеристиками систем (2.26), (2.27). Регулярные системы (2.23) с нетривиальной алгеброй $A_0$; ($\dim A_0 \geq 1$) обладают свойством декомпозиции системы (2.6).

Согласно [149], регулярная система (2.23) допускает $q$-параметрическую группу симметрий по состоянию (2.28) (q $\leq n$) с операторами $Y_k$, $k = \overline{1,q}$, в том и только в том случае, если неособенным преобразованием $x \leftrightarrow y, z$ системе (2.23) можно придать вид

$$\dot{y} = \tilde{f}_i(t, y, u), \ i = \overline{1, n-q},$$
$$\dot{z}_k = \sum_{l=1}^{q} f_{lk}(z) f_l(t, y, u), \ k = \overline{1, q}, \quad (2.29)$$

причем для квадратной матрицы $\| f_{lk}(z) \|$ необходимо выполнение

$$\det \| f_{lk}(z) \| \neq 0,$$

$$\left[ Z_i, Z_j \right] = \sum_{k=1}^{n} C_{ij}^{k} Z_k, \qquad i, j = \overline{1, q},$$

где

$$Z_l = \sum_{k=1}^{q} f_{lk}(z) \frac{\partial}{\partial z_k}, \ l = \overline{1, q}.$$

Числа $C_{ij}^{k}$ совпадают со структурными постоянными группы симметрий (2.28) и соответствующей алгебры $A_0$.

Приведение системы (2.23) к указанному виду (2.29) эквивалентно процессу приведения преобразованием $x \leftrightarrow y, z$ базисных операторов $Y_j$, $k = \overline{1, q}$, алгебры $A_0$ к виду:

$$Y_l = \sum_{k=1}^{q} \varepsilon_l^k(z) \frac{\partial}{\partial z^k}, \ l = \overline{1, q}, z \in \mathbb{R}^q.$$

Таким образом, методику группового анализа нелинейных управляемых систем, допускающих симметрии локальных областей фазовых траекторий можно представить в виде алгоритма, состоящего из следующих шагов.

*1. Вычисление инфинитезимального оператора (2.7).*

*2. Вычисление операторов (2.8) В-системы.*

*3. Составление матрицы из коэффициентов операторов $X_i$; вычисление определителя матрицы.*

*4. Дополнение базиса операторов до F-системы.*

*5. Проверка конечнопараметричности группы (2.28).*

*6. Вычисления коэффициентов оператора симметрий:*

$$Y = \theta(t, x, y) \frac{\partial}{\partial x} + \mu(t, x, y) \frac{\partial}{\partial y},$$

*соответствующего (2.25).*

*7. Построение эквивалентной системы.*

*8. Вычисление решений эквивалентной системы.*

*9. Построение группы симметрий фазовых траекторий 2.28.*

*10. Анализ управляемости и возможности построения моделей по наблюдаемым частным решениям.*

Рассмотрим 2 примера, иллюстрирующие описанную технику.

**Пример 1.** Рассмотрим систему
$$\dot{x} = ux + v + by^2,$$
$$\dot{y} = u, \tag{2.30}$$

где $x$, $y$ — переменные состояния, $u$, $v$ — управления, $b = \text{const}$. Системе соответствует оператор (2.7):

$$X(u, v) = \frac{\partial}{\partial t} + (ux + v + by^2)\frac{\partial}{\partial x} + u\frac{\partial}{\partial y}$$

и операторы (2.8) B-системы

$$X_0 = X(0, 0) = \frac{\partial}{\partial t} + (by^2)\frac{\partial}{\partial x},$$

$$X_1 = X(1, 0) = \frac{\partial}{\partial t} + (x + by^2)\frac{\partial}{\partial x} + \frac{\partial}{\partial y},$$

$$X_2 = X(0, 1) = \frac{\partial}{\partial t} + (1 + by^2)\frac{\partial}{\partial x}.$$

Определитель матрицы, составленной из коэффициентов операторов $X_i$, при любых значениях параметров $a$ и $b$ равен 1.

Условие (2.9) выполняется в случае:
$$f_0 = 1 - u - v,$$
$$f_1 = u,$$
$$f_2 = v.$$

Заметим, что B-система $X_i$, является и F-системой, т. к. операторы $X_i$, линейно выражается любой оператор $X(u, v)$, а также каждый оператор в пространстве переменных $t, x, y$, в том числе коммутаторы операторов $X_i$. Нетривиальные первые интегралы у системы (2.30) отсутствуют, система (2.30) допускает конечномерную алгебру $A_0$ симметрий (конечно-параметрическую группу (2.28)).

Для вычисления коэффициентов $\theta(t, x, y)$, $\mu(t, x, y)$ оператора симметрий (2.22):

$$Y = \theta(t, x, y)\frac{\partial}{\partial x} + \mu(t, x, y)\frac{\partial}{\partial y},$$

преобразуем его к эквивалентной системе операторов

$$X_0^* = X_0 \quad = \frac{\partial}{\partial t} + by^2\frac{\partial}{\partial x},$$

$$X_1^* = X_1 - X_0 = x\frac{\partial}{\partial x} + \frac{\partial}{\partial y},$$

$$X_2^* = X_2 - X_0 = \frac{\partial}{\partial x}.$$

Используем полученные уравнения, а также тот факт, что (2.26), (2.27) для коэффициентов выполняются тождественно, т. е. условий на функции

θ(*t*, *x*, *y*), μ(*t*, *x*, у) не накладывают. Одно из уравнений $[X_2^*, Y] = 0$ системы (2.26) имеет вид:

$$\frac{\partial \theta}{\partial x}\frac{\partial}{\partial x} + \frac{\partial \mu}{\partial x}\frac{\partial}{\partial y} = 0,$$

что эквивалентно системе относительно коэффициентов оператора

$$\frac{\partial \theta}{\partial x} = 0, \qquad \frac{\partial \mu}{\partial x} = 0.$$

С учетом этого уравнение $[X_2^*, Y] = 0$ системы (2.26) принимает вид:

$$\frac{\partial \theta}{\partial x}\frac{\partial}{\partial x} + \frac{\partial \mu}{\partial y}\frac{\partial}{\partial y} - \theta\frac{\partial}{\partial x} = 0,$$

или

$$\frac{\partial \theta}{\partial x} = 0, \qquad \frac{\partial \mu}{\partial x} = 0,$$

и результату для коэффициентов оператора: θ = ν(*t*)$e^y$, μ(*t*).

Следовательно, оставшееся уравнение $[X_2^*, Y] = 0$ системы (2.26) принимает вид:

$$\dot{\mu}(t)\frac{\partial}{\partial x} + \dot{\nu}(t)e^y\frac{\partial}{\partial y} - 2\{by\mu(t)\}\frac{\partial}{\partial x} = 0,$$

что эквивалентно системе

$$\dot{\mu}(t) - 2\{by\mu(t)\} = 0, \qquad \dot{\nu}(t)e^y = 0,$$

которая после «расщепления» по переменным $t, x, y$ и очевидных сокращений приводит к требованию выполнения равенств:

$$\dot{\mu}(t) = 0, \qquad b\mu(t) = 0, \qquad \dot{\nu}(t) = 0. \tag{2.31}$$

Определим группу симметрий фазовых траекторий системы (2.29), к которым сводится соответствующей заменой переменных система (2.30).

Из (2.31) следует ν = $C_1$ = const, μ = 0, а с учетом θ = ν$e^y$: θ = $C_1 e^y$, μ = 0, т. е. алгебра симметрий $A_0$ одномерна с базисным оператором

$$Y_1 = e^y\frac{\partial}{\partial x}.$$

Вычисление по полученному оператору, однопараметрической группы симметрий (2.28), дает результат:

$$\tilde{x} = x + \tau e^y,$$
$$\tilde{y} = y.$$

Замена переменных $x, y \leftrightarrow z = xe^y, y$ приводит систему к иерархическому виду (2.29):

$$\dot{y} = u,$$
$$\dot{z} = (v + by^2)e^{-y}.$$

Таким образом, замена $x, y \leftrightarrow z = xe^y, y$ вычисляется при переводе оператора (2.30) к выпрямленному виду $Z = \partial/\partial z$.

**Пример 2.** Для системы

$$\dot{x} = ux + v + ax^2,$$
$$\dot{y} = u,$$  (2.32)

($a$ = Const) оператор (2.7) имеет вид

$$X(u, v) = \frac{\partial}{\partial t} + (ux + v + ax^2)\frac{\partial}{\partial x} + u\frac{\partial}{\partial y}.$$

Операторы (2.8) B-системы

$$X_0 = X(0, 0) = \frac{\partial}{\partial t} + (ax^2)\frac{\partial}{\partial x},$$

$$X_1 = X(1, 0) = \frac{\partial}{\partial t} + (x + ax^2)\frac{\partial}{\partial x} + \frac{\partial}{\partial y},$$

$$X_2 = X(0, 1) = \frac{\partial}{\partial t} + (1 + ax^2)\frac{\partial}{\partial x}.$$

Условие (2.9) выполняется в случае:

$$f_0 = 1 - u - v,$$
$$f_1 = u,$$
$$f_2 = v.$$

Видно, что B-система $X_i$, является и F-системой, т. к. операторы $X_i$, линейно выражается любой оператор $X(u, v)$, а также каждый оператор в пространстве переменных $t, x, y$, в том числе коммутаторы операторов $X_i$. Нетривиальные первые интегралы у системы (2.32) отсутствуют, система допускает конечномерную алгебру $A_0$ симметрий.

Для вычисления коэффициентов $\theta(t, x, y)$, $\mu(t, x, y)$ оператора симметрий преобразуем к эквивалентной системе операторов

$$X_0^* = X_0 \qquad = \frac{\partial}{\partial t} + ax^2\frac{\partial}{\partial x},$$

$$X_1^* = X_1 - X_0 = x\frac{\partial}{\partial x} + \frac{\partial}{\partial y},$$

$$X_2^* = X_2 - X_0 = \frac{\partial}{\partial x}.$$

Из полученных уравнений следует, что (2.26), (2.27) для коэффициентов выполняются тождественно, т. е. условий на функции $\theta(t, x, y)$, $\mu(t, x, y)$ не накладывают.

Первое из уравнений $[X_2^*, Y] = 0$ системы (2.26) имеет вид:

$$\frac{\partial \theta}{\partial x}\frac{\partial}{\partial x} + \frac{\partial \mu}{\partial x}\frac{\partial}{\partial y} = 0,$$

что эквивалентно системе относительно коэффициентов оператора

$$\frac{\partial \theta}{\partial x} = 0, \qquad \frac{\partial \mu}{\partial x} = 0.$$

С учетом этого уравнение $[X_2^*, Y] = 0$ системы (2.26) принимает вид:

$$\frac{\partial \theta}{\partial x}\frac{\partial}{\partial x} + \frac{\partial \mu}{\partial y}\frac{\partial}{\partial y} - \theta\frac{\partial}{\partial x} = 0,$$

или

$$\frac{\partial \theta}{\partial x} = 0, \qquad \frac{\partial \mu}{\partial x} = 0,$$

и результату для коэффициентов оператора: $\theta = \nu(t)e^y$, $\mu(t)$.

Следовательно, оставшееся уравнение $[X_2^*, Y] = 0$ системы (2.26) принимает вид:

$$\dot{\mu}(t)\frac{\partial}{\partial x} + \dot{\nu}(t)e^y\frac{\partial}{\partial y} - 2ax\nu(t)e^y\frac{\partial}{\partial x} = 0,$$

или

$$\dot{\mu}(t) - 2ax\nu(t)e^y = 0, \qquad \dot{\nu}(t)e^y = 0,$$

которая после очевидных преобразований приводит к требованию выполнения равенств:

$$\dot{\mu}(t) = 0, \qquad a\nu(t) = 0, \qquad \dot{\nu}(t) = 0. \qquad (2.33)$$

Определим группу симметрий фазовых траекторий системы (2.32), к которым сводится соответствующей заменой переменных система (2.33).

Получим группы симметрий фазовых портретов.

Из (2.32) следует, что $\nu = 0$, $\mu = C_2$,:

$$\theta = \nu e^y = 0, \quad \mu = C_2,$$

т. е. алгебра симметрий $A_0$ одномерна с базисным оператором

$$Y_2 = \frac{\partial}{\partial y}.$$

Соответствующая однопараметрическая группа симметрий фазовых траекторий:

$$\tilde{x} = x,$$
$$\tilde{y} = y + \tau_2.$$

Групповые системы, несмотря на ярко выраженную групповую структуру, могут допускать группу симметрий по состоянию, состоящую только из тождественного преобразования. Это имеет место для случая, когда количество параметров в группе симметрий по состоянию совпадает с размерностью пространства состояний ($q = n$) [148].

В [148] доказано, что регулярная $n$-мерная система с управлением (2.23) допускает $n$-параметрическую группу симметрий по состоянию (2.28) в том и только в том случае, если (2.23) является L-системой (2.14) при $r = n$.

Следовательно, выбором базисных операторов $Y_1$ в алгебре симметрии $A_0$ или операторов $X_1$, определяющих L-систему, можно добиться совпадения структурных постоянных $C_{ij}^k$ у группы симметрий (2.28) и у группы сдвигов вдоль решений (2.17). То есть, произвольной групповой системе (2.14) можно поставить в соответствие (неоднозначно) L-систему, а, следовательно, и группу симметрий.

Пусть для (2.14) выполняется
$$n > r.$$
Тогда пространство состояний $\mathbb{R}^n(x)$ расслаивается на инвариантные поверхности системы (2.14):
$$w_1(x) = c_1, ..., w_{n-r}(x) = c_{n-r}.$$
Существует неособенная замена переменных
$$x_1, ..., x_n \leftrightarrow z_1, ..., z_r, w_1, ..., w_{n-r},$$
которая ставит в соответствие групповой системе (2.14) L-систему:
$$\dot{z}_k = \sum_{l=1}^{r} \overline{f_{lk}}(z) u_l, \quad k = \overline{1, r},$$
определяющую поведение групповой системы (2.14) на каждой инвариантной поверхности (2.36).

Пусть для групповой системы (2.14) выполняется
$$n > r.$$
Тогда добавлением к (2.14) уравнений
$$\dot{x}_{n+i} = \sum_{l=1}^{r} f_{l,n+i}(x) u_l, \qquad i = \overline{1, r-n}.$$
можно добиться того, что расширенная система (2.14), является L-системой с теми же структурными постоянными $C_{ij}^k$, что и в (2.16) [148].

## 2.3. Аппарат непрерывных симметрий дискретных моделей

Раздел посвящен разработке основ аппарата групп симметрий дискретных систем с управлением [274, 273].

В математической физике определено новое геометрическое направление, связанное с приложением групп Ли к конечно-разностным уравнениям, сеткам, разностным функционалам. Результаты группового анализа дискретных структур [45, 183] показывают, что наличие симметрии у разностных моделей так же, как и в классическом случае инвариантности дифференциальных уравнений, приводит к понижению порядка и интегрируемости обыкновенных разностных уравнений, к наличию инвариантных решений у уравнений в частных разностных производных, к существованию разностных законов сохранения у инвариантных вариационных задач. При построении разностных моделей, сохраняется непрерывная симметрия исходных дифференциальных уравнений. Вместе с тем, наличие дискретных геометрических моделей имеет практическую значимость применительно к исследованию сечений Пуанкаре.

Нелокальность объектов, необходимых для описания разностных уравнений, сеток, разностных функционалов с необходимостью требует привлечения бесконечномерных пространств или пространств последовательностей [45].

Рассмотрим пространство $Z$ последовательностей $(t, x_1, x_2, ..., u)$, где $u$ — независимая переменная, $x$ — зависимая переменная, $x_1, x_2, ...,$ — дифференциальные переменные; Под $z$ будем подразумевать любое конечное число координат вектора $(t, x_1, x_2, ...)$, под $z^i$ — $i$-ю его координату.

В пространстве $Z$ зададим отображение $D$ (дифференцирование), действующее по правилу:

$$D(t) = 1, \quad D(x_1) = x_2, ..., \quad D(x_{s-1}) = x_s, \ s = 1, 2, ...$$

Пусть $A$ — пространство аналитических функций $F(z)$ от конечного числа переменных $z$ (при этом разные функции $F(z)$, входящие в $A$, могут зависеть от разного набора переменных из $(t, x_1, x_2, ...)$, но сам набор конечен всегда). Отождествляя $D$ с действием линейного дифференциального оператора первого порядка

$$D = \frac{\partial}{\partial t} + x_1 \frac{\partial}{\partial x_2} + ... + x_{s-1} \frac{\partial}{\partial x_s} + ..., \qquad (2.34)$$

тем самым распространим дифференцирование $D$ на функцию из $A$, при этом $D(F(z)) \in A$.

Рассмотрим последовательности формальных степенных рядов от параметра $a$:

$$f^i(z, a) = \sum_{k=0}^{\infty} A_k^i(z) a^k, \ i = 1, 2, ..., \qquad (2.35)$$

где $A_k^i(z) \in A$, причём $A_i^0 = z^i$; $z^i$ — $i$-я координата вектора из $Z$.

Пространство последовательностей формальных степенных рядов (2.35)

$$\left( f^1(z,a), f^2(z,a), ..., f^s(z,a), ... \right)$$

обозначим через $\tilde{Z}$. Последовательности $(t, x_1, x_2, ...)$ являются частным случаем последовательности рядов (2.35): $Z \subset \tilde{Z}$.

В $\tilde{Z}$ по определению зададим операции сложения, умножения на число и произведения формальных рядов, совпадающими с соответствующими операциями для сходящихся рядов:

$$\alpha \left( \sum_{k=0}^{\infty} A_k^i a^k \right) + \beta \left( \sum_{k=0}^{\infty} B_k^i a^k \right) = \sum_{k=0}^{\infty} \left( \alpha A_k^i + \beta B_k^i \right) a^k,$$

$$\sum_{k_1=0}^{\infty} A_{k_1}^i a^{k_1} \left( \sum_{k_2=0}^{\infty} B_{k_2}^i a^{k_2} \right) = \sum_{k=0}^{\infty} \left( \sum_{k_1+k_2=k}^{\infty} A_{k_1}^i B_{k_2}^i \right) a^k, \qquad (2.36)$$

$$i = 1, 2, ..., \alpha, \beta = const,$$

а также операции дифференцирования рядов (2.35):

$$D\left(\sum_{k=0}^{\infty} A_k^i(z)a^k\right) = \sum_{k=0}^{\infty} D\left(A_k^i(z)\right)a^k, \qquad (2.37)$$

$$\frac{\partial}{\partial a}\left(\sum_{k=0}^{\infty} A_k^i(z)a^k\right) = \sum_{k=1}^{\infty} k A_k^i(z)a^{k-1},$$

$$\left.\frac{\partial}{\partial a}\left(\sum_{k=0}^{\infty} A_k^i(z)a^k\right)\right|_{a=0} = A_1^i(z), \quad i = 1, 2, \ldots.$$

Фигурирующие выше равенства степенных рядов означают совпадение коэффициентов рядов при соответствующих степенях $a$. Например, ряд (2.35) равен 0, если все $A_k^i = 0, k = 0, 1, 2, \ldots$.

В $\tilde{Z}$ рассмотрим преобразования, определяемые рядами (2.35):

$$z^{i*} = f^i(z, a), \quad i = q, 2, \ldots, \qquad (2.38)$$

Переводящие последовательность $z^i$ в $z^{i*}$.

Введенные операции над рядами (2.35) позволяют рассматривать степени от рядов вида (2.35), мономы, полиномы, и аналитические функции (или формальные степенные ряды) от конечного числа переменных $z^i$. Тем самым определена суперпозиция преобразований вида (2.38)

$$z^{i**} = f^i(z^*, b) = \sum_{k=0}^{\infty} A_k^i(z^*)b^k = \sum_{k=0}^{\infty} A_k^i\left(f^i(z^*, b)\right)b^k, \quad i = 1, 2, \ldots.$$

Такая суперпозиция, вообще говоря, выводит однопараметрические ряды (2.35) из $\tilde{Z}$. Рассмотрим только ряды (2.35), и соответствующие преобразования (2.38), структура коэффициентов которых обеспечивает замкнутость в $\tilde{Z}$ преобразований (2.38):

$$z^{i**} = f^i(z^*, b) = f^i(z, (a+b)) = \sum_{k=0}^{\infty} A_k^i(z)(a+b)^k, \quad i = 1, 2, \ldots. \ (2.39)$$

Свойство (2.39) формальных рядов (2.35) означает, что преобразования (2.36) образуют формальную однопараметрическую группу в $\tilde{Z}$.

Свойство (2.39) эквивалентно экспоненциальному представлению степенных рядов [124]:

$$z^{i*} = f^i(z, b) = e^{aX}(z^i) \equiv \sum_{s=0}^{\infty} \frac{a^s}{s!} X^{(s)}(z^i), \quad i = 1, 2, \ldots, \qquad (2.40)$$

где $X$ — инфинитезимальный оператор группы,

$$X = \xi^i(z)\frac{\partial}{\partial x_i}, \qquad (2.41)$$

$$\xi^i(z) = \left.\frac{\partial f^i(z, a)}{\partial a}\right|_{a=0}, \quad i = 1, 2, \ldots, \xi^i(z) \in A.$$

Представление (2.40) эквивалентно определению формальной группы (2.39). Для этого рассмотрим суперпозицию преобразований вида

$$z^{i*} = e^{bX}\left(e^{aX}(z^i)\right) = \sum_{l=0}^{\infty}\frac{b^l}{l!}X^{(l)}\left(\sum_{s=0}^{\infty}\frac{a^s}{s!}X^{(s)}(z^i)\right) =$$

$$= \sum_{k=0}^{\infty}\left(\sum_{s+l=k}^{\infty}\frac{a^s b^l}{s!l!}X^{(l=s)}(z^i)\right) = \sum_{k=0}^{\infty}\frac{(a+b)^k}{k!}X^{(k)}(z^i) = e^{(a+b)X}(z^i).$$

Представление (2.40) эквивалентно (2.39). Кроме того, очевидно, что

$$f^i(z,a)\Big|_{a=0} = z^i, \ f^i\left(f^i(z,a),(-a)\right) = z^i.$$

Таким образом, для формальных однопараметрических групп справедливо экспоненциальное представление через инфинитезимальный оператор так же, как и для классических локальных групп Ли преобразований.

Заметим, что экспоненциальное представление (2.40) дает рекуррентную цепочку для коэффициентов формальной группы (2.38)

$$A_k^i(z) = \frac{1}{k}X\left(A_{k-1}^i(z)\right), \ k=1,2,...,i=1,2,..., \qquad (2.42)$$

а также удобную для вычисления коэффициентов формальных рядов (2.38) формулу

$$A_k^i(z) = \frac{1}{k!}X^{(k)}(z^i), \ k=1,2,...,i=1,2,....$$

В теории формальных групп показывается [45], что касательное векторное поле

$$\xi^i(z) = \frac{\partial f^i(z,a)}{\partial a}\bigg|_{a=0}$$

связано с формальными рядами (2.35) с помощью уравнений Ли:

$$\begin{aligned}\frac{\partial f^i}{\partial a} &= \xi^i(f), \\ f^i(z,a)\Big|_{a=0} &= z^i, i=1,2,...,\end{aligned} \qquad (2.43)$$

т. е. последовательность формальных рядов $(f^1, f^2, ...)$, образующих группу с касательным полем $\xi^i(z)$, удовлетворяет системе (2.43), и, обратно, для любой последовательности функций $(\xi^1(z), \xi^2(z),...), \xi^i(z) \in A$, решение системы (2.43) образует формальную однопараметрическую группу.

Таким образом, на формальные однопараметрические группы распространяется такая же связь группы и оператора, какая присутствует в классических локальных группах Ли. В случае, когда формальные степенные ряды сходятся и образуют достаточно гладкие дифференцируемые функции, мы имеем дело с группами Ли точечных или касательных преобразований. Таким образом, точечные и контактные преобразования составляют часть формальных групп, но не исчерпывают их целиком. Дополнительным классом к группам точечных и касательных преобразований являются группы «высших симметрии» или группы Ли–Беклунда [58].

Для формальных групп, так же, как и для групп Ли точечных и касательных преобразований, можно определить понятие инварианта и инвариантного многообразия.

Как известно, локально-аналитическая функция $F(z)$ от конечного числа переменных называется инвариантом формальной группы, если $F(z^*) = F(z)$ для любых преобразований группы (2.38).

Для того чтобы $F(z) \in A$ была инвариантом, необходимо и достаточно, чтобы

$$X\, F(z) = 0,$$

где $X$ — оператор группы (2.41).

Многообразие, заданное в $\tilde{Z}$ с помощью функции $\varphi(z) \in A$:

$$\varphi(z) = 0$$

инвариантно, если для всех решений и всех преобразований формальной группы справедливо

$$\varphi(z^*) = 0.$$

Критерий инвариантности многообразия также записывается с помощью оператора группы [45]:

$$X\varphi(z)\big|_{\varphi(z)=0} = 0.$$

Среди формальных групп, преобразования которых описываются формальными степенными рядами (2.35), особое место занимают точечные и контактные группы, а также высшие симметрии. Если первые два класса преобразований можно рассматривать в конечномерной части $\tilde{Z}$, то любая нетривиальная высшая симметрия может быть реализована лишь во всем бесконечномерном пространстве $\tilde{Z}$.

Рассмотрим группу преобразований (2.38) в пространстве $Z$ формальных рядов с инфинитезимальным оператором, совпадающим с оператором (полного) дифференцирования:

$$D = \frac{\partial}{\partial t} + x_1 \frac{\partial}{\partial x_2} + \ldots + x_{s-1} \frac{\partial}{\partial x_s} + \ldots$$

Для простоты рассмотрим случай одной независимой переменной $x$ и одной зависимой переменной $u$.

Преобразования (2.38) этой группы в соответствии с экспоненциальным представлением (2.40) определены действием оператора $T_a \equiv e^{aD}$:

$$z^{i*} = e^{aD}(z^i) \equiv \sum_{s=0}^{\infty} \frac{a^s}{s!} D^{(s)}(z^i).$$

Точка $z^* \in \tilde{Z}$ имеет следующие координаты:

$$t^* = T_a(x) = t + a$$

...

$$x_k^* = T_a(x_k) = \sum_{s=0}^{\infty} \frac{a^s}{s!} x_{k+s},$$

...

преобразования, рассмотренные на поверхности $x = x(t)$, представляют собой разложение в формальные ряды Тейлора функции $x = x(t)$ в точке $(x + a)$, поэтому группа преобразований с оператором полного дифференцирования была названа группой тейлоровского сдвига или группой Тейлора [124].

В теории высших симметрий или групп Ли–Беклунда [58] среди преобразований вида (2.38), составляющих группу, выделяются преобразования, которые сохраняют определение и геометрический смысл производных $(x_1, \ldots)$ в $\tilde{Z}$.

Группа Тейлора сохраняет инвариантность системы и поэтому является группой высших симметрий.

Более того, группа Тейлора является нетривиальной группой высших симметрий, т. е. не является продолжением в $\tilde{Z}$ группы точечных или касательных преобразований. Это следует из того, что последовательность уравнений Ли для определения конечных преобразований группы Тейлора по оператору группы $D$ не имеет решения в конечномерной части $\tilde{Z}$. Таким образом, разложения в формальные ряды Тейлора образуют однопараметрическую группу лишь в бесконечномерном пространстве $\tilde{Z}$, т. е. для динамических систем.

## 2.4. Группы симметрий фазового пространства

Рассмотрим модель дискретной динамической системы:
$$x(t+1) = \psi(x(t),\ u(t)), \tag{2.44}$$

где $x(t)$ — $n$-мерный вектор состояний системы из многообразия $X$, которое обычно отождествляется с $\mathbb{R}^n$; $u(t)$ — $r$-мерный вектор управлений из множества допустимых управлений $U \subseteq \mathbb{R}^r$; $x(t+1) \in T_x X$; $T_x X$ — касательное пространство к $X$ в точке $x$, определяемое допустимыми управлениями; $t$ — дискретное время. Функции $\psi$ удовлетворяют принятым в п.1.2 допущениям о гладкости.

Для практического применения также будем рассматривать линеаризованную вдоль программной траектории систему
$$\Delta x(t+1) = A(t)\Delta x(t) + B(t)\Delta u(t), \tag{2.45}$$

где $A(t) = D_x F(\overline{x}(t), \overline{u}(t))$ — матрица Якоби, которая определяет свойства устойчивости по Ляпунову; $B(t) = D_u F(\overline{x}(t), \overline{u}(t))$ — матрица, определяющая вынужденные колебания, вызванные управляющими процессами; $D$ — оператор дифференцирования.

Введем следующие определения [274].

В динамической системе имеет место преобразование симметрии, если нелинейное динамическое уравнение сохраняет свою структуру при линейных преобразованиях $g\colon x \to x' = g(x)$ в пространстве состояний.

Формально, уравнение динамической системы (2.44), допускает группу симметрий $G$, если отображение $\psi$ коммутирует по всем групповым операциям:
$$\psi(g(x), u) = g(\psi(x, u)), \quad \forall g \in G, \forall x \in X,$$
или, другими словами, группа $G$ делает функцию $\psi(x, u)$ инвариантной по первому аргументу. Группа $G$ может состоять из базовых преобразований симметрии физического интервала — преобразование сдвига, вращение и др. [8], которые обычно являются однопараметрическими.

Предположим, что преобразования, переводящие решения системы (2.44) в программную траекторию принадлежат некоторой подгруппе $G_{\bar{x}}$ группы G
$$g(\bar{x}) = \bar{x}, \quad \forall g \in G_{\bar{x}}.$$

При линеаризации структура симметрий эволюционного уравнения (2.44) не исчезает, а заменяется связанной динамической симметрией. Это связано с понятием алгебры Ли, представляющей собой в определенном смысле локальную линеаризацию группы Ли.

Рассмотрим стационарные линейные системы, т. е. матрицы $A$ и $B$ явно не зависят от времени.

Используя определение, и то, что преобразования симметрии линейные, для линейной аппроксимации при произвольных $g \in G_{\bar{x}}$ [275]:
$$\begin{aligned}\bar{x} + g(A\Delta x) &= g(\bar{x}) + g(A\Delta x) = g(\bar{x} + A\Delta x) = \\ &= g(\psi(\bar{x}, \bar{u}) + A\Delta x) = g(\psi(\bar{x} + \Delta x, \bar{u})) = \\ &= \psi(g(\bar{x} + \Delta x), \bar{u}) = \psi(g(\bar{x}) + g(\Delta x)), \bar{u}) = \\ &= \psi(\bar{x} + g(\Delta x), \bar{u}) = \psi(\bar{x}, \bar{u}) + Ag(\Delta x) = \\ &= \bar{x} + Ag(\Delta x),\end{aligned}$$
т. е. при преобразованиях $g$ уравнение сохраняет свою структуру.

Согласно [275], определим, что группа $L$ называется полной группы симметрий линеаризованного уравнения (2.45) автономной системы ($\Delta u(t) = 0$):
$$g(A\Delta x) = Ag(\Delta x); \forall g \in L.$$

Таким образом, группа $L$ описывает динамическую симметрию системы, линеаризованную около программной траектории $\bar{x}$, включает все преобразования $g \in G_{\bar{x}}\colon G_{\bar{x}} \subseteq L$. В практических приложениях можно принять, что $L$ обычно совпадет с $G_{\bar{x}}$ [278]. Это обосновывается тем, что на основе экспериментального исследования невозможно определить группу $G_{\bar{x}}$, которая исчерпывает динамические симметрии системы или группа $L$ содержит «скрытые» симметрии [31]. Однако можно показать [277], что множество ограничений при определении управляющих параметров может быть получено на основе произвольной однопараметрической подгруппы $L' \subset L$.

В касательном пространстве $T_xX$, под действием преобразований $g$ из произвольной подгруппы $L'$ полной динамической группы симметрии $L$, образуется группа, которая может быть заданна в виде оператора $T$ в матричном представлении:

$$(g(x))_i = (T(g)x)_i = \sum_{j=1}^{n} T_{ij}(g) x_j, \quad \forall x \in T_xX. \qquad (2.46)$$

Нестационарные модели могут быть рассмотрены как совокупность стационарных систем, каждая из которых определенна на некотором временном интервале. При этом подходе определение группы симметрий $L$ можно дать аналогично стационарным системам. Полагая, что траектории $\bar{x}(1), \bar{x}(2), ..., \bar{x}(\tau)$ имеет период $\tau$, и симметрия точки $\bar{x}(t)$ в программной траектории описывается группой $G_{\bar{x}(t)} \subseteq G$, можно записать

$$g(\bar{x}(t+1)) = g(F(\bar{x}(t), \bar{u})) =$$
$$= F(g(\bar{x}(t), \bar{u})) = F(\bar{x}(t), \bar{u}) = \bar{x}(t+1)$$

для каждого $g \in G_{\bar{x}(t)}$.

Свойства симметрии всех точек в программной траектории те же, что и у группы симметрии $G_{\bar{x}}$, которая может быть однозначно определена для произвольной точки $\bar{x}(t)$, $G_{\bar{x}} = G_{\bar{x}(t)}$. Матрица динамики линеаризованной системы будет иметь блочно-диагональный вид:

$$\bar{J}_k(t) = \begin{bmatrix} \bar{J}_k^1(t) & & \\ & \ddots & \\ & & \bar{J}_k^q(t) \end{bmatrix}.$$

где матрица $J_k$ определенна для каждого $k$-го интервала стационарности.

Приведенные ниже результаты легко обобщаются для нестационарных динамических систем в контексте такого определения.

Как известно, для линейных систем управления можно подобрать такой базис, при котором матрица динамики дифференциальных уравнений записывается в каноническом виде: диагональном, канонической жордановой форме, форме Фробениуса и т. п. В этом смысле имеет значение получение формы описания преобразований в групповых координатах, тем более что для групп преобразований вид инфинитезимальной образующей зависит от выбранного базиса операторов.

По определению алгебра Ли, отвечающая группе Ли — пространство левоинвариантных векторных полей на группе Ли с операцией $[\cdot,\cdot]$ — скобкой Ли (коммутатором векторных полей).

Известно, что группа является группой симметрий уравнения, если она переводит любое решения этого уравнения в некоторое решение того же уравнения. Необходимым и достаточным условием этого является выполнение равенства:
$$[A,T]=0.$$
или
$$T(g)A = AT(g), \quad \forall g \in L' \subseteq L,$$
где $T$ определяется соотношением (2.46).

Группа $T$ может быть определена базисным набором преобразований (все преобразования симметрии могут быть выражены через базовые) в виде
$$T = p_1 T^1 \oplus p_2 T^2 \oplus ... \oplus p_q T^q \qquad (2.47)$$
с $n = p_1 d_1 + p_2 d_2 + ... + p_q d_q$; где $p_r$ — число эквивалентных представлений $T_r$ в декомпозиции (2.46), и $q$ — общее число инфинитезимальных образующих в базисе.

Аналогично (2.47) может быть разложено само касательное пространство $T_x X$ на сумму инвариантных подпространств $\mathcal{L}_{L'}^{r\alpha}$, таких, что $T(g)x \in \mathcal{L}_{L'}^{r\alpha}, \forall x \in \mathcal{L}_{L'}^{r\alpha}$ и $\forall g \in L'$:
$$T_x X = \mathcal{L}_{L'}^1 \oplus \mathcal{L}_{L'}^2 \oplus ... \oplus \mathcal{L}_{L'}^q, \qquad (2.48)$$
где
$$\mathcal{L}_{L'}^r = \mathcal{L}_{L'}^{r1} \oplus \mathcal{L}_{L'}^{r1} \oplus ... \oplus \mathcal{L}_{L'}^{rp_r}, \qquad (2.49)$$
и $\alpha = \overline{1,p_r}$ — индексы возможных инвариантных подпространств, которые вписываются в ту же группу эквивалентного $T^r$. Это имеете место, и в случае, если декомпозиция (2.48) уникальная, а декомпозиция (2.49) — нет, а так же при $p_r = 1$. Обозначим базисные векторы $e_i^{r\alpha}, i = \overline{1,d_r}$, в каждом инвариантном подпространстве $\mathcal{L}_{L'}^{r\alpha}$ и выберем те базисные векторы, которые соответствуют $T^r$, то есть —
$$T(g)e_i^{r\alpha} = \sum_{j=1}^{d_r} T_{ij}^r(g) e_i^{r\alpha}, \quad \forall g \in L'.$$

Для однопараметрических преобразований $T^r$, универсальное условие ортогональности между базисными векторами $e_i^{r\alpha}$ может быть установлено в соответствии с [175]:
$$\left(e_i^{r\beta} \cdot e_j^{s\alpha}\right) = \delta_{r,s}\delta_{i,j}\left(e_i^{r\beta} \cdot e_i^{r\alpha}\right).$$

Кроме того, для $p_r > 1$, декомпозиция (2.49) может всегда выполнена таким образом, чтобы $\left(e_i^{r\beta} \cdot e_i^{r\alpha}\right) = \delta_{\alpha,\beta}$. Это требование оставляет некоторую свободу на выборе инвариантного подпространств $\mathcal{L}_{L'}^{\alpha}$. Набор базисных векторов $\{e_i^{r\alpha}\}$, ($r = \overline{1, q}, \alpha = \overline{1, p_r}, \ i = \overline{1, d_r}$) — ортонормальный. Определим ортогональную матрицу $P$:

$$P = \begin{bmatrix} P^1 \\ \vdots \\ P^q \end{bmatrix}, \quad P^r = \begin{bmatrix} P_1^r \\ \vdots \\ P_{d_r}^1 \end{bmatrix}, \quad P_i^r = \begin{bmatrix} \left(e_i^{r1}\right)^T \\ \vdots \\ \left(e_i^{rp_r}\right)^T \end{bmatrix}.$$

Согласно известной теореме [30] элементы произвольной матрицы (в конкретном случае матрицы $A$) инвариантны относительно любого группового преобразования

$$T(g)A = AT(g), \ \forall g \in L' \subseteq L,$$

если удовлетворяет условию:

$$\left(e_i^{r\beta} \cdot A e_j^{s\alpha}\right) = \delta_{r,s} \delta_{i,j} \left(e_i^{r\beta} \cdot A e_i^{r\alpha}\right),$$

где скалярное произведение

$$\left(\overline{\Lambda}^r\right)_{\alpha\beta} \equiv \left(e_i^{r\alpha} \cdot A e_i^{r\beta}\right) \tag{2.50}$$

не зависит от индекса $i = \overline{1, d_r}$ (но зависит от декомпозиции (2.49)).

В результате в групповых координатах матрицу Якоби можно представить в блочно диагональном виде:

$$\overline{A} = PAP^{-1} = \begin{bmatrix} \overline{A}^1 & & \\ & \ddots & \\ & & \overline{A}^q \end{bmatrix},$$

где каждый блок $\overline{A}^r$ сам является блочно диагональным

$$\overline{A}^r = \begin{bmatrix} \overline{\Lambda}^1 & & \\ & \ddots & \\ & & \overline{\Lambda}^r \end{bmatrix},$$

и состоит из $d_r$ идентичных $p_r \times p_r$ блоков $\overline{\Lambda}^r$ с матричными элементами, определемыми скалярным произведением (2.50).

# ГЛАВА 3. РЕДУКЦИЯ НА ЦЕНТРАЛЬНОЕ МНОГООБРАЗИЕ СИСТЕМ, ДОПУСКАЮЩИХ ГРУППЫ СИММЕТРИЙ

Рассмотрено поведение фазовых портретов на инвариантном торе; сформулированы и доказаны теоремы для редукции систем, допускающих группы симметрий, на устойчивое и неустойчивое многообразие; сформулировано обобщение теоремы о центральном многообразии на основе однопараметрических преобразований.

Глава содержит описание результатов, посвященных обоснованию выбора вида модели и фазового пространства, на котором будет происходить разработка методов моделирования и параметрической идентификации по экспериментальным данным.

### *3.1.1. Использование групповых свойств для построения эквивалентных отображений*

Рассмотрим вопросы исследования глобальной структуры орбит динамической системы, которые не зависят от выбора системы координат, т. е. допускают преобразования симметрии. С глобальной точки зрения замена координат представляет собой некоторый диффеоморфизм (в случае гладкой структуры) или гомеоморфизм (в топологической ситуации) фазовых пространств. Таким образом, можно ввести естественные отношения эквивалентности между динамическими системами, связанные с различными классами замен координат, и интерпретировать проблему описания структуры орбит как задачу классификации динамических систем с точностью до этих отношений эквивалентности.

Рассмотрим дискретный случай при условии существования $\mathbb{C}^m$-эквивалентности, определяемой диффеоморфизмом $h$. Это означает [66], что $f_1 : M \to M$ совпадает с $f_2 : N \to N$ с точностью до некоторой $\mathbb{C}^m$-замены координат. Это отношение является естественным отношением эквивалентности для динамики со структурной точки зрения, и в свете рассматриваемой задачи. Пусть $p$ — периодическая точка периода $n$ для $f_1$. Очевидно, для всякого отображения $f_2$, $\mathbb{C}^m$-эквивалентного $f_1$, выполнено $f_2(h(p))^n = h(f_1(p))^n = h(p)$, так что точка $q = h(p)$ является периодической точкой для $f_2$ того же периода $n$. Кроме того, если $m \geq 1$, то для всякой, не обязательно периодической, точки $x$ и для каждого $n$ выполнено равенство:

$$D_x f_1^n = D_{g \circ hx} h^{n-1} \circ D_{hx} f_2 h^{n-1} \circ D_x h.$$

Учитывая свойства эквивалентности траекторий на торе [143], можно показать, что всякая периодическая точка $p$ отображения $f_1$, спектр которой не содержит единицы, определяет некоторые $\mathbb{C}^1$-модули. Так как такие периодические орбиты отделены друг от друга, их спектры могут быть возмущены независимо, по крайней мере, для любой конечной совокупности точек. Таким образом, модули, полученные из различных периодических орбит, в определенном смысле независимы.

С другой стороны, по крайней мере, в некоторых случаях, локально спектр является полным инвариантом относительно гладкой сопряженности. Различные подходы к проблеме локальной гладкой сопряженности показаны в [66].

Для глобальной структуры орбит существуют способы построения конструкции независимых модулей, связанных с периодическими орбитами. В [29] показано, что в случае бесконечно большого числа периодических орбит, как у растягивающего отображения и гиперболического автоморфизма тора, имеется бесконечно много инвариантов локальной $\mathbb{C}^1$-эквивалентности. Для этих двух случаев, которые представляют собой самые простые примеры гиперболических систем [126], спектры периодических орбит образуют полную систему инвариантов для $\mathbb{C}^1$- и $\mathbb{C}^\infty$-эквивалентностей в окрестности указанных отображений соответственно. Также известен $\mathbb{C}^1$-сопряжения отображений тора [18]. Удовлетворительное описание множества возможных значений для собственных значений всех периодических точек остается открытой проблемой.

Разнообразные модули дают существенную, хотя и не полную информацию о гладкой эквивалентности в окрестности вращения $R_a$. Каждое вещественно аналитическое сжатие сохраняет определенную единственным образом аффинную структуру. Для рассматриваемого дискретного отображения ψ, две структуры, определенные вблизи концов отрезка, встречаются в середине. Функции перехода между двумя структурами в любой фундаментальной области $[a, \psi(a)]$ порождают бесконечномерное пространство модулей ψ [66]. Практически это можно интерпретировать так, что существуют такие замены координат, что ψ станет аффинным отображением из $[0, \psi^{-1}(0)]$ в $[0, a]$ и из $[\psi(a), 1]$ в $[\psi^2(a), 1]$. Координаты определяются единственным образом с точностью до двух множителей, по одному на каждый конец. Тогда отображение $[\psi(a), 1] \to [\psi^2(a), 1]$, может быть нормализовано, так что $\psi_a : [0, 1] \to [0, 1]$:

$$\psi_a = \frac{\psi(a + t(\psi(a) - a)) - \psi(a)}{\psi^2(a) - \psi(a)},$$

который может быть расширен на всю действительную прямую по формуле

$$\psi_a(T + k) = \psi_a(T) + k, \quad k \in \mathbb{Z}.$$

Таким образом, два диффеоморфизма $\psi_1$ и $\psi_2$, для которых $\psi_1(0) = \psi_2(0) = 0$ являются эквивалентными, если $\psi_2(T) = \psi_1(T + s) - \psi_1(s)$ для некоторого *s* из [0, 1]. Построенное выше отображение перехода зависит от множителей, определяющих линеаризацию ψ в окрестности концов отрезка, а также от выбора базовой точки *a*. Очевидно, что изменение линеаризации не меняет отображение перехода, если *a* изменено соответствующим образом. Изменение базовой точки приводит к замене отображения перехода на эквивалентное.

Понятие гладкой эквивалентности в соответствии с [66] может быть легко перенесено на случай непрерывного времени. Эквивалентность потоков есть сопряженность потоков как дифференцируемых действий группы $\mathbb{R}$ вещественных чисел. Структура орбит потоков, в отличие от случая систем с дискретным временем, имеет два различных аспекта: (1) относительное поведение точек на различных орбитах и (2) эволюция данного начального условия вдоль орбиты с течением времени. Имеется естественный способ изменять данный поток, сохраняя первый аспект структуры его орбит, а именно вовсе не изменять его орбит.

Для двух потоков $\varphi^t$ и $\varepsilon^t$ дадим определение замены времени с использованием векторные полей инфинитезимальной образующей группы симметрий:

$$\xi = \left.\frac{d\varphi^t}{dt}\right|_{t=0} \text{ и } \eta = \left.\frac{d\varepsilon^t}{dt}\right|_{t=0}.$$

Из единственности решений дифференциальных уравнений следует, что нули векторного поля являются неподвижными точками соответствующего потока. Таким образом, мы получаем, что $\xi(x) = 0$, тогда и только тогда, когда $\eta(x) = 0$. Кроме того, если *x* не является неподвижной точкой, то касательные векторы к кривым $\{\varphi^t(x)\}$ и $\{\varepsilon^t(x)\}$, не обращаются в нуль и имеют одно и то же направление.

Естественно попытаться описать все замены времени данного потока по модулю тривиальных замен. Эта проблема по существу эквивалентна проблеме описания пространства всех достаточно гладких положительных функций с точностью до прибавления функций, которые представляют собой производные других гладких функций по направлению потока. В [275] предложено решение проблемы с помощью естественных модулей, а именно периодов периодических орбит. Это верно, например, для специальных потоков, соответствующих гиперболическому автоморфизму тора.

Определение $\mathbb{C}^m$-модуля может использовано для потоков двумя способами, предполагающими сохранение значений либо на эквивалентных потоках, либо на орбитально эквивалентных потоках. Заметим, различие между эквивалентностью и орбитальной эквивалентностью потоков, состоит в том, что в обоих случаях образ периодической орбиты потока есть периодическая орбита его образа. Однако даже из $\mathbb{C}^0$-эквивалентности потоков следует, что период такой орбиты сохраняется, в то время как при орбитальной эквивалентности может произойти изменение этого периода. Например, типичная замена времени изменяет период данной орбиты, и такие изменения могут производиться независимо для различных орбит. Таким образом, периоды периодических орбит являются модулями даже для $\mathbb{C}^0$-эквивалентности потоков, и в случаях типа надстройки гиперболического автоморфизма тора $F$ имеется бесконечно много таких модулей. Взаимоотношения между этими модулями и модулями, соответствующими собственным значениям линеаризации этого отображения, далеко не тривиальны.

Имеет смысл при определении эквивалентности рассматривать сохранение структурной устойчивости для потоков. Один из способов определения базируется на эквивалентность всех возмущений [16]. Не являясь полностью вырожденным, это требование редко выполняется; например, при наличии периодических орбит их периоды являются модулями в таком смысле. Будем считать, что из локальной топологической эквивалентности следует сохранение структурной устойчивости, при этом преобразующий гомеоморфизм может быть достаточно близким к тождественному для малых возмущений.

Для всех приведенных понятиях этого раздела компактность соответствующих фазовых пространств несущественна. Кроме того, можно естественным образом модифицировать эти определения для случаев, когда для некоторых точек динамическая система определена только на конечном отрезке времени, как это имеет место, например, в окрестности гиперболической неподвижной точки линейного отображения. Такое обобщение ведет к понятиям локальной и полулокальной структурной устойчивости.

Для произвольного сжимающего отображения фазовое пространство может не иметь гладкой структуры, так что приведенные понятия прямо не применимы. Однако сжимающее отображение, определенное в маленьком диске в евклидовом пространстве, структурно устойчиво, так же как и гиперболическое линейное отображение в окрестности неподвижной точки [66].

Отображения поворота не являются структурно устойчивыми. Поскольку топологическая сопряженность сохраняет периодические орбиты, а преобразование поворота на иррациональный угол не может быть сопряжено с преобразованием поворота, для которого соответствующее число рационально. Но так как и множество рациональных чисел, и множество иррациональных чисел плотны, то среди произвольно малых возмущений поворота на рациональный угол можно найти поворот на иррациональный угол, и наоборот. Аналогичное утверждение верно и для сдвига на торе.

## 3.2. Метод моделирования систем, редуцированных на инвариантное многообразие в локальной области

### 3.2.1. Исследования инвариантного многообразия

Согласно теореме Гробмана-Хартмана [135], в окрестности грубого состояния равновесия система

$$\dot{x} = Ax + f(x) \qquad (3.1)$$

где

$$f(0) = 0, \quad f'(0) = 0,$$

топологически эквивалентна линеаризованной системе

$$\dot{y} = Ay. \qquad (3.2)$$

Известно, что в некоторой малой окрестности положения равновесия $O$ существует гладкая замена переменных

$$y = x + \varphi(x), \qquad (3.3)$$

где $\varphi(0) = 0$ и $\varphi'(0) = 0$, позволяющая привести систему (3.1) к виду (3.2). При гладкой замене переменных сохраняются собственные значения ($\lambda_1$, ..., $\lambda_n$) матрицы $A$; более того, при замене переменных типа (3.3), т. е. локально близкой к тождественной, сохраняется и сама матрица $A$. При приведении исходной нелинейной системы к линейному виду возникает ряд трудностей, главная из которых обусловлена наличием резонансов.

По определению, множество $\{\lambda_1, ..., \lambda_n\}$ собственных значений ($\lambda_1$, ..., $\lambda_n$) матрицы $A$ называется *резонансным*, если существует линейная зависимость (резонанс)

$$\rho_k = (m, \lambda) = \sum_{j=1}^{n} m_j \lambda_j,$$

где $m = (m_1, ..., m_n)$ — строка таких неотрицательных целых чисел, что порядок резонанса $|m| = \sum_{j=1}^{n} m_j \geq 2$.

Важной характеристикой колебательных систем являются характеристические показатели или показатели Ляпунова, связанные с собственными значениями матрицы линеаризованного уравнения вдоль периодической траектории (мультипликаторами):

$$\lambda_k = \frac{1}{\tau} \ln \rho_k.$$

Для практического применения важно, что показатели Ляпунова, являясь инвариантами, могут быть вычислены на основании экспериментально полученного временного ряда [92]. Автором в соавторстве с Воловичем М. Е. [35] разработаны методики по расчету показателей Ляпунова по временным рядам [248, 248, 250], которые применены для различных классов задач [252, 267, 271, 273, 277, 278, 279].

В окрестности положения равновесия $\mathbb{C}^\infty$-систему можно привести к одной из двух нормальной форм: либо к нормальной форме Пуанкаре
$$\dot{y} = Ay,$$
либо к нормальной форме Дюлака
$$\dot{y} = A(y) + R(y),$$
где $R(y)$ — полином конечного порядка, составленный из резонансных мономов.

С точки зрения нелинейной динамики значительный интерес представляет линеаризация вблизи седла, поскольку седло может иметь двояко-асимптотические траектории, принадлежащие как устойчивому, так и не устойчивому многообразию. Такие траектории называются гомоклиническими петлями. В случае состояния «седло-фокус», из одной гомоклинической петли может возникнуть бесконечное множество периодических траекторий. В локальной области можно ограничиться рассмотрением линеаризованной системы, однако при рассмотрении глобальных бифуркаций требуется рассмотрение конечно-параметрического семейства систем [143].

Рассмотрим семейство динамических систем
$$\dot{x} = f(x, u), \qquad (3.4)$$
где $x \in \mathbb{R}^n$; $u \in U \subseteq \mathbb{R}^p$; $f$ — $\mathbb{C}^r$-гладкая относительно всех своих аргументов функция, определенная на некоторой области $D \times U$, где $D \subseteq \mathbb{R}^n$. Положим, что при $u = u^0$ семейство (3.4) имеет экспоненциально устойчивую неподвижную точку $O_0(x = x_0)$. Т. е. корни характеристического уравнения
$$\det |A_0 - \lambda I| = 0$$
соответствующей линеаризованной системы
$$\dot{x} = A_0 x$$
лежат слева от мнимой оси. Здесь $A_0 = \left(\partial f(x_0, u^0)/\partial x\right)$.

Поскольку $\det |A_0| \neq 0$, то, согласно теореме о неявной функции, существует такое малое $\delta > 0$, что при $|u - u^0| < \delta$ система (3.4) имеет состояние равновесия $O_u(x = x(u))$, близкое к точке $O_0$. При этом неподвижная точка $O_u$ остается устойчивой для всех $|u - u^0| < \delta_0 \leq \delta$, так как корни характеристического уравнения
$$\det |A(u) - \lambda I| = 0,$$
где $A(u) = \left(\partial f(x(u), u)/\partial x\right)$, непрерывно зависят от $u$. Рассмотрим произвольно выбранное управление $u^1$, удовлетворяющую условию $|u^1 - u^0| < \delta_0$. Повторяя рассуждения, можно найти новую окрестность $|u - u^1| < \delta_1$, в которой система (3.4) имеет устойчивое состояние равновесия, и так далее. В итоге в пространстве параметров можно построить максимальное открытое множество $\Upsilon$, которое называется областью устойчивости состояния равновесия $O_u$. При этом может оказаться, что область устойчивости имеет разветвленную структуру.

Граница Γ области устойчивости ϒ соответствует случаю, когда несколько характеристических показателей положения равновесия $O_u$ (скажем $\lambda_1, ..., \lambda_m$) лежат на мнимой оси, тогда как остальные собственные значения $\lambda_{m+1}, ..., \lambda_n$ расположены в открытой левой полуплоскости. Таким образом, при фиксированном значении параметра на границе система в окресности негрубого состояния равновесия принимает вид

$$\dot{y} = By + f_1(x, y),$$
$$\dot{x} = Ax + \psi_0(x, y), \tag{3.5}$$

где $x \in \mathbb{R}^m$; $y \in \mathbb{R}^{n-m}$; spectr $B = \{\lambda_{m+1}, ..., \lambda_n\}$ (здесь $\operatorname{Re} \lambda_j < 0, j = \overline{m+1, n}$); spectr $A = \{\lambda_1, ..., \lambda_m\}$ (здесь $\operatorname{Re} \lambda_j = 0, j = \overline{1, m}$); $\psi_0$ и $f_1$ — $\mathbb{C}^r$-гладкие функции, которые вместе с их первыми производными обращаются в нуль в начале координат.

Для описания поведения траекторий вблизи точки $O_u$ недостаточно анализа только линеаризованной системы, необходимо также учитывать нелинейные члены. Ляпунов назвал такие случаи критическими и получил для них ряд условий устойчивости положения равновесия.

В теории локальных бифуркаций, фундаментальную роль в которой играет теорема о центральном многообразии Тураева, включает в себя решение задач потери устойчивости, а также поведение при переходе через границу Γ области устойчивости.

Для решения указанных задач необходимо исследовать систему с управлением $u$:

$$\dot{y} = By + f_1(x, y, u),$$
$$\dot{x} = Ax + \psi_0(x, y, u), \tag{3.6}$$

где $u$ принимает значения, близкие к некоторой критической величине параметра $u^*$ (далее полагаем, что $u^* = 0$).

Теорема о центральном многообразии Тураева [143] формулируется следующим образом.

Пусть в системе (3.6) $f_1$, $\psi_0 \in \mathbb{C}^r$, где $1 \leq r < \infty$. Тогда существует окрестность ϒ состояния равновесия $O$, которая при всех достаточно малых $u$ содержит $\mathbb{C}^r$-гладкое инвариантное центральное многообразие $W^C$, задаваемое уравнением

$$y = \phi(x, u), \tag{3.7}$$

где

$$\phi(0, 0) = 0, \qquad \frac{\partial \phi}{\partial x}(0, 0) = 0.$$

Все траектории, не покидающие окрестности ϒ, лежат в центральном многообразии [143].

Существование центрального многообразия позволяет свести решение задач, связанных с критическими случаями, к исследованию $m$-мерной системы

$$\dot{x} = Bx + \psi_0(x, \phi(x,u), u). \qquad (3.8)$$

Размерность последней системы равна количеству характеристических показателей, лежащих в критический момент времени на мнимой оси, и не зависит от размерности исходной системы ($\dim = n$), которая может быть неограниченно большой. Редукция произвольной системы (3.8) большой размерности к системе (3.6) меньшей размерности значительно упрощает исследование и определяет идентифицируемость таких систем.

Центральное инвариантное многообразие обладает лишь конечной гладкостью, так что даже в случае, когда исходная система является аналитической, соответствующая редуцированная система теряет аналитическую структуру. Следовательно, результаты, полученные при исследовании аналитических систем малой размерности, не могут быть непосредственно применены при изучении критических случаев. Необходимо рассмотреть инвариантные характеристики, связанные с неединственностью центрального многообразия.

Для применения теории центрального многообразия необходимо рассматривать другой геометрический объект — сильно устойчивое инвариантное слоение. Его существование позволяет локально привести систему (путем $\mathbb{C}^{r-1}$-гладкой замены переменных) к наиболее простой и удобной треугольной форме

$$\begin{aligned}\dot{y} &= (B + F_1(u, y, u))y, \\ \dot{x} &= Ax + \Psi_0(x, u),\end{aligned} \qquad (3.9)$$

где $\Psi_0(x, u) \equiv \psi_0(x, \phi(x,u), u)$; $F_1 \in \mathbb{C}^{r-1}$; $F_1(0, 0, 0) = 0$. Таким образом, поведение «критических» переменных $x$ в малой окрестности негрубого состояния равновесия не зависит от других переменных и одинаково с поведением на центральном многообразии. Для переменных $y$ имеет место экспоненциальное сжатие (поскольку спектр матрицы $B$ лежит строго слева от мнимой оси).

Аналог теоремы о центральном многообразии верен и в общем случае, то есть для состояний равновесия, некоторые из характеристических показателей которых лежат справа от мнимой оси. Таким образом, в этом случае качественное исследование локальных бифуркаций также можно свести к изучению системы меньшей размерности. Аналогичен результат при изучении поведения решений на границе области устойчивости периодических траекторий, но в условиях существования границы областей устойчивости двух различных типов [143]:

1) бифурцирующая периодическая траектория существует, когда параметр лежит на границе;
2) периодическая траектория не существует на границе.

# ГЛАВА 2. АППАРАТ ГРУПП СИММЕТРИЙ ДЛЯ ИССЛЕДОВАНИЯ МОДЕЛЕЙ УПРАВЛЯЕМЫХ СИСТЕМ

В главе изложены вопросы применения аппарата групп симметрий для анализа систем с управлением; разработан аппарат непрерывных групп симметрий дискретных моделей; определены основные результаты для анализа фазовых портретов систем с нелинейной динамикой.

## 2.1. Групповой анализ нелинейных систем с управлением

Использование аппарата групп симметрий для моделирования динамики управляемых систем по экспериментальным данным является оригинальным подходом, в связи, с чем необходимо проведение исследований в рамках группового анализа с целью его эффективного использования в аппарате геометрического моделирования.

Развитие дифференциальной геометрии и теории групп Ли [49, 101, 103, 124, 134, 138] позволило выделить в отдельное научное направление теорию групп преобразований решений дифференциальных уравнений — теорию групп симметрий. Софус Ли [203], классифицируя преобразования, связывающие частные решения дифференциальных уравнений, получил, что преобразования образуют группу (единичный элемент — тождественное преобразование). Развитие методов группового анализа нелинейных динамических получило в работах Л. В. Овсянникова [105, 106] Н. Х. Ибрагимова [58], П. Олвера [212], а также в работах по исследованию управляемых систем Ю. Н. Павловского [109], В. И. Елкина [50], Г. Н. Яковенко [150] и др. [см., напр. обзор 77]. Анализ результатов позволил разработать геометрические основы исследования управляемых систем, отраженных в публикациях автора [259–265, 268, 273, 274, 275, 280].

Системы, в динамическом поведении которых наблюдаются предельные циклы и аттракторы, допускают конечнопараметрические группы преобразований решений. Как правило, у таких систем множество преобразований симметрии не вкладывается в конечнопараметрическую группу. Для систем с управлением множество преобразований сдвигов вдоль решений имеет функциональную мощность [149], а единственным преобразованием симметрии является тождественное. Важными свойствами систем, допускающих группы симметрий, является возможность редукции к системам меньшей размерности [43, 50, 125] и возможности интегрирования [52, 69, 82].

Пусть система

$$\dot{x} = f(t, x), \, x \in \mathbb{R}^n \qquad (2.1)$$

в некоторой области удовлетворяет условиям существования и

Для состояний равновесия границ второго типа нет, в то время как периодические траектории могут исчезать при достижении момента бифуркации. Практическое значение в рамках решаемой задачи имеют границы первого типа. В момент бифуркации можно построить секущую к критической периодической траектории (которая существует по предположению) и изучить поведение траекторий соответствующего отображения Пуанкаре вблизи бифуркационной неподвижной точки. После этого теория центрального многообразия применяется так же, как в случае состояний равновесия. Это позволит развить единый подход, действительный как для состояния равновесия, так и для периодических траекторий. Помимо динамических приложений теорем об инвариантных многообразиях, полученные результаты можно использовать, например, при разработке алгоритмов управления для проведения движения системы в окресности седла к специальной форме. Для этого используются сильно устойчивое и неустойчивое многообразия.

### 3.2.2. Метод редукции систем на центральное многообразие

Рассмотрим $n$-мерную систему дифференциальных уравнений (3.5) в малой окрестности негрубого состояния равновесия $O$. Пусть система зависит от допустимых управлений (3.6). Проведем построение центральных многообразий в соответствии с [42].

Пусть система (3.6), где функции $f$ и $\psi$ вместе со своими первыми производными по $(x, y)$ непрерывны по $u$ и $f_1(0, 0, 0) = 0$, $\psi_0(0, 0, 0) = 0$. $(f_1, \psi_0)_{(x,y)}(0,0,0) = 0$, при любых малых $u$ имеет $m$-мерное $\mathbb{C}^r$-гладкое инвариантное локальное центральное многообразие $W_{loc}^C : y = \phi(x,u)$ (здесь функция $\phi$ вместе со всеми ее производными по $x$ непрерывно зависит от $u$), которое касается пространства в точке $O$ при $u = 0$, $\phi(0, 0) = 0$, $\partial\phi(0, 0)/(\partial x) = 0$.

В этих условиях при всех $u$ центральное многообразие содержит все траектории, которые целиком лежат в малой окрестности точки $O$.

В случае, когда правая часть системы (3.6) гладко зависит от $u$, центральное многообразие также гладко зависит от $u$. Таким образом, если функции $f$ и $\psi$ являются $\mathbb{C}^r$-гладкими относительно $(x, y, u)$, то функция $\phi$ (график которой $y = \phi(x, u)$ представляет собой многообразие $W_{loc}^C$) может быть выбрана $\mathbb{C}^r$-гладкой относительно $(x, u)$. Этот результат получается формальным добавлением к системе (3.6) уравнения

$$\dot{u} = 0.$$

Если рассматривать $(x, u)$ в качестве нового состояния $x$, то вид расширенной системы будет аналогичен системе (3.5). Применяя для данного случая теорему Тураева, получаем центральное многообразие, которое $\mathbb{C}^r$-гладко зависит от новой переменной *x*, то есть является $\mathbb{C}^r$-гладким относительно (*x*, *u*).

Центральное многообразие $W^C$ не обязательно является $\mathbb{C}^\infty$-гладким, даже если система имеет гладкость $\mathbb{C}^\infty$. Если при любом значении *r* можно применить теорему о центральном многообразии, то выполнено следующее утверждение [143]. Если исходная система является $\mathbb{C}^\infty$-гладкой, то при любом конечном значении *r* существует окрестность $\Upsilon_r$ начала координат, где многообразие $W^C_{loc}$ является $\mathbb{C}^r$-гладким. Указанные окрестности могут сжиматься до нуля, однако при изменении *u* состояние равновесия *O* может сохраняться, но характеристические показатели точки *O*, лежащие на мнимой оси при *u* = 0, могут сместиться при *u* ≠ 0, скажем, влево. Эти показатели соответствуют ведущим собственным значениям соответствующей линеаризованной системы. Следовательно, при ненулевом *u* центральное многообразие будет совпадать с некоторым ведущим многообразием, которое имеет лишь конечную гладкость.

При *u* = 0 выполняется следующее достаточное условие гладкости [143]. Если при *u* = 0 каждая траектория на центральном многообразии $\mathbb{C}^\infty$-гладкой системы стремится к состоянию равновесия *O* либо при $t \to +\infty$, либо при $t \to -\infty$, то центральное многообразие является $\mathbb{C}^\infty$-гладким.

Следствие теоремы о центральном многообразии заключается в том, что при исследовании локальных бифуркаций негрубого состояния равновесия *O* (то есть при изучении множества траекторий, никогда не выходящих за пределы малой окрестности точки *O*, и зависимости этого множества от *u*) систему можно ограничить на центральное многообразие $W^C$ и исследовать систему (3.8). Здесь есть неопределенность, вызванная тем, что центральное многообразие определяется системой неоднозначно. Поэтому редуцирование исследований на центральное многообразие требует доказательства и разработки специализированной методики.

На основании теории центрального многообразия, для любых двух центральных многообразий $y = \phi_1(x)$ и $y = \phi_2(x)$ при каждом $x_0$ — таком, что для некоторого $y_0$ точка $(x_0, y_0) \in N$, — функция $\phi_1$ вместе со всеми своими производными совпадает с $\phi_2(x)$:

$$\left.\frac{d^k \phi_1}{dx^k}\right|_{x=x_0} = \left.\frac{d^k \phi_1}{dx^k}\right|_{x=x_0}, \quad k = \overline{1, r}.$$

Применяя приведенный результат к точке *O*, получаем, что все производные функции ϕ, график которой задает центральное многообразие, определены однозначно в начале координат. Следовательно, несмотря на то, что центральное многообразие не единственно, группа симметрий, порожденная разложением Тейлора редуцированной системы, определена однозначно [143].

Другим важным утверждением является теорема о гладкой сопряженности [143], согласно которой, для любых двух локальных центральных многообразий $W^{C1}$ и $W^{C2}$ существует $\mathbb{C}^{r-1}$-гладкая замена переменных $x$, отображающая траектории первой редуцированной системы

$$\dot{x} = Ax + \psi_0(x, \phi_1(x,u), u)$$

на траектории второй редуцированной системы

$$\dot{x} = Ax + \psi_0(x, \phi_2(x,u), u).$$

Приведенная формулировка результата фактически утверждает, что в динамике на центральных многообразиях одной и той же системы нет существенных качественных различий. То есть, система на центральном многообразии представляет собой достаточно определенный идентифицируемый объект. Разложение редуцированной системы в ряд Тейлора может быть найдено различными способами. Инвариантность многообразия $y = \phi(x)$ означает, в силу системы (3.5), что

$$\frac{\partial \phi}{\partial x}\left(Bx + \psi_0(x, \phi(x,u), u)\right) = A\phi(x) + f_1(x, \phi(x,u), u).$$

Разлагая функции, входящие в данное уравнение, в формальный ряд по степеням $x$, можно последовательно найти все коэффициенты ряда Тейлора функции $\phi$. После этого можно определить по экспериментальным данным правой части редуцированной системы (3.8).

Далее использована теорема [143], в которой показано существование $\mathbb{C}^{r-1}$-гладкого преобразования координат ($\mathbb{C}^1$-близкое к тождественному вблизи начала координат), приводящее систему (3.1) к виду

$$\begin{aligned} \dot{y} &= (B + F_1(u, y, u))y, \\ \dot{x} &= Ax + \Psi_0(x, u), \end{aligned} \qquad (3.10)$$

где $\Psi_0(x, u) \equiv \psi_0(x, \phi(x, u), u)$; $\Psi_0 \in \mathbb{C}^{r-1}$, $F_1 \in \mathbb{C}^r$;

$$F_1(0, 0, 0) = 0, \ \Psi_0(0,0,0) = 0, \ \Psi'_0(0) = 0.$$

Здесь поверхность $\{y = 0\}$ является инвариантным центральным многообразием, т. е. распрямлением многообразия $W^C_{loc}$, представляющего собой $\mathbb{C}^r$-преобразование. Один порядок гладкости теряется, потому что на самом деле в теореме достигнуто гораздо большее: локальная эволюция переменных $x$ теперь полностью не зависит от $y$. Заметим, что, преобразование координат является $\mathbb{C}^{r-1}$-гладким, функция $\Psi_0$ является $\mathbb{C}^r$-гладкой, т. е. она совпадает с нелинейной частью ограничения (3.8) исходной системы на $\mathbb{C}^r$-гладкое центральное многообразие. Таким образом, для любой траектории системы (3.10) поведение переменных $x$ такое же, как на центральном многообразии, для переменных $y$ при $t \to +\infty$ имеет место экспоненциальное сжатие к $y = 0$. Последнее утверждение можно проверить, на основе асимптотической экспоненциальной устойчивости состояний равновесия [143].

Заметим, что результат о гладкой сопряженности следует непосредственно из [143]. Если система (3.10) имеет центральное многообразие, отличное от $\{y = 0\}$, то редуцированная система задается вторым уравнением системы (3.10); то есть для системы в треугольной форме (3.10) ограничения на любые два центральные многообразия являются тривиально сопряженными. Вследствие того, что преобразование координат, приводящее систему к этой форме, является $\mathbb{C}^{r-1}$-гладким, в случае, когда система не приведена к такой форме, мы имеем $\mathbb{C}^{r-1}$-сопряженность.

Геометрическая интерпретация указанных свойств следующая. Очевидно, что если система приведена к треугольной форме (3.10), то образ любой поверхности $\{x = \text{const}\}$ при сдвиге на время $t$ содержится в поверхности того же вида при любом $t$ (до тех пор, пока траектории остаются в малой окрестности точки $O$). Следовательно, слоение малой окрестности состояния равновесия $O$ на поверхности постоянного значения $x$ является инвариантным относительно системы (3.10). Замена координат, возвращающая систему (3.10) к исходному виду (3.5), отображает поверхности $\{x = \text{const}\}$ в поверхности вида
$$x = \xi + \eta(y, \xi), \tag{3.11}$$
где $\xi$ — координата $x$ пересечения поверхности с центральным многообразием; $\mathbb{C}^{r-1}$-гладкая функция $\eta$ вместе с ее первыми производными обращается в нуль в начале координат (заметим, что $\eta \equiv 0$ на всем многообразии $W^C$).

Преобразование, отображающее поверхности $\{x = \text{const}\}$ в поверхности, задаваемые уравнением (3.11), является диффеоморфизмом, т. о. уравнение (3.11) определяет слоение малой окрестности начала координат на поверхности, соответствующие фиксированному значению $\xi$. То есть, для каждой точки $(x, y)$ существует единственное $\xi$, при котором поверхность, соответствующая данному значению $\xi$, проходит через точку $(x, y)$. Такая поверхность является слоем слоения: каждая точка в малой окрестности состояния равновесия $O$ может принадлежать только одному слою. Поскольку слои параметризованы точками на многообразии $W^C_{loc}$, центральное многообразие является базой слоения. Слоение $\{x = \text{const}\}$ инвариантно относительно системы (3.10), т. е. его образ (3.11) является инвариантным слоением системы (3.5). При произвольном значении $t$ сдвиг на время $t$ любого слоя лежит в одном слое того же слоения до тех пор, пока траектория остается в малой окрестности точки $O$. После распрямления центрального многообразия приведение к треугольной форме (3.8) осуществляется просто преобразованием $x \mapsto \xi(x, y)$ (обратным к (3.11)): новой переменной $x$ служит $x$-координата проекции точки вдоль слоев инвариантного слоения на центральное многообразие.

Инвариантность слоения означает, что эволюция новой координаты $x = \xi$ не зависит от $y$. Таким образом, по существу, установлено существование слоения вида (3.11), трансверсального центральному многообразию и инвариантного относительно системы (3.5).

### 3.3. Обобщение теоремы о центральном многообразии для систем, допускающих группы симметрий

Рассмотрим общий случай, когда и справа от мнимой оси есть характеристические показатели состояния равновесия. Система вблизи точки $O$ принимает вид

$$\begin{aligned}
\dot{y} &= By + f_1(x,y,z), \\
\dot{z} &= Cz + f_2(x,y,z), \\
\dot{x} &= Ax + \psi_0(x,y,z),
\end{aligned} \quad (3.12)$$

где $x \in \mathbb{R}^m, y \in \mathbb{R}^k, z \in \mathbb{R}^{n-m-k}$; spectr $B = \{\lambda_1,...,\lambda_m\}$, $\operatorname{Re}\lambda_j = 0$, $(j = \overline{1,m})$; spectr $A = \{\lambda_{m+1},...,\lambda_k\}$, $\operatorname{Re}\lambda_j < 0$, $(j = \overline{m+1,k})$; spectr $C = \{\lambda_{k+1},...,\lambda_n\}$, $\operatorname{Re}\lambda_j > 0$, $(j = \overline{k+1,n})$; $\mathbb{C}^r$-функции $f_1, f_2$ и $\psi_0$ вместе со своими первыми производными обращаются в нуль в начале координат. Правые части системы могут зависеть от управления $u$ либо непрерывно (в таком случае гладкие многообразия, рассматриваемые ниже, непрерывно зависят от $u$), либо гладко. В последнем случае $u$ включается в число «центральных» переменных $x$ и, таким образом, исследуемые далее многообразия и слоения будут иметь гладкость по $u$, равную гладкости по $x$.

Докажем действие теоремы о центральном многообразии для случая, когда системы допускают группы симметрий.

**Теорема 3.1 (Обобщение теоремы Тураева о центрально-устойчивом многообразии для систем, допускающие группы симметрий [284]).** *В малой окрестности состояния равновесия $O$ существует $(m + k)$-мерное инвариантное центрально-устойчивое многообразие $W_{loc}^{sC} : \phi^{sC}(x, y)$ класса $\mathbb{C}^r$, которое содержит точку $O$ и касается в этой точке подпространства $\{z = 0\}$. Многообразие $W_{loc}^{sC}$ включает в себя все траектории, остающиеся в малой окрестности точки $O$ при всех положительных значениях времени. Хотя центрально-устойчивое многообразие определено неоднозначно, для любых двух многообразий $W_1^{sC}$ и $W_2^{sC}$ функции $\phi_1^{sC}$ и $\phi_2^{sC}$ определяют одну и ту же группу симметрий в точке $O$ (и в каждой точке, траектория которой остается в малой окрестности точки $O$ при всех значениях $t \geq 0$), и, следовательно, могут быть идентифицированы на основании экспериментального исследования.*

**Доказательство.** Доказательство основы этой теоремы приводится в [143]. Заметим, что, если система является $\mathbb{C}^\infty$-гладкой, центрально-устойчивое многообразие имеет, вообще говоря, только конечную гладкость: при любом конечном значении $r$ существует окрестность $\Upsilon_r$ точки $O$, в которой многообразие $W^{sC}$ является $\mathbb{C}^r$-гладким.

Требуется доказать, что система допускает группу симметрий, $\phi_1^{sC}$ и $\phi_2^{sC}$ определяют одну и ту же группу симметрий в точке O.

Не нарушая общности, рассмотрим систему второго порядка. Пусть система, определенная на многообразии $W_1^{sC}$ порождает однопараметрическую группу преобразований, порожденная оператором

$$A = \xi_1(x)\frac{\partial}{\partial x_1} + \xi_2(x)\frac{\partial}{\partial x_2},$$

с инфинитезимальным оператором

$$X = \eta_1(x)\frac{\partial}{\partial x_1} + \xi_2(x)\frac{\partial}{\partial x_2}.$$

Необходимо показать, что оператор

$$\tilde{A} = \tilde{\xi}_1(x',\tau)\frac{\partial}{\partial x_1} + \tilde{\xi}_2(x',\tau)\frac{\partial}{\partial x_2}$$

системы, определенной на $W_2^{sC}$ является эквивлентным ($\tau$ — групповой параметр).

Запишем преобразование группы в виде рядов Ли (операторной экспоненты):

$$x'_1 = e^{\tau X}x_1,$$
$$x'_2 = e^{\tau X}x_2,$$
$$x_1 = e^{\tau X}x'_1,$$
$$x_2 = e^{\tau X}x'_2.$$

Представим оператор $\tilde{A}$ в прежних координатах. Для этого вычислим:
$$A = \left(\tilde{A}x_1\right)\Big|_{\dot{x}=x'(x)}\frac{\partial}{\partial x_1} + \left(\tilde{A}x_2\right)\Big|_{\dot{x}=x'(x)}\frac{\partial}{\partial x_2} = \left(\tilde{A}e^{-\tau X}x'_1\right)\Big|_{\dot{x}=x'(x)}\frac{\partial}{\partial x_1} + \left(\tilde{A}e^{-\tau X}x'_2\right)\Big|_{\dot{x}=x'(x)}\frac{\partial}{\partial x_2},$$
отсюда
$$\left(\tilde{A}e^{-\tau X}x'_1\right)\Big|_{\dot{x}=x'(x)} = \xi_1(x),$$
$$\left(\tilde{A}e^{-\tau X}x'_2\right)\Big|_{\dot{x}=x'(x)} = \xi_2(x).$$

Так как $\xi_{1,2}$ не зависят от $\tau$, поскольку определяется оператором $X$, имеем
$$\frac{d}{d\tau}\left(\tilde{A}e^{-\tau X}x'_1\right) = 0,$$

следовательно

$$\frac{\partial \tilde{A}}{\partial \tau} e^{-\tau X} x'_1 - \tilde{A} X e^{-\tau X} x'_1 + X \tilde{A} e^{-\tau X} x'_1 = 0.$$

Аналогичная формула имеет место и для второй координаты, которые вместе образуют дифференциальные уравнения:

$$\frac{\partial \tilde{A}}{\partial \tau} = \tilde{A} X - X \tilde{A} = [\tilde{A}, X]$$

с начальным условием

$$\tilde{A}(x', \tau)\big|_{\tau=0} = A(x').$$

Решение полученной задачи Коши можно получить, разложив оператор $\tilde{A}(x', \tau)$ в ряд Тейлора по степеням $\tau$:

$$\tilde{A}(x', \tau) = \tilde{A}(x') + \tau \frac{\partial \tilde{A}}{\partial \tau}\bigg|_{\tau=0} + \frac{\tau^2}{2!} \frac{\partial^2 \tilde{A}}{\partial \tau^2}\bigg|_{\tau=0} + \dots,$$

таким образом, имеем:

$$\frac{\partial \tilde{A}}{\partial \tau}\bigg|_{\tau=0} = [A, X].$$

Аналогично

$$\frac{\partial^2 \tilde{A}}{\partial \tau^2}\bigg|_{\tau=0} = [[A, X], X].$$

Окончательно ряд примет вид, в форме Хаусдорфа:

$$\tilde{A} = A + \tau[A, X] + \frac{\tau^2}{2!}[[A, X], X], \dots.$$

По базовой теореме Тураева функции $\phi_1^{sC}$ и $\phi_2^{sC}$ определяют одинаковое разложение по ряду Тейлора, что позволяет сделать вывод о коммутировании полей:

$$[A, X] = 0,$$

т. е.

$$\tilde{A} = A.$$

Таким образом, теорема применительно к система, допускающих группы симметрий доказана.

При обращении времени $t \to -t$ матрицы $A$, $B$ и $C$ переходят в $-A$, $-B$ и $-C$ соответственно. Таким образом, часть спектра характеристических показателей, соответствующая переменным $z$, теперь находится слева от мнимой оси, а часть спектра, соответствующая переменным $y$ — справа от нее. К системе, полученной из системы (3.12) путем обращения времени, снова можно применить теорему о центрально-устойчивом многообразии и получить следующую теорему о центрально-неустойчивом многообразии.

**Теорема 3.2 (Обобщение теоремы Тураева о центрально-неустойчивом многообразии для систем, допускающие группы симметрий [284]).** *В малой окрестности состояния равновесия О существует (n – k)-мерное $\mathbb{C}^r$-гладкое инвариантное многообразие $W_{loc}^{usC} : y = \phi^{usC}(x, z)$, содержащее точку О и касающееся в этой точке подпространства $\{y = 0\}$. Центрально-неустойчивое многообразие включает в себя все траектории, остающиеся в малой окрестности точки О при всех отрицательных значениях времени. Для любых двух многообразий $W_1^{usC}$ и $W_2^{usC}$ функции $\phi_1^{usC}$ и $\phi_2^{usC}$ определяют одну и ту же группу симметрий в точке О (так же, как и в каждой точке, траектория которой остается в малой окрестности точки О при всех значениях $t \leq 0$), , и, следовательно, могут быть идентифицированы на основании экспериментального исследования.*

*Когда система $\mathbb{C}^\infty$-гладкая, центрально-неустойчивое многообразие имеет в общем случае только конечную гладкость, но если каждая траектория многообразия $W_{loc}^{usC}$ стремится к состоянию равновесия при $t \to -\infty$, то многообразие $W_{loc}^{usC}$ является $\mathbb{C}^\infty$-гладким.*

Логика доказательства аналогична предыдущей теореме с учетом рассмотренных во второй главе особенностей группового анализа дискретных систем.

Пересечение центрально-устойчивого и центрального неустойчивого многообразий является $\mathbb{C}^r$-гладким *m*-мерным инвариантным центральным многообразием $W_{loc}^{C} = W_{loc}^{sC} \bigcap W_{loc}^{usC}$, определяемым уравнением вида $(y, z) = \phi^C(x)$. Функция $\phi^C$ вместе со всеми производными однозначно определена во всех точках множества *N*. В частности, разложение в ряд Тейлора функции $\phi^C$ в точке *O* однозначно определяется системой.

Система (3.12), ограниченная на центрально-устойчивое многообразие, имеет вид

$$\begin{aligned}\dot{y} &= By + f_1(x, y, \phi^{sC}(x, y)), \\ \dot{x} &= Ax + \psi_0(x, y, \phi^{sC}(x, y)),\end{aligned} \quad (3.13)$$

аналогичный системе (3.5). Следовательно, в данном случае применима теорема Тураева, а именно: для центральных многообразий все же выполняется условие гладкой сопряженности. Таким образом, при изучении локальных бифуркаций систему также можно ограничить на центральное многообразие. Более того, между динамикой на различных центральных многообразиях одной системы нет значительной разницы.

Распрямление центрально-устойчивого и центрально-неустойчивого многообразий, а также распрямление сильно устойчивого и сильно неустойчивого инвариантных слоений на этих многообразиях приводит к следующей теореме, аналогичной теореме, являющейся следствием теоремы Тураева [143].

**Теорема 3.3 [284].** *При помощи $\mathbb{C}^{r-1}$-гладкого преобразования систему* (3.12) *можно локально привести к виду*

$$\dot{y} = (A + F_1(x, y, z))y,$$
$$\dot{z} = (C + F_2(x, y, z))z, \qquad (3.14)$$
$$\dot{x} = Bx + \Psi_0(x) + \Psi_1(x, y, z)y + \Psi_2(x, y, z)z,$$

*где $\Psi_0$ — $\mathbb{C}^r$-гладкая функция, которая вместе со своей первой производной обращается в нуль при $x = 0$; $F_{1,2}$ являются $\mathbb{C}^{r-1}$-функциями, которые обращаются в нуль в начале координат; $\Psi_{1,2} \in \mathbb{C}^{r-1}$; $\Psi_1(x, y, 0) = \Psi_2(x, 0, z) = 0$. При этом $\Psi_i(\cdot)$ определяются группой симметрий.*

Здесь локальное центрально-неустойчивое многообразие задается уравнением $\{y = 0\}$, локальное центрально-устойчивое многообразие — $\{z = 0\}$, а локальное центральное многообразие — $\{y = 0, z = 0\}$. Сильно устойчивое слоение состоит из поверхностей $\{x = \text{const}, z = 0\}$, а слои сильно неустойчивого слоения имеют вид $\{x = \text{const}, y = 0\}$.

Аналогичная теория строится для негрубых периодических траекторий. Изучение динамики в малой окрестности периодической траектории сводится к рассмотрению отображения Пуанкаре на секущей; точка $O$ пересечения траектории с секущей является неподвижной точкой отображения Пуанкаре.

Пусть система имеет размерность $(n + 1)$; таким образом, секущая $n$-мерна. Пусть $m$ мультипликаторов периодической траектории лежат на единичной окружности, $k$ мультипликаторов лежат строго внутри единичной окружности, а остальные $(n - m - k)$ мультипликаторов строго больше 1 по абсолютной величине. Отображение Пуанкаре вблизи неподвижной точки $O$ записывается в виде

$$\overline{y} = By + f_1(x, y, z),$$
$$\overline{z} = Cz + f_2(x, y, z), \qquad (3.15)$$
$$\overline{x} = Ax + \psi_0(x, y, z),$$

где $x \in \mathbb{R}^m$, $y \in \mathbb{R}^k$, $z \in \mathbb{R}^{n-m-k}$; spectr $A = \{\lambda_1, ..., \lambda_m\}$, $|\lambda_j| = 1$, $(j = \overline{1, m})$; spectr $B = \{\lambda_{m+1}, ..., \lambda_k\}$, $|\lambda_j| < 1$, $(j = \overline{m+1, k})$; spectr $B = \{\lambda_{m+1}, ..., \lambda_k\}$, $|\lambda_j| > 1$, $(j = \overline{m+1, k})$; $f_1$, $f_2$ и $\psi_0$ — $\mathbb{C}^r$-гладкие функции, которые вместе с их первыми производными обращаются в нуль в начале координат. Предполагаем, что правые части отображения (а также их производные) могут непрерывно зависеть от управлений $u$. В этом случае многообразия и слоения, рассматриваемые ниже, будут непрерывно зависеть от $u$ вместе со всеми их производными.

Распрямлением инвариантных многообразий и инвариантных слоений получаем теорему, полностью аналогичную приведенной выше.

Теорему о центральном многообразии можно сформулировать следующим образом.

**Теорема 3.4 [284].** *При помощи $\mathbb{C}^{r-1}$-гладкого преобразования систему* (3.15) *можно локально привести к виду*

$$\overline{y} = \left(A + F_1(x, y, z)\right) y,$$
$$\overline{z} = \left(C + F_1(x, y, z)\right) z, \qquad (3.16)$$
$$\overline{x} = Bx + \Psi_0(x) + \Psi_1(x, y, z) y + \Psi_2(x, y, z) z,$$

*где $\Psi_0$ — $\mathbb{C}^r$-гладкая функция, которая вместе со своей первой производной обращается в нуль при $x = 0$; $F_{1,2}$ являются $\mathbb{C}^{r-1}$-функциями, которые обращаются в нуль в начале координат; $\Psi_{1,2} \in \mathbb{C}^{r-1}$; $\Psi_1(x, y, 0) = \Psi_2(x, 0, z) = 0$. При этом $\Psi_i(\cdot)$ определяются группой симметрий.*

Здесь локальное центрально-неустойчивое многообразие задается уравнением $\{y = 0\}$, локальное центрально-устойчивое многообразие — $\{z = 0\}$, а локальное центральное многообразие — $\{y = 0, z = 0\}$. Сильно устойчивое слоение состоит из поверхностей $\{x = \text{const}, z = 0\}$, а слои сильно неустойчивого слоения имеют вид $\{x = \text{const}, y = 0\}$.

С учетом введенной во второй главе полной группы симметрий, из приведенных выше теорем можно показать, что функции $\Psi_i(\cdot), (i = 0, 1, 2)$ определяются групповым анализом фазовых портретов, на основе конечнопараметрических преобразований графиков и соответствующей алгеброй Ли.

Применительно к системам, допускающих группы симметрий и построение их моделей, редуцированных на центральное многообразие результаты можно сформулировать следующим образом.

**Теорема 3.5 [275, 278].** *В локальной области качественное динамическое поведение систем с нелинейной динамикой может описываться редуцированной на центральное многообразие моделью*

$$\dot{x} = Ax(t) + \Psi_0(x), \qquad (3.17)$$

*где $\Psi_0(x, t)$ — $\mathbb{C}^r$-гладкая функция ($\Psi_0(x_0, t) = \Psi_0'(x_0, t) = 0, x_0 = 0$), которая определяется преобразованием симметрии, допускаемой реконструируемой по экспериментальным данным минимальной инвариантной системой.*

**Теорема 3.7 [275, 278].** *В локальной области качественное динамическое поведение дискретных систем с нелинейной динамикой может описываться редуцированной на центральное многообразие моделью*

$$x(t+1) = Bx(t) + \Psi_0(x, t), \qquad (3.18)$$

*где $x(t+1)$ — $n$-мерный диффеоморфизм; $\Psi_0(x, t)$ — $\mathbb{C}^r$-гладкая функция ($\Psi_0(x_0, t) = \Psi_0'(x_0, t) = 0, x_0 = 0$), которая определяется преобразованием* *симметрии, допускаемой реконструируемой по экспериментальным данным минимальной инвариантной системой.*

Таким образом, в результате применения приведенных теорем, качественное исследование в локальной области поведения нелинейной системы, допускающей группы симметрий, сводится к системе меньшей размерности, редуцированной на центральное многообразие. На основании анализа модели можно сделать вывод о возможности ее параметрической идентификации на основании наблюдаемых процессов систем с нелинейной динамикой.

# ГЛАВА 4. МОДЕЛИРОВАНИЕ СИСТЕМ С НЕЛИНЕЙНОЙ ДИНАМИКОЙ ПО ЭКСПЕРИМЕНТАЛЬНЫМ ДАННЫМ

Проведена классификация методов оценки характеристических показателей по временным рядам; изложен модифицированный метод восстановления аттракторов нелинейных систем; разработан метод построения идентифицируемых моделей; приводятся примеры моделирования реальных систем по экспериментальным данным.

## *4.1. Оценка показателей Ляпунова по временному ряду*

### 4.1.1. Методика использования свойств показателей Ляпунова для моделирования

Оценка показателей Ляпунова предлагается как обобщение подхода к исследованию устойчивости нелинейных систем на случай траектории общего вида [91, 107]. При исследовании временных рядов восстанавливаются линеаризованные модели нестационарной неавтономной системы. Поэтому некоторые усредненные аналоги собственных значений при исследовании позволяют получить ряд существенных оценок [177, 250]. Как и в случае неподвижной точки, набор показателей Ляпунова не всегда полностью характеризует устойчивость соответствующей траектории. Тем не менее, он несет существенную информацию о системе.

Пусть задана непрерывная динамическая система
$$\dot{x} = f(x),$$
$$x(0) = x_0$$
или ее дискретный аналог
$$x_{k+1} = \psi(x_k),$$
$$x(0) = x_0.$$

Исследуем изменение $x(t)$ (соответственно, $x_k$), если начальным данным дать бесконечно малое приращение $\delta x$, т. е. рассмотреть бесконечно близкую траекторию $x(t) + \delta x(t)$ или разность $\delta x(t)$ (соответственно, $\delta x_k$). Для широкого класса систем решение дифференцируемо по начальным данным для конечных значений $t$, поэтому
$$\delta x(t) = \Phi(t)\delta x,$$
где $\Phi(t)$ — матрица производных решения по начальным данным:
$$\Phi(t)_{ij} = \frac{\partial x_i(t)}{\partial x_{0j}}.$$

Для линейных систем Ф(*t*) совпадает с нормированной фундаментальной матрицей.

По заданному начальному возмущению δ*x* можно найти δ*x*(*t*), решая соответствующую линейную систему

$$\delta dx/dt = Df(x(t))\delta x, \quad \delta x(0) = \delta x_0$$

или $\delta_{xk+1} = D\psi(x_k)\delta x_k, \delta x_0 = \delta X$.

В силу линейности амплитуда решения несущественна, важен только «коэффициент прироста» решения за время *t*, поэтому от бесконечно малых величин δ можно перейти к конечным и (можно, например, считать, что δ*x* = ε*s*, а бесконечно малая амплитуда ε, входящая множителем, как в правую, так и в левую части уравнений, сокращается). Таким образом, исследование устойчивости приводит к линейным системам

$$ds/dt = A(t)s, \; A(t) = Df(x(t)), s(0) = S$$

и

$$s_{k+1} = B_k s_k, B_k = D\psi(x_k), s_0 = S.$$

Начальное возмущение *S*(δ*x* = ε*S*) определяет направление, в котором мы выбираем бесконечно близкую траекторию в точке *X*.

Заметим, что, строго говоря, векторы *x* и *s* принадлежат к разным пространствам: *x* принадлежит фазовому пространству динамической системы, а *s* — касательному пространству в точке *x*.

Для заданных систем, определенных на $\mathbb{R}^n$, решение удобно выразить через нормированную фундаментальную матрицу Ф, которая удовлетворяет уравнениям

$$d\Phi/dt = A(t)\Phi, \Phi(0) = I$$

и

$$\Phi_{k+1} = B_k \Phi_k, \Phi_0 = I.$$

Тогда *s*(*t*) = Ф(*t*)*S* (*s*$_k$ = Ф$_k$*S*). В общем случае для одних направлений *S* близкие траектории будут экспоненциально удаляться, для других — экспоненциально сближаться, для третьих расстояние остается примерно тем же или меняется медленнее, чем экспоненциально. Для неподвижной точки, когда $A(t) = A = \text{const}$ ($B_k = B = \text{const}$), эти случаи соответствуют собственным значениям с $Re\,\nu > 0$, $Re\,\nu < 0$, $Re\,\nu = 0$ ($|\mu| > 1, |\mu| < 1, |\mu| = 1$), где ν и μ — собственные значения соответственно *A* и *B*. Для того, чтобы охарактеризовать ситуацию в общем случае, А. М. Ляпунов ввел так называемый характеристический показатель решения *s*(*t*):

$$\lambda(s) = \overline{\lim_{t \to \infty}} t^{-1} \ln |s(t)|.$$

Рассмотрим, чему будут равны значения λ(*s*) в случае неподвижной точки для системы $dx/dt = f(x)$, $x(t) = X = \text{const}$. Предположим, что все собственные значения $\nu_i$ матрицы A вещественны, различны и пронумерованы в порядке убывания: $\nu_1 > \nu_2 > ... > \nu_n$. Обозначим соответствующие им собственные вектора через $r^{(i)}$, а $r^{(i)} = v_i r^{(i)}$, $\|r^{(i)}\| = 1$. Тогда $r^{(i)}$ образуют базис в касательном пространстве в точке $X$, в общем случае неортогональный. Любое решение линейной системы $Ds/Dt = Df(x(t)) - s(t)$ можно представить как комбинацию базисных решений $s_i(t) = e^{v_i t} r^{(i)}$, отвечающих начальным данным $s_i(0) = r^{(i)}$. Если $s = \sum_i c_i r^{(i)}$, то

$$s(t) = \sum_i c_i s_i(t) = \sum_i c_i e^{v_i t} r^{(i)}.$$

Пусть $j$ — номер, такой что $c_1 = c_2 = ... = c_{j-1} = 0$, $c_j \neq 0$. Тогда очевидно, что

$$s(t) = \sum_{i=j} c_i e^{v_i t} r^{(i)} = e^{v_j t}\left( c_j r^{(i)} + \sum_{i=j+1} c_i e^{(v_i - v_j)t} r^{(i)} \right)$$

и

$$\lambda(u) = \lim_{t \to \infty} \frac{1}{t} \ln \|s(t)\| = v_i + \lim_{t \to \infty} \frac{1}{t} \ln \left\| c_j r^{(i)} + \sum_{i=j+1} c_i e^{(v_i - v_j)t} r^{(i)} \right\| = v_j.$$

Таким образом, характеристический показатель может принимать только $N$ различных значений $\{v_1, v_2, ..., v_n\}$ в зависимости от начальных данных.

В общем случае у матрицы $A$ могут быть кратные и комплексные собственные значения. В этом случае соответствующих вещественных собственных векторов может не быть, а будут минимальные инвариантные подпространства $\mathbb{R}^j$ (для любого $u \in \mathbb{R}^j$, $Au \in \mathbb{R}^j$) размерности

$$d_j = \text{Dim } \mathbb{R}^j \ (i = \overline{l, m}), \sum_{s=1}^{m} d_s = n.$$

Действительную матрицу $A$ можно при помощи преобразований подобия $A' = CAC^{-1}$ привести к блочно-диагональному виду. У такой матрицы $k$-мерное инвариантное пространство, но всего один собственный вектор.

Каждому подпространству $\mathbb{R}^j$ будут отвечать собственные значения $v$ с одинаковыми действительными частями $Re\,v$, и для всех $s \in \mathbb{R}^j$ характеристический показатель будет равен именно этому значению $\Lambda = Re\,v$. Размерность $d_j = \dim \mathbb{R}^j$ называют кратностью соответствующего показателя $\lambda$. В дальнейшем мы будем предполагать, что в ряду показателей каждый из них встречается столько раз, какова его кратность, так что всего показателей ровно $n$ штук, а упорядочены они по убыванию: $\lambda_1 < \lambda_2 < ... < \lambda_n$. Если $s$ имеет ненулевые проекции на несколько $\mathbb{R}^j$, то $\lambda(s)$ будет равен наибольшему среди показателей для этих подпространств, т.е. показателю $\lambda_i$ с наименьшим номером $i$. Заметим, что кратность показателей может быть больше кратности собственных значений матрицы $A$, поскольку показатели действительные, а собственные значения, вообще говоря, комплексные.

В данном случае также можно выбрать базис линейно независимых векторов $r^{(i)}$, так чтобы каждый вектор принадлежал одному из $\mathbb{R}^j$ (но не все из них могут быть собственными векторами $A$), а для решений с начальными данными $s_i(0) = r^{(i)} \lambda(s_i) = \lambda_i$.

Рассмотрим случай линейных систем общего вида с переменными коэффициентами:

$$ds/dt = A(t)s , \ s \in \mathbb{R}^j , \ \overline{\lim_{t \to \infty}} t^{-1} \int_0^t \|A(\tau)\| d\tau < \infty.$$

Вместо выделения инвариантных подпространств Ляпунов ввел понятие нормальной системы линейно независимых решений: система решений $\{s_1(t), ..., s_n(t)\}$ называется нормальной, если для любой другой линейно независимой системы $\{s'_i(t)\}$ $\lambda(s'_i(t)) \geq \lambda(s_i(t))$. Сам набор показателей $\lambda_1, ..., \lambda_n$ (среди них могут быть совпадающие, т. е. кратные показатели), не зависит от выбора нормальной системы решений и является характеристикой данной системы. Для любого решения $s(t) \neq 0$ показатель $\lambda(s)$ может принимать только одно из этих значений.

В случае отображений вместо $Rev$ будет использоваться $|\mu|$ для собственных значений матрицы $B$. Аналогично для циклов систем ОДУ показатели Ляпунова равны логарифмам модулей множителей Флоке [6].

Среди всего набора показателей Ляпунова важен старший $\lambda_1$ [126]. Поскольку для почти всех начальных данных $S$ будет иметь ненулевую проекцию на направление $r^{(1)}$, то типичной ситуацией будет $\lambda(s) = \lambda_1$. Чтобы получить меньшее значение показателя $\lambda(s) = \lambda_i$ необходим специальный выбор начальных данных.

Кроме обычных показателей $\lambda(s)$, характеризующих одно решение, т. е. растяжение или сжатие в одном направлении, используют показатели порядка $m > 1$, характеризующие изменение $n$-мерных фазовых объемов ($m \leq n$). Пусть $s_1(t), s_2(t), ... , s_m(t)$ — линейно независимые решения, а $\text{Vol}(s_1(t), s_2(t), ... , s_m(t))$ — объем образуемого ими $n$-мерного параллелепипеда,

$$\text{Vol}(s_1, ...., s_m) = \sqrt{\det \begin{Vmatrix} (s_1, s_1) & ... & (s_1, s_m) \\ ... & \ddots & ... \\ (s_m, s_1) & ... & (s_m, s_m) \end{Vmatrix}}.$$

Тогда показателем Ляпунова порядка $m$ называется

$$k_m(s_1, ...., s_m) = \overline{\lim_{t \to \infty}} \frac{1}{t} \ln \text{Vol}(s_1, ...., s_m).$$

Подобно тому, как типичным значением $\lambda(s)$ является $\lambda_1$ типичным значением $k_m$ является $\lambda_1 + \lambda_2 + ... + \lambda_n$. Чтобы добиться иных значений, необходим специальный выбор начальных данных.

Среди всех $k_m$ выделяется показатель $n$-го порядка, для которого можно получить дополнительные результаты. Дело в том, что любой набор $n$ линейно независимых решений образует фундаментальную матрицу $\Phi(t)$.

Объем Vol($s_1(t), s_2(t), \ldots, s_m(t)$) = $|W(t)|$, где $W(t)$ – определитель Вронского, для которого $d\ln|W|/dt = tr\,A(t)$. Отсюда получаем, что

$$k_n = \overline{\lim_{t \to \infty}} \frac{1}{t} \int_0^t (tr A(\tau)) d\tau.$$

Если существует точный предел, то $k_n = <trA>$, т. е. среднему по времени от следа матрицы $A$.

В случае отображения фундаментальная матрица удовлетворяет уравнению $\Phi_{k+1} = B_k \Phi_k$, откуда $W_{k+1} = W_k \times \det B_k$. Поэтому

$$k_n = \overline{\lim_{t \to \infty}} \frac{1}{t} \sum_{k=1}^t \ln|\det B_k|,$$

т. е. как правило $k_n = <\ln|\det B|>$.

В соотношения для $k_n$ параметр $s$ не входит, поэтому этот показатель можно рассматривать как характеристику всей линейной динамической системы в целом. Согласно его значению динамические системы подразделяют на консервативные, сохраняющие $n$-мерные фазовые объемы ($k_n = 0$) и диссипативные ($k_n < 0$).

Для нелинейной системы $dx/dt = F(x)$ и ее траектории $x(t)$ набор ЛП для систем будет характеризовать устойчивость траектории соответствующей ей линейной системы. В этой связи выделяют класс так называемых правильных по Ляпунову систем. Система $ds/dt = A(t)s$ является правильной, если существует предел

$$\overline{\lim_{t \to \infty}} \frac{1}{t} \int_0^t tr A(\tau) d\tau = k_n = \lambda_1 + \lambda_2 + \ldots + \lambda_n.$$

Аналогично, дискретная система является правильной, если существует предел

$$\overline{\lim_{t \to \infty}} \frac{1}{t} \sum_{k=1}^t \ln|\det B_k| = k_n = \lambda_1 + \lambda_2 + \ldots + \lambda_n.$$

Для неправильных систем предел может не существовать или может нарушаться равенство. Однако на практике неправильные по Ляпунову системы обычно не встречаются и соотношения эти всегда полагаются справедливыми. Для правильных систем доказано, что если $\lambda_1 < 0$, то траектория $x(t)$ асимптотически устойчива, если $\lambda_1 > 0$ — неустойчива.

Набор показателей Ляпунова иногда называют спектром показателей Ляпунова соответствующей динамической системы.

Отметим важное свойство показателей Ляпунова. Если уравнения движения динамической системы инвариантны относительно некоторого преобразования, характеризующегося непрерывно изменяющимся инвариантом, то у системы обычно есть нулевой показатель Ляпунова, связанный с этим преобразованием.

Пусть $x(\alpha, t)$ — однопараметрическое семейство траекторий, причем $\varphi^t(x)$ явно от $\alpha$ не зависит. Тогда $x(\alpha, t) = \varphi^t(x(\alpha, 0))$, а

$$\partial x(\alpha, t)/\partial \alpha = D\varphi^t(x(\alpha, 0)) \cdot \partial x(\alpha, 0)/\partial \alpha = \Phi(t) \cdot \partial x(\alpha, 0)/\partial \alpha,$$

т. е. $s(t) = \partial x(\alpha,t)/\partial \alpha$ — решение линеаризованной системы для начальных данных $S = \partial x(\alpha,0)/\partial \alpha$. Если справедливо соотношение

$$c_1 \leq \left\|\frac{\partial x(\alpha,t)}{\partial \alpha}\right\| \leq c_2,$$

где $c_1$ и $c_2$ не зависят от времени, то характеристический показатель для этого решения будет равен нулю. Следовательно, в спектре показателей Ляпунова динамической системы должен быть нулевой показатель.

Частным случаем этой ситуации является сдвиг по времени для автономных систем $dx/dt = F(\boldsymbol{x})$. Если $x(t)$ решение, то $x(t + \alpha)$ — тоже решение. Так как $\partial x/\partial \alpha = dx/dt = F(x)$, то условие существования нулевого показателя сводится просто к требованиям, чтобы на траектории не было неподвижных точек. Таким образом, для аттракторов ОДУ, отличных от неподвижных точек, должен быть по крайней мере один нулевой показатель Ляпунова.

С точки зрения не линеаризованной, а исходной динамической системы показатели Ляпунова характеризуют скорость разбегания бесконечно близких траекторий, а показатели высших порядков — скорость изменения бесконечно малых фазовых объемов.

Динамические системы, для которых $n$-мерный фазовый объем уменьшается, называются диссипативными. Если фазовый объем сохраняется, то такие системы носят название консервативных. У консервативных систем всегда существует хотя бы один закон сохранения. Наличие закона сохранения часто влечет существование соответствующего ему нулевого показателя Ляпунова.

Для диссипативных динамических систем сумма показателей Ляпунова всегда отрицательна, $k_n < 0$.

Если от системы дифференциальных уравнений $Dx/Dt = F(x)$ с набором показателей Ляпунова $\lambda_1, \lambda_2, \ldots, \lambda_n$ перейти к отображению $x_{k+1} = f(x_k)$, $x_k = x(t)$, $x_{k+1} = x(t+\tau) \equiv \varphi^\tau(x(t))$, то показателями Ляпунова для этого отображения будут $\lambda'_i = \lambda_i \tau$. Аналогично, если от отображения перейти к некоторой его степени $f^m(x)$, то для нового отображения показатели будут в $m$ раз больше, $\lambda_i = m\lambda_i$.

Если от системы дифференциальных уравнений перейти к сечению Пуанкаре и соответствующему отображению на единицу меньшей размерности, то в спектре показателей для отображения не будет нулевого показателя, «отвечающего» за сдвиг вдоль траектории, остальными показателями будут $\lambda'_i = \lambda_i \langle\tau\rangle$, где $\langle\tau\rangle$ — среднее время возвращения на плоскость Пуанкаре.

При обращении времени (но для той же самой инвариантной меры $\mu_i$) «типичными» показателями Ляпунова будут $-\lambda_1, -\lambda_2, \ldots, -\lambda_n$. Однако при этом вместо аттрактора, к которому притягиваются траектории при $t \to \infty$, нужно рассматривать неустойчивое множество — репеллер, к которому траектории притягиваются при $t \to -\infty$.

По показателям Ляпунова можно многое сказать о динамической системе, о наблюдаемом режиме, о размерности аттрактора, если таковой имеется, и об энтропии динамической системы. Динамическому хаосу отвечает неустойчивость каждой отдельной траектории, т.е. наличие хотя бы одного положительного показателя Ляпунова. Для странного нехаотического аттрактора $\lambda_1 = 0$.

Регулярные периодические или квазипериодические режимы не имеют в спектре положительных показателей, а для $k$ независимых частот имеют $k$ (для ОДУ) или $k-1$ (для отображений) нулевых показателей. Поэтому для случая дифференциальных уравнений у цикла один нулевой показатель, у тора — два, у 3-тора — три и т.д. Для отображений у цикла нулевых показателей обычно нет, у тора — 1, у 3-тора — 2 и т. д. Эта закономерность легко объяснима: когда аттрактором является «хорошее множество», $m$-мерное многообразие, $n$-мерный фазовый объём должен сохраняться. Притяжение к аттрактору требует, чтобы фазовые объёмы больших размерностей сжимались. Это и отражено в ляпуновском спектре.

Тем не менее, количество независимых частот можно выяснить не всегда, так как нулевые показатели могут быть связаны и с наличием сохраняющихся величин. Для диссипативных систем наличие законов сохранения, вообще говоря, нетипично, однако соответствующие примеры существуют.

Знание показателей Ляпунова позволяет оценить и фрактальную размерность аттрактора

$$k_k = \sum_{i=1}^{k} \lambda_i .$$

Пусть $k$ — такое число, что $k_1, k_2, ..., k_k \geq 0$, а $k_{k+1} < 0$, т. е. $k$-мерный фазовый объём не уменьшается, а $k+1$-мерный — сокращается. Для аттракторов-многообразий фазовый объём, отвечающий размерности аттрактора, сохраняется. Для хаотических аттракторов обычно получается так, что $k_k > 0$, а $k_{k+1} < 0$, и целой размерности, обладающей таким свойством, не существует. Однако можно попытаться найти подходящую дробную размерность. Для этого аппроксимируем зависимость $k_j = k(j)$ кусочно-линейной функцией $k(j) = aj+b$. Эта функция обязательно обратится в ноль в некоторой промежуточной точке $j = d_L$, $k < d_L < k + 1$. Это значение и принимается за оценку размерности аттрактора, которая получила название ляпуновская размерность [90]. Получим для неё выражение. Как говорилось выше

$$ak + b = k_k,\ a(k + 1) + b = k_k - |\lambda_{k+1}|,$$

т. е. $a = -|\lambda_{k+1}|$, $b = k_k + k|\lambda_{k+1}|$, а для ляпуновской размерности получаем соотношение —

$$|\lambda_{k+1}|d_L + k_k + k|\lambda_{k+1}| = 0$$

или

$$d_L = k + \frac{k_k}{|\lambda_{k+1}|} .$$

Доказано несколько теорем [195, 196, 210], согласно которым $d_L$ дает оценку сверху для хаусдорфовой размерности аттрактора, правда иногда вместо обычных показателей Ляпунова используют так называемые «глобальные показатели Ляпунова», которые больше или равны обычным. Соответствующая оценка размерности также будет больше.

Пусть у динамической системы $j$ строго положительных показателей $\lambda_i > 0$. Для энтропии $K_1$ существует строгая оценка

$$K_1 \leq \sum_{i=1}^{j} \lambda_i.$$

Однако на практике обычно считают, что выполняется точное равенство

$$K_1 = \sum_{i=1}^{j} \lambda_i.$$

Для прочих энтропии $K_q$ аналогичных оценок нет, но поскольку чаще всего все $K_q$ при не слишком больших $q$ близки, то приближенно можно пользоваться точным равенством.

### 4.1.2. Методы расчета показателей Ляпунова

Пусть задана система $x_{k+1} = f(x_k)$. После линеаризации получаем линейную систему $s_{k+1} = B_k s_k$. Тогда набор показателей Ляпунова будет зависеть от базовой траектории $x(t)$, т.е. будет характеризовать траекторию, а не аттрактор. Фундаментальное значение показателям Ляпунова придала мультипликативная эргодическая теорема, доказанная В. И. Оселедцем [107]. Согласно ей, показатели Ляпунова совпадают почти для всех траекторий по инвариантной мере $\mu$. Этот факт использовался во многих методах [196, 208, 219, 227].

Обозначим $x_0 = x$, и пусть $\Phi_k$ — фундаментальная матрица линеаризованной системы. Мультипликативная эргодическая теорема предполагает, что траектория $x_k$ может быть продолжена до бесконечности в обе стороны, т.е. при $k \to \infty$, и при каждом $x_k$ существует $\Phi_k = D\varphi^k(x)$. Это, вообще говоря, справедливо не для всех точек $x$. Поэтому будем предполагать, что $x$ принадлежит инвариантному множеству, которое является носителем некоторой инвариантной меры $\mu$, т. е. преобразование $f$ сохраняет меру $\mu$. Матрицы $\Phi_k$ предполагаются невырожденными, а матрицы $B_k = Df(x_k)$ будем полагать ограниченными. В этих условиях: (1) для почти всех $x$ по мере $\mu$ существуют точные значения показателей Ляпунова всех порядков при $t \to \pm\infty$, т.е. линеаризованная система является правильной по Ляпунову; (2) значения показателей $\lambda_i$ совпадают для почти всех $x$ по мере $\mu$; (3) касательное пространство $T_x M(x_k)$ в каждой точке $x_k$ распадается на прямую сумму подпространств $R_i(x_k)$, так что если

$$s(0) \in R_i(x_0),$$

то

$$\lim_{t\to\infty}\frac{1}{|t|}\ln\|u(t)\| = \pm\lambda_i.$$

Подпространства инвариантны в том смысле, что если $x_{k+1} = f(x_k)$, то $R_i(x_{k+1}) = B_k R_i(x_k)$.

Таким образом, почти для всех точек $x$ по мере $\mu$ линеаризованная система оказывается правильной по Ляпунову. Такие точки также называют правильными. Согласно теореме, это свойство оказывается типичным по инвариантной мере.

Показатели Ляпунова являются средними по мере от некоторого функционала, зависящего от $x$. Предположим, что все показатели различны, т.е. все подпространства $R_i(x_k)$ одномерны. Тогда в касательных пространствах можно выбрать базисы $\{R^{(i)}(x_k)\}$, причем так, что они тоже будут инвариантными:

$$r^{(i)}(x_{k+1}) = \frac{B_k r^{(i)}(x_k)}{\|B_k r^{(i)}(x_k)\|}.$$

Такой базис называют базисом Оселедца.

Рассмотрим решение линеаризованной системы $s_{k+1} = B_k s_k$, начальные данные для которой $S_0 = R^{(i)}(x_0)$. Тогда, последовательно подставляя $R^{(i)}(x_{k+1}) \cdot \|B_k R^{(i)}(x_k)\|$ вместо $B_k R^{(i)}(x_k)$, получим

$$u_k = B_{k-1}B_{k-2}...B_1 B_0 r^{(i)}(x_0) = B_{k-1}B_{k-2}...B_1 r^{(i)}(x_1) \cdot \|B_0 r^{(i)}(x_0)\| =$$
$$= B_{k-1}r^{(i)}(x_{k-1}) \cdot \|B_{k-2}r^{(i)}(x_{k-2})\| \cdot ...... \cdot \|B_1 r^{(i)}(x_1)\| \cdot \|B_0 r^{(i)}(x_0)\|,$$

откуда

$$\frac{1}{k}\ln\|u_k\| = \frac{1}{k}\sum_{j=0}^{k-1}\ln\|B_j r^{(i)}(x_j)\|.$$

Поскольку линеаризованная система является правильной почти для всех $x_0$ по мере $\mu$,

$$\lambda_i = \lim_{t\to\infty}\frac{1}{k}\sum_{j=0}^{k-1}\ln\|B_j r^{(i)}(x_j)\| = \langle\ln\|Df(x_j)r^{(i)}(x_j)\|\rangle = \langle\ln\|Df(x)r^{(i)}(x)\|\rangle_\mu.$$

Однако инвариантные подпространства $W_i = \mathrm{span}\{R^{(1)}(x),...,R^{(i)}(x)\}$, численно найти удается при помощи алгоритма Бенеттина [91]. В нем получаются другие вектора $e^{(i)}(x)$, такие что образуемые ими подпространства оказываются теми же самыми,
$$\mathrm{span}\{R^{(1)}(x), ..., R^{(i)}(x)\} = \mathrm{span}\{e^{(1)}(x),...,e^{(i)}(x)\}.$$
Они оказываются тесно связаны с другим вариантом мультипликативной эргодической теоремы, предложенным Д. Рюэллем [120].

Переменные $s_k$ можно выразить через фундаментальную матрицу $\Phi_k = Df^k(x)$ и начальные данные $S$: $s_k = \Phi_k S$. Выражение, входящее в определение показателя Ляпунова, можно переписать следующим образом:

$$\frac{1}{2k}\ln(u_k,u_k) = \frac{1}{2k}\ln\left(U,\Phi_k\Phi^*_k U\right) = \ln\left(\left(U,\Phi^*_k\Phi_k U\right)^{1/2k}\right).$$

Как показал Рюэлль, если преобразование $f(x)$ сохраняет меру $\mu$, а на $Df(x)$ наложены те же условия, что и в теореме Оселедца, то для почти всех $x$ по мере $\mu$ существует предельная матрица

$$G_\infty(x) = \lim_{k\to\infty}\left(\Phi^*_k\Phi_k\right)^{\frac{1}{2k}}.$$

Логарифмы собственных значений этой матрицы $\lambda_i = \ln l_i$ — это набор показателей Ляпунова, отвечающих точке $x$, (т.е. для почти всех $x$ собственные значения $l_i$ совпадают. Собственные вектора $g^{(i)}(x)$ этой матрицы образуют ортонормированный базис в касательном пространстве к точке $x$. Их связь с векторами $r^{(i)}(x)$ довольно очевидна: они получаются, если векторы $\{r^{(i)}(x)\}$ ортонормировать, начиная с последнего:

$$g^{(n)}(x) = r^{(n)}(x),$$
$$g^{(n-1)}(x) = c_{n-1,n-1}r^{(n-1)}(x) + c_{n-1,n}r^{(n)}(x),$$
$$\ldots\ldots\ldots\ldots\ldots\ldots\ldots,$$
$$g^{(j)}(x) = \sum_{i=j}^{n} c_{j,i} r^{(i)}(x), c_{j,j} \neq 0,$$
$$\ldots\ldots\ldots\ldots\ldots\ldots\ldots,$$

где коэффициенты $c_{i,j}$ определяются из условия ортонормированности базиса $\{g\}$. (Если ортонормировать начиная с первого, то начальным данным $S = g^{(i)}$ в общем случае будет отвечать показатель не $\lambda_i$, а $\lambda_1$ поскольку будет ненулевая проекция на направление $r^{(1)}(x)$).

Если отображение $f$ обратимо, то для обратного отображения $f^1$ и траектории, продолженной в обратную сторону до бесконечности ($x_{k-1} = f^1(x_k)$, $S_{k-1} = Df^1(x_k)s_k$), можно построить аналогичную матрицу

$$G_{-\infty}(x) = \lim_{k\to\infty}\left(\Phi^*_k\Phi_k\right)^{\frac{1}{2|k|}}.$$

Ее собственные значения $l'_i = l_i^{-1} = e^{-\lambda_i}$, а собственные вектора $e^{(i)}(x)$ получаются ортогонализацией Грамма-Шмидта системы $\{r^{(i)}(x)\}$, но начиная с первого.

### 4.1.3. Разработка алгоритмов оценки показателей Ляпунова по временному ряду

Существующие в настоящее время алгоритмы оценки можно разделить на 2 класса: матричные методы [182, 226] и методы аналога [239, 91, 7].

<u>Матричные методы</u>. Алгоритмы, связанные с восстановлением в каком-либо виде уравнений движения, аппроксимацией матрицы $Df$ и расчетом показателей называют матричными. Алгоритмы основаны на построении локальных матриц Якоби (матрица $B_k$ в системе $s_k+1 = B_k s_k$) для каждой точки реконструированного аттрактора, после чего для нахождения показателей (можно попытаться оценить весь спектр) используют численные методы, например метод Беннетина [162].

1. Задаем $n$ ортонормированных векторов $v^{(i)}_0$, $i = 1,...,n$, присваиваем $\sigma^{(i)}_0 = 0$, $t_0 = 0$, $s^{(i)}_0 = v^{(i)}_0$, а также определяем шаг перенормировки $\Delta t$.

2. Находим $x_{k+1} = f(x_k)$, $s^{(i)}_k = Df(x_k)$.

3. Ортогонализуем систему векторов $s^{(i)}_{k+1}$ и получаем $s'^{(i)}_{k+1}$:

$s'^{(1)}_{k+1} = s^{(1)}_{k+1}$,

$s'^{(2)}_{k+1} = s^{(2)}_{k+1} + a_{ij} s^{(1)}_{k+1}$, $(s'^{(j)}_{k+1}, s^{(i)}_{k+1}) = 0$,

... ... ... ... ... ... ... ... ... ...,

$$s'^{(i)}_{k+1} = s^{(i)}_{k+1} + \sum_{i=1}^{j-1} a_{j,i} s^{(i)}_{k+1}, a_{j,j} \neq 0, \left(s'^{(j)}_{k+1}, s'^{(i)}_{k+1}\right) = 0, i < j.$$

4. Увеличиваем $\sigma^{(i)}$ на логарифм нормы соответствующего вектора: $\delta\sigma^{(i)}_{k+1} = \ln\left\|u'^{(i)}_{k+1}\right\|, \sigma^{(i)}_{k+1} = \sigma^{(i)}_k + \delta\sigma^{(i)}_{k+1}$; увеличиваем $t$: $t_{k+1} = t_k + \Delta t$. В качестве текущей оценки показателя Ляпунова можно использовать $\lambda'_i(t) = \sigma^{(i)}_{k+1} / t_{k+1}$.

5. Нормируем систему векторов $s'^{(i)}_{k+1}$, получаем ортонормированный базис на следующем шаге $v^{(i)}_{k+1} = u'^{(i)}_{k+1} / \left\|u'^{(i)}_{k+1}\right\|$ и заносим его снова в вектора $s$: $s'^{(i)}_{k+1} = v^{(i)}_{k+1}$.

6. Повторяем пункты 2–5 заданное число раз.

7. Получаем окончательную оценку показателя Ляпунова:

Ортогонализация осуществляется процедурой Грамма–Шмидта, которая в матричном виде называется $QR$-разложением на ортогональную матрицу $Q$, столбцы которой образуют ортонормированный базис $\{v^{(i)}_k\}$, и верхнюю треугольную матрицу $R$, тогда $\delta\sigma^{(i)}_k = \ln R_{i,k}$. Поскольку, $v^{(i)}_k \to e^{(i)}(x_k)$ то с использованием векторов $e^{(i)}(x_k)$ можно написать выражение для показателей Ляпунова:

$$\lambda_i = \lim_{N \to \infty} \frac{1}{N} \sum_{k=1}^{N} \ln\left|\left(e^{(i)}(x_{k+1}), Df(x_k) e^{(i)}(x_k)\right)\right|.$$

В описанном методе не обязательно вычислять все показатели. Если использовать $m < n$ векторов $s^{(i)}$ (или $v^{(i)}$), то будут получены $m$ наибольших показателей.

Методы аналога.

1. Метод Волфа [239]. Первым шагом произвольная точка траектории $z_0$ (в реконструированном фазовом пространстве) принимается за начальную и ищется соседняя ближайшая к ней точка $z^0_0$. Расстояние между этими двумя точками $\| L^0_0 \|$.

При хаотической динамике со временем это расстояние растет. Если следующее значение $\|L^0_1\| > \|L^0_0\|$, то оно отбрасывается и ищется новая точка $z^1_1$, соседствующая с $z_1$ и лежащая по возможности в том же направлении, что и $z^0_1$. Для поиска точки, удовлетворяющей этому условию можно определить скалярное произведение $S = \left(\dfrac{L^0_1}{\|L^0_1\|}, \dfrac{L^1_1}{\|L^1_1\|}\right)$, величина которого должна быть как можно ближе к единице.

Так как $L^i_j$ описывает поведение малого возмущения, его длина должна быть по возможности малой, чтобы линеаризованная вдоль траектории система хорошо описывала эволюцию. С другой стороны она не должна быть настолько малой, чтобы стать сравнимой с уровнем шумов. Кроме того, необходимо чтобы $z_0$ и $z^0_0$ принадлежали разным траекториям, иначе не будут получены положительные первые показатели Ляпунова $\lambda_1$.

Если эти условия выполняются, то старший показатель Ляпунова определяется из выражения:

$$\lambda_1 = \dfrac{1}{t_M - t_0} \sum_{j=0}^{M-1} \log\left(\dfrac{L^j_{j+1}}{L^j_j}\right),$$

где ($M$–1) число смен соседних траекторий.

Если выбрать основание для логарифма равное двум, то первый показатель $\lambda_1$ измеряется в единицах бит/шаг во времени.

Методы аналога не требуют смены траекторий. Наибольшую известность получили методы Канца [196] и Розенштейна [221].

2. <u>Метод Розенштейна</u> прост для реализации и показывает хорошую скорость расчета, однако, результатом его работы является не численное значение $\lambda_1$, а некоторая функция от времени:

$$y(i, \Delta t) = \dfrac{1}{\Delta t}\langle \ln d_j(i)\rangle, \quad d_j(i) = \min_{x_j} \|x_j - x'_j\|,$$

где $x_j$ — рассматриваемая точка, а $x'_j$ — один из ее «соседей». Алгоритм основан на связи $d_j$ и показателей Ляпунова: $d_j(i) \approx e^{\lambda_1(i\Delta t)}$. Для оценки используется ближайший сосед рассматриваемой точки. Старший показатель Ляпунова предлагается вычислить как угол наклона ее наиболее линейного участка. Нахождение такого участка, оказывается нетривиальной задачей, а иногда такой участок и вовсе указать не удается.

3. <u>Метод Канца</u> основан на соотношении $d_j(i) \approx e^{\lambda_1(i\Delta t)}$ и вычислении ЛП по углу наклона наиболее линейного участка некоторой функции вида:

$$S(\varepsilon_0, j) = \left\langle \ln\left(\dfrac{1}{\aleph_n} \sum_{x'_j \in \aleph_n} \|x_j - x'_j\|\right)\right\rangle.$$

Усреднение берется по всем ближайшим соседям $x_j$ в окрестности точки равной $\varepsilon$.

## 4.2. Разработка алгоритмов оценки инвариантных характеристик

Для практического использования, кроме приведенных выше характеристических показателей выбраны инвариантные характеристики, для которых разработаны соответствующие расчетные алгоритмы.

<u>Выбор временного интервала</u>. Используется методика, основанная на теории информации и использовавшая первый минимум взаимной информации для $x(t)$ и $x(t+\tau)$. Для этого по временному ряду изготовляются гистограммы, аппроксимирующие распределение $x(t)$ (оно же будет и для $x(t+\tau)$) и совместное распределение $x(t)$ и $x(t+\tau)$. Далее по построенным гистограммам рассчитываются энтропии и взаимная информация.

$$S = -\sum_{ij} p_{ij}(\tau) \ln \frac{p_{ij}(\tau)}{p_i p_j},$$

где $p_i$ — вероятность нахождения точки в $i$–том интервале, а $p_{ij}(\tau)$ — совместная вероятность, попадания $x(t)$ в $i$-й интервал и попадания $x(t+\tau)$ в $j$-й.

<u>Оценка корреляционной размерности по временному ряду</u>. Для динамической системы очень важным является исследование структуры аттракторов. Странные аттракторы нелинейных динамических систем имеют самоподобную структуру, поэтому для них удобно применять качественное оценивание, идентификацию масштабных свойств, которые могут быть измерены с помощью фрактальных размерностей. Это позволяет оценить геометрическую структуру аттракторов и ввести меру для числа степеней свободы динамической системы [29].

Корреляционная размерность является частным случаем так называемой генеральной размерности, для определения которой пространство вложения, имеющее размерность E$D$, разбивается на ячейки $V_i$ размером $\varepsilon$, $i = 1, ..., m$. Пусть вероятность того, что какая-то точка аттрактора находится в ячейке $V_i$ есть $p_i$. Тогда генеральная размерность определяется:

$$D_q = \lim_{R \to 0} \frac{1}{q-1} \frac{\log_2 \left( \sum_{i=1}^{N} p_i^q \right)}{\log_2(\varepsilon)}, \quad \text{для } q = 0,$$

$$D_0 = \lim_{\varepsilon \to 0} \frac{\log(m)}{\log(1/\varepsilon)}, \quad \text{для } q \to 1$$

Здесь информационная размерность D1 и информация $I(\varepsilon)$ определяются соотношениями:

$$D_1 = \lim_{\varepsilon \to 0} \frac{I(\varepsilon)}{\log_2(1/\varepsilon)}; \quad I(\varepsilon) = -\sum_{i=1}^{N} p_i \log_2(p_i).$$

Для $q = 2$ получается выражение для так называемой корреляционной размерности, используемой для определения размерности объектов, трудно поддающихся или не поддающихся аналитическому описанию (например, аттракторы, построенные методом задержек по опытным данным):

$$D_2 = \lim_{\varepsilon \to 0} \frac{\log_2 C(\varepsilon)}{\log_2 \varepsilon},$$

где $C(\varepsilon)$ — корреляционный интеграл:

$$C(\varepsilon) = \lim_{m \to \infty} \frac{1}{m^2} \sum_{\substack{i,j=1 \\ i \neq j}}^{m} H(\varepsilon - \| z_i - z_j \|);$$

$\varepsilon$ — радиус сферы, для которого определяется число точек $M(\varepsilon)$, оказавшихся внутри сферы; $H$ — функция Хевисайда.

При условии достаточно большого времени наблюдения $T$ значениями

$$C(\varepsilon) \approx \frac{1}{m^2} \sum_{i,j=1}^{m} H(\varepsilon - \| z_i - z_j \|), \quad D_2 \approx \frac{\log_2 C(\varepsilon)}{\log_2 \varepsilon}$$

определяется число точек $M(\varepsilon)$ фазовой траектории, реконструированной из временного ряда, оказавшихся внутри сферы радиусом $\varepsilon$.

Таким образом, на основании приведенного анализа результатов можно сформировать алгоритмическое обеспечение, которое реализовано М. В. Воловичем [248–250].

**Пример вычислений**. Рассмотрим временной ряд, порожденный системой:

$$\begin{cases} \dot{x} = -y - z \\ \dot{y} = x \\ \dot{z} = 0.375 \cdot (y - y^2) - 0.23 \cdot z \end{cases}$$

1. Расчет времени задержки иллюстрирует рис. 4.1. Расчетное значение $\tau = 37$.

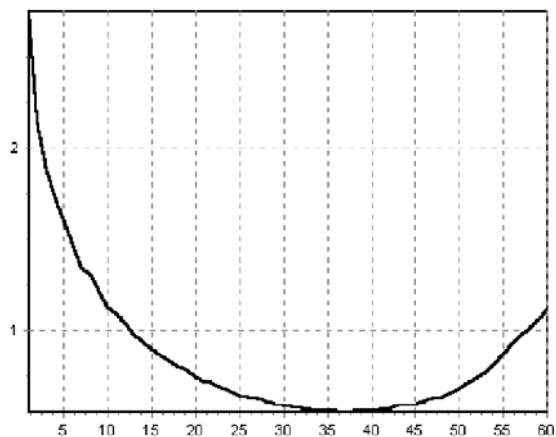

Рис. 4.1. Расчет времени задержки

2. Оценка старшего показателя Ляпунова проводится с использованием 3-х методов. Результаты сведены в табл. 4.1.

Таблица 4.1

| Метод оценки | Значение старшего показателя Ляпунова |
|---|---|
| Метод Вольфа | 0,008912 |
| Метод Канца | 0,009171 |
| Метод Розенштейна | 0,009653 |

Окончательный результат получен как усреднение по всем трем методам

$$\lambda_{max} \approx 0{,}009245.$$

3. Оценка спектра показателей Ляпунова с использованием алгоритма, основанного на методе Беннетина, значение с первого по третий показатели Ляпунова имеют соответствующее значения: 0,008675; 0,000059; –0,028752.

4. Вычисление энтропии системы показано на рис. 4.2.

5. Оценка корреляционной размерности приведена на рис. 4.3.

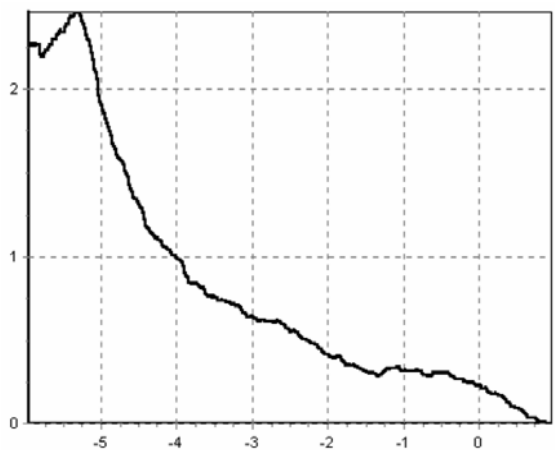

Рис. 4.2. Результат вычисления энтропии

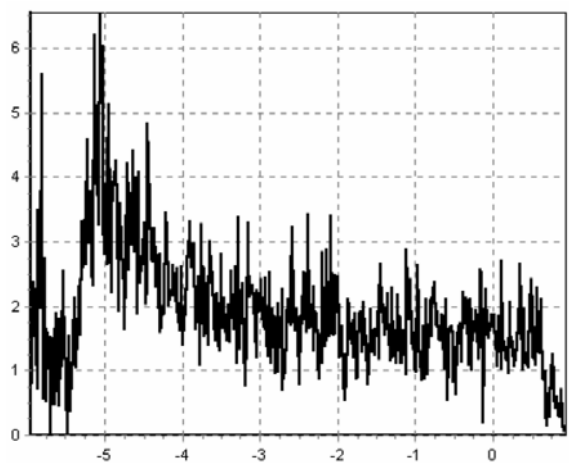

Рис. 4.3. Вычисление корреляционной размерности

## 4.3. Модифицированный метод реконструкции аттракторов для систем, допускающих группы симметрий

Первой работой по реконструкции странного аттрактора по временным рядам была публикация результатов по гидродинамике [214]. В статье показано, что можно получить удовлетворительную геометрическую картину странного аттрактора небольшой размерности, если вместо переменных $x$, входящих в уравнения динамической системы $dx/dt = F(x)$, использовать $m$-мерные вектора, получаемые из элементов временного ряда по тому же принципу, что и в задачах авторегрессии

$$z_i = \{x_i, x_{i+1}, \ldots, x_{i+m-1}\}.$$

В том же году Ф. Такенс доложил о своей теореме, опубликованной годом позже [232]. Именно она лежит в основе всех алгоритмов анализа временных рядов методами нелинейной динамики [9, 108, 155, 198, 207, 215, 249 и др.].

Пусть $M^k$ — $k$-мерное многообразие. Когда такое многообразие реализуется в виде поверхности $L^k$ в $n$-мерном пространстве, которая не пересекается сама с собой, то говорят, что оно вложено в $\mathbb{R}^n$. Само вложение можно представить себе как дифференцируемую векторную функцию $F$, определенную на $M^k$, для которой отображение $M^k \to L^k$ является взаимно однозначным и существует обратная дифференцируемая функция $F^{-1}$, отображающая $L^k$ обратно в $M^k$. То есть $L^k = F(M^k)$. Функция $F^{-1}$ определена только на $L^k$, в противном случае она не будет однозначной. Выбирая разные $F$ и $n$, можно получить различные представления одного и того же многообразия.

Пусть на многообразии $M^k$ (или на какой-либо поверхности $L^k$ диффеоморфной ему) определена векторная функция, нужное количество раз дифференцируемая и отображающая $M^k$ в $m$-мерное евклидово пространство $\mathbb{R}^n$.

Пусть $M^k$ — как минимум дважды дифференцируемое многообразие, а $g(x)$ — некоторая дважды дифференцируемая функция, отображающая $M^k \to \mathbb{R}^n$, для которой матрица производных $\partial g_i / \partial x_j$ имеет ранг $k$. Последнее условие необходимо, чтобы при отображении не получился объект меньшей размерности; скажем, плоскость не отображалась в одномерную кривую, т. е. ранг отображения должен быть равен $k$. Такое отображение будет давать погружение многообразия $M^k$ в $\mathbb{R}^n$ при условии, что $m \geq 2k + 1$ (теорема Уитни [136]). Погружение локально аналогично вложению, но может содержать самопересечения, а потому глобально невозможно определить обратное отображение. Например, если в качестве многообразия рассматривать окружность, то на плоскости эллипс будет вложением, а восьмерка — только погружением. Точке пересечения восьмерки будут соответствовать две различные точки окружности. Поэтому теоремы Уитни оказалось недостаточно для обоснования методов обработки временных рядов.

Пусть задана динамическая система $\varphi^t(x)$ с фазовым пространством P. Будем считать, что числа, образующие временной ряд, являются значениями некоторой «наблюдаемой» — скалярной функции состояния динамической системы $x(t)$:

$$x_i = h(x(t_i)) = h(\varphi^t(x_0)).$$

В качестве многообразия M, фигурирующего в теореме Уитни, может использоваться либо само фазовое пространство $P$, либо какое-либо инвариантное многообразие $M^d$ из P.

Пусть временной шаг между элементами временного ряда равен $\tau$, а вектора $x(t_i)$ будем обозначать $x_i$. Тогда

$$\begin{aligned} x_i &= h(x_i) \equiv \Phi_0(x_i), \\ x_{i+1} &= h(x_{i+1}) = h(\varphi^\tau(x_i)) \equiv \Phi_1(x_i), \\ x_{i+2} &= h(x_{i+2}) = h(\varphi^{2\tau}(x_i)) \equiv \Phi_2(x_i), \\ &\ldots \ldots \ldots \ldots \ldots \ldots \ldots \\ x_{i++m-1} &= h(x_{i++m-1}) = h(\varphi^{(m-1)\tau}(x_i)) \equiv \Phi_{m-1}(x_i), \\ x_{i+m} &= h(x_{i+m}) = h(\varphi^{m\tau}(x_i)) \equiv \Phi_m(x_i) \end{aligned}$$

Все компоненты вектора $z$ связаны с одним и тем же состоянием динамической системы $x_i$. Следовательно, существует векторная функция, которую, следуя Такенсу, обозначим $\Lambda$, отображающая вектора $x_i \in M^d$ в точки $m$-мерного евклидова пространства $\mathbb{R}^m$,

$$z_i = \Lambda(x_i), x_i \in M^d, \; z_i \in \mathbb{R}^m.$$

В теореме предполагается, что $M^d$, $h$ и $\varphi^t$ по крайней мере дважды дифференцируемы, а для всех неподвижных точек и циклов с периодами $k\tau$, $k < d$, предполагается, что у них все собственные значения простые и не равны 1, а $h(x)$ для них различны. Тогда теорема Такенса утверждает, что случаем общего положения, т. е. типичным свойством отображения $\Lambda$ будет то, что при $m \geq 2d + 1$ оно будет давать вложение $M^k$ в $\mathbb{R}^m$. Образ $M^k$ в $\mathbb{R}^m$ будем обозначать $L^d$: $L^d = \Lambda(M^d)$, и, согласно теореме, в типичном случае у него не должно быть самопересечений.

«Вложение» в данном случае будет означать, что:

1. Функция $\Lambda$ будет дифференцируема и будет иметь обратную дифференцируемую $\Lambda^{-1}$, определенную на $L^D$: $M^d = \Lambda^{-1}(L^d))$.

2. Каждой траектории динамической системы будет соответствовать ее образ в $z$-пространстве. Причем для образов будут иметь место те же свойства, что и для исходных траекторий, в частности, через каждую точку $L^d$ будет проходить одна и только одна $z$-траектория.

3. На $L^d$ можно определить динамическую систему.

$$\begin{aligned} x_i &= \Lambda^{-1}(z_i), x_{i+1} = \varphi^\tau(x_i), \\ z_{i+1} &= \Lambda(x_{i+1}) = \Lambda(\varphi^\tau(\Lambda^{-1}(z_i))) \equiv \Psi(z_i), z_i \in L^d \end{aligned}$$

Отображение $\Psi$ переводит $L^d$ в $L^d$, а вне поверхности $L^d$ — $\Psi$ не определено. Если оставить только последнюю компоненту этого соотношения, получим другой вариант, который можно записать в виде «отображения с запаздыванием» или «нелинейной авторегрессии»

$$x_i = F(x_{i-1}, \ldots, x_{i-m}).$$

Таким образом, имеется два отображения:

$$x_{i+1} = \varphi^\tau(x_i) \equiv \Phi(x_i),\ x_i \in M^d,\ \Phi: M^d \to M^d$$

и

$$z_{i+1} = \Psi(z_i),\ z_i \in L^d,\ \Psi: L^d \to L^d.$$

Их можно рассматривать как отображения, связанные невырожденной и обратимой заменой переменных $z = \Lambda(x)$ или как различные представления одного и того же отображения. Следовательно, характеристики, инвариантные относительно невырожденной замены, у обеих систем должны совпадать. К ним относятся фрактальные размерности аттрактора, набор обобщенных энтропий и все $d$ показателей Ляпунова. Поэтому указанные свойства можно пытаться определять по экспериментальным данным, не зная всех переменных динамической системы. Можно пытаться восстановить (аппроксимировать) и саму функцию $\Phi(z)$.

Таким образом, теорема Такенса подводит строгую математическую основу под идеи нелинейной авторегрессии. Практическая реализация идей реконструкции часто сталкивается с проблемами. Возникают они из-за того, что длина обрабатываемого ряда всегда ограничена, во-первых, возможностями хранения информации, во-вторых, скоростью обработки, и, в-третьих, стационарностью исследуемого объекта — важно знать, в течение какого времени мы можем полагать, что исследуем одну и ту же динамическую систему (как только изменится $\varphi^t(x)$, вектора $z$ начнут строиться по-другому). Пусть имеется временной ряд из $N$ чисел, которые являются значениями некоторой наблюдаемой, характеризующей одну и ту же динамическую систему. Тогда реконструированные $z$-вектора дадут $N-m$ точек на поверхности $L^d \in \mathbb{R}^n$, по которым надо будет судить о динамической системе $\Psi$ и ее аттракторе. Объем информации, который можно извлечь из этого множества точек, вообще говоря, зависит от свойств поверхности (насколько она искривлена, закручена и т. п.) и от свойств функции $\Psi(z)$ (насколько велики ее производные). Так как точек конечное число, то существует некоторое характерное расстояние $l$ между точкой и ее ближайшим соседом. Меньшие масштабы будут неразрешимы для данного временного ряда. Если на масштабах порядка $l$ поверхность $L^d$ сильно искривлена, а функция $\Psi(z)$ сильно изменяется, то методы нелинейной динамики будут, скорее всего, бесполезны. Эта же проблема в несколько ином виде встречается, например, в задачах цифровой обработки сигналов (теорема Котельникова). Считается, что если временной интервал между отсчетами равен $\Delta t$, то частоты больше чем $1/2\Delta t$ разрешить невозможно.

Однако в задачах реконструкции свойства $L^d$ и $\Psi(z)$ априорно неизвестны, поэтому аналогичных оценок (скажем, кривизна или производная, не превышающие $\sim l^{-1}$) сделать невозможно. Можно только разумно распорядиться несколькими свободными параметрами. Чаще всего это $m$ и $\tau$.

Свойства $L^d$ и $\Psi(z)$ зависят от динамической системы $\varphi$, наблюдаемой $h$, задержки $\tau$ и размерности векторов $m$ («размерность вложения»).

Метод восстановления аттрактора системы был модифицирован автором применительно к системам, допускающих группы симметрий [275, 278, 279]. На основании теоремы Кинга-Стеварта [199] о вложениях, для систем, допускающих симметрии, наблюдаемый выход был векторной, а не скалярной функцией состояния системы $s(t)$:

$$y(t) = W(s(t)), \qquad (4.1)$$

отображая пространство состояний идентифицируемой системы в $m$-мерное евклидово пространство. Состояние системы может аналогично представлено вектором координат фазового пространства на основании соответствующих временных интервалов задержек

$$x(t) = \left[ y(t+T_1), ..., y(t+T_{n_e}) \right]^T,$$

где размерность вложенного пространства $n = m\, n_e$. Так как вид эволюционных уравнений неизвестен, то возможность построения функции $W$ можно установить на базе основных преобразований — групп сдвига, которые для систем с управлением обладают функциональной мощностью [276].

Линеаризуем выход (4.1) в окрестности неизменяемого во времени состояния $\bar{s}$, и, обозначая смещение $\Delta y(t) = G(s(t)) - G(\bar{s})$, получим

$$\Delta y(t) = C(s(t)), \qquad (4.2)$$

где матрица $C$ — определена таким образом, что $C = D_s G(\bar{s})$.

Динамическая система или пара $(A, C)$ должна быть наблюдаемой, т. е. в течение любого времени начальное состояние $\Delta s(t_i) = \Delta s_i$ может быть определено из измерения управляющих возмущений $\Delta u(t)$ и выхода $\Delta y(t)$.

Группа симметрий $T$, в соответствии с определениями, введенными во второй главе, может быть определена базисным набором преобразований в виде:

$$T = p_1 T^1 \oplus p_2 T^2 \oplus ... \oplus p_q T^q, \qquad (4.3)$$

где $n = p_1 d_1 + p_2 d_2 + ... + p_q d_q$; $p_r$ — число эквивалентных представлений $T_r$ в декомпозиции, и $q$ — общее число инфинитезимальных образующих в базисе. Аналогично (4.3) может быть разложено само касательное пространство $T_xX$ на сумму инвариантных подпространств $\mathfrak{L}_{L'}^{r\alpha}$, таких, что

$$T(g)x \in \mathfrak{L}_{L'}^{r\alpha}, \forall x \in \mathfrak{L}_{L'}^{r\alpha} \text{ и } \forall g \in L';$$

$$T_xX = \mathfrak{L}_{L'}^1 \oplus \mathfrak{L}_{L'}^2 \oplus ... \oplus \mathfrak{L}_{L'}^q, \tag{4.4}$$

где $\mathfrak{L}_{L'}^r = \mathfrak{L}_{L'}^{r1} \oplus \mathfrak{L}_{L'}^{r1} \oplus ... \oplus \mathfrak{L}_{L'}^{rp_r}$; $\alpha = \overline{1, p_r}$ — индексы возможных инвариантных подпространств, которые вписываются в группу $T^r$.

На основании группового анализа систем, редуцированных на центральное многообразие впервые получен следующий результат.

**Теорема 4.1** [274, 279]. *Если в системе нет случайных вырождений, и группа T содержит не более одной копии каждого элемента декомпозиции (4.4) представления группы симметрий, то для реконструкции динамических систем в окрестности состояния s(t) необходимо, чтобы число (m) измеряемых скалярных выходных сигналов $y_i(t)$, равнялось размерности конечнопараметрической алгебры $A_0$, отвечающей группе симметрий графиков фазовых траекторий.*

***Следствие*** [274, 279]. *Для скалярного выходного сигнала наблюдаемая колебательная система может быть реконструирована на центральном инвариантном многообразии на основании однопараметрической алгеброй Ли, а функция $\Psi_0$ порождена групповой операцией сдвига.*

Динамическая система или пара $(A, C)$ называется наблюдаемой [99] если, в течение любого времени, начальное состояние $\Delta s(t_i) = \Delta s_i$ может быть определено из измерения управляющих возмущений $\Delta u(t)$ и выхода $\Delta y(t)$. Как известно, понятие наблюдаемости дуальное к понятию управляемости — условие наблюдаемости для пары $(A, C)$ эквивалент условия управляемости для пары $(A^*, C^*)$. Наличие группы симметрий непосредственно подразумевает, что свойства симметрии матриц $A$ и $A^*$ по существу идентичны (одинаковы структуры спектров собственных значений, жордановых нормальных форм, и т. п.). В результате, известное ранговое условие управляемости линейных систем, накладываемое на матрицу $B$ должно также удовлетворяться и для матрицы $C^*$. Иными словами наблюдаемость линеаризованной системы является эквивалентом наблюдаемости собственных векторов матрицы Якоби исходной системы.

### 4.4. Метод моделирования нелинейных систем по экспериментальным данным

Полученные выше результаты позволили обосновать и обобщить предложенные автором в работах [250–255] модели нелинейных систем, допускающие группы симметрий, т. е. на основании временного ряда идентифицировать параметры модели, редуцированной на центральное многообразие

$$\dot{x} = Ax + \Psi_0(x,t);$$
$$y = Cx. \tag{4.5}$$

На основе доказанных утверждений разработан оригинальный метод моделирования по экспериментальным данным, представляющий собой реконструкцию систем, редуцированных на центральное многообразие. Наиболее эффективно практическое применение метода для повышения качества функционирования существующих систем. Как правило, в этом случае — наблюдаемые в результате эксперимента процессы являются грубыми периодическими траекториями. Алгоритм метода можно представить в виде 4 этапов.

*1. Оценка инвариантных характеристик и реконструкция аттрактора.* Включает оценку размерности минимального инерциального множества; применение методов вычисления по временному ряду выбранных характеристик и построение аттрактора системы с учетом сформулированной теоремы (3) третьей главы.

*2. Вычисление преобразований, допускаемых системой.* При этом идентифицируются наблюдаемые по экспериментальным данным замены координат, переводящие одну область решений в другую, с учетом существования симметрии сдвига. Эта процедура возможна, в случае наблюдаемости системы, описывающей переход от оператора (2.2) к группе (2.1), а также определяется линейностью алгебры Ли.

*3. Определение вида функции* $\Psi_0(x,t)$ с учетом ее коммутирования с инфинитезимальным оператором, т. е. на основании теоремы Ли и формулы Хаусдорфа.

*4. Идентификация параметров системы в форме, предложенной в теореме 1 или теореме 2* с помощью метода наименьших квадратов. Для приведенных в работе примеров использовался метод наименьших квадратов, реализованный Л. Льюнгом в System Identification Toolbox for MATLAB.

На основании полученных результатов сформулируем методику вычисления функций $\Psi_0(x,t)$.

Сформировано и апробировано на большом количестве задач моделирование тепловых процессов. Для этих процессов можно выделить характерные особенности на реконструированном фазовом портрете — четко видны периодические (квазипериодические) траектории, следовательно, имеет место симметрия сдвига. Учет симметрии сдвига в функции $\Psi_0(x,t)$, как правило, определяет хорошую адекватность модели временному ряду. Даже выбор в качестве линейной функции времени $\Psi_0(x,t)$ дает адекватность процесса около 50%, что является хорошей оценкой для восстановленных эволюционных уравнений нелинейных систем.

Таким образом, для задач моделирования тепловых процессов сформированы следующие методологические рекомендации.

1. Реконструкция аттрактора (или фазового портрета) системы.
2. Выделение в фазовом пространстве областей с периодическими траекториями.
3. Вычисление преобразования (автором использовались автрегрессионные полиноминальные модели, представляющие собой группу Тейлора). С точки зрения системного анализа это преобразование можно интерпретировать, как «внутреннее» параметрическое управление автоколебательной системы.
4. Если полученное преобразование имеет адекватность более 75%, то делается допущение о достаточной согласованности преобразования. Следовательно, выбранные локальные области являются топологически эквивалентными.
5. Построение системы вида (4.5) с учетом выполнения уравнений Ли.

При моделировании ситуации седло-фокус, а также других возможных ситуаций, первым этапом предполагается, что система периодическая, что, как показано во второй главе, как правило, имеет место в системах с управлением в рассматриваемом классе технических задач. И искомая функция $\Psi_0(x,t)$ представляет собой декомпозицию преобразования сдвига и найденного преобразования. Для моделирования подобных явлении использовались группы преобразований, допускаемые известными системы исходя из физических явлений.

Приведем примеры реконструкции уравнений по модельным примерам и данным экспериментов реальных технических систем. Для тестирования использовались данные, полученные при компьютерном моделировании известных систем в среде Simulink [272], так и данные, полученные в результате экспериментального исследования реальных систем.

**Пример 1.** Рассмотрим временной ряд, порожденный линейной системой (модель приведена на рис. 4.4). Источником установившихся колебаний служит единичный сигнал. Сложность классического анализа этой системы заключается в наличии двух чисто мнимых полюсов.

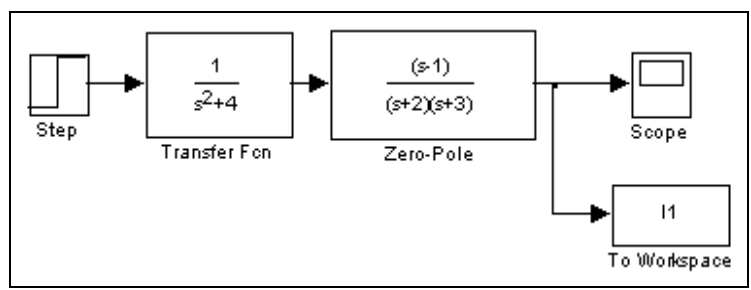

Рис. 4.4. Simulink-модель системы примера 1

Сравнение исходного ряда с динамическим поведением идентифицированной модели приведено на рис. 4.5. По сути, полученная модель представляет собой генератор колебаний с заданными параметрами. Размерность полученной модели равна 4. В качестве функции $\Psi_0(x,t)$ использовалась линейная функция.

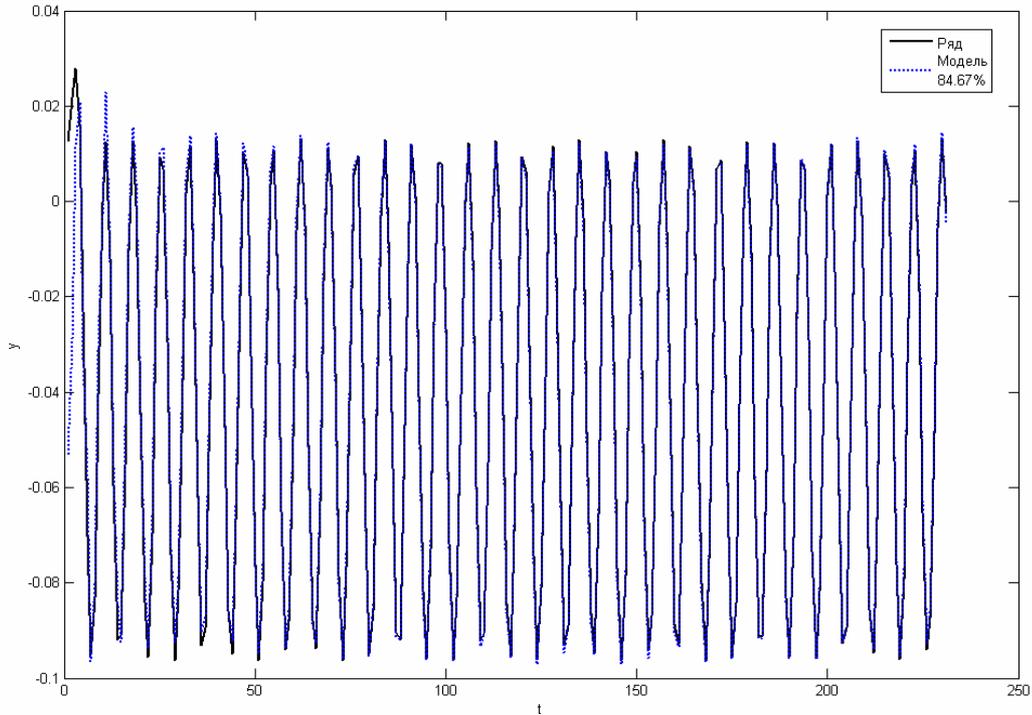

Рис. 4.5. Сравнение исходного ряда с динамическим поведением идентифицированной модели

**Пример 2.** Пусть временной ряд порожден системой:
$$\dot{x} = y$$
$$\dot{y} = -x + x^2 - 0,05\sin 2t.$$

Отображение Пуанкаре со структурой резонансов этой системы приведено на рис. 4.6.

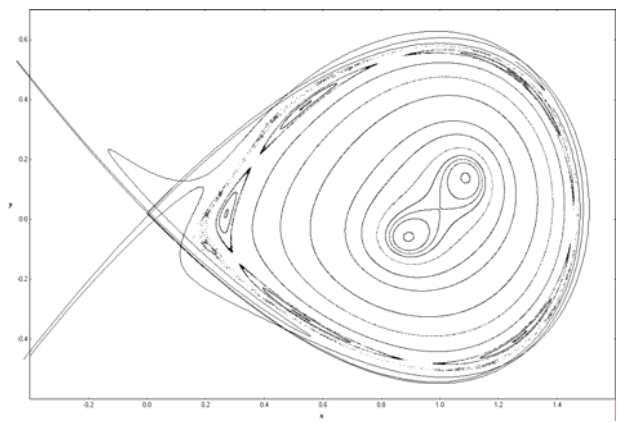

Рис. 4.6. Отображение Пуанкаре

Для идентификации (как «черный ящик») рассмотрим временной ряд, порожденный Simulink-моделью (рис. 4.7) при начальных условиях $x(0) = 0; y(0) = 0,042$. Динамика системы приведена на рис. 4.8.

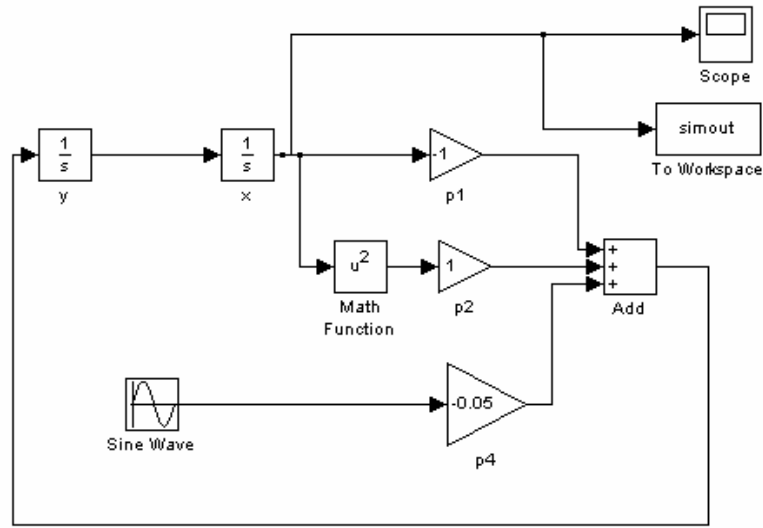

Рис. 4.7. Simulink -модель системы

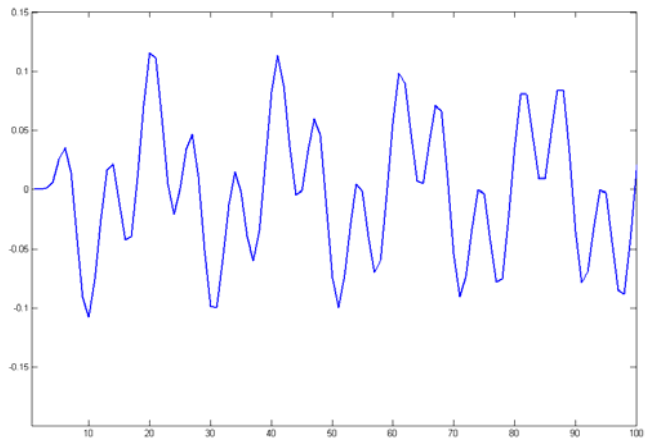

Рис. 4.8. Временной ряд, порожденный системой

В результате параметрической идентификации системы (4.1) с помощью Identification Toolbox, в условиях предположении о симметрии сдвига получены следующие матрицы уравнения системы:

$$A = \begin{bmatrix} 0,7707 & -0,3206 & 0,542 & -0,004127 & -0,03038 \\ 0,5014 & 0,8746 & -0,18 & 0,05212 & 0,02569 \\ -0,3528 & 0,2858 & 0,6789 & -0,5847 & 0,1485 \\ -0,2553 & 0,2566 & 0,4667 & 0,7665 & 0,00253 \\ -0,1971 & -0,016 & -0,00074 & -0,0691 & -0,0922 \end{bmatrix}.$$

$$\Psi_0 = (0,6757; -0,4970; 0,4360; 0,8414)^T (4t^2 + 3t);$$

$$C = (0,4123; -0,1394; 0,1550; -0,0662).$$

На рис. 4.9 приведено сравнение динамики модели с исходным временным рядом. Адекватность модели 84%, при этом качественное поведение обоих систем совпадает.

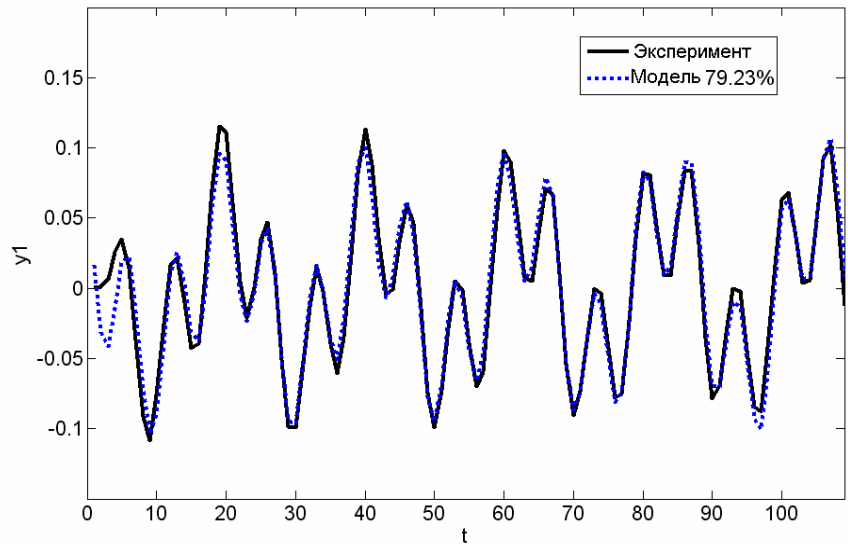

Рис. 4.9. Сравнение динамики модели и системы (сплошная линия — динамика исходной системы, пунктирная — динамика модели)

**Пример 3.** Имеется два дискретных процесса, измеренные на выходе нелинейной системы (рис. 4.10). Процессы получены из системы:

$$x_1(t+1) = 1.25 x_1(t) \cdot (1 - x_2);$$
$$x_2(t+1) = 1.3 x_2(t) \cdot (1 - x_1) + 0.1(x_1(t) - x_2(t)).$$

В результате вычислений — размерность пространства состояний 4, задержка — 14. Для проверки существования аттрактора — реконструируем фазовый портрет системы (рис. 4.11). Видно, что система хаотическая. На рис. 4.12 представлено сравнение динамического поведения модели с исследуемых экспериментальных данных для момента времени от 200 до 250. Высокая оценка адекватности (91% — для первого, 80% — для второго процесса) объясняется тем, что пример был взят модельный, и в нем отсутствуют шумы и ошибка измерений. Матрицы идентифицируемой системы:

$$A = \begin{pmatrix} -0{,}9755 & -0{,}18879 & -0{,}1167 & 0{,}0013756 \\ -0{,}11943 & 0{,}0011553 & 0{,}96261 & -0{,}001488 \\ 0{,}18096 & -0{,}95403 & -0{,}005751 & -0{,}2764 \\ 0{,}017716 & -0{,}11469 & 0{,}24917 & 0{,}20189 \end{pmatrix};$$

$$\Psi(t) = \begin{pmatrix} 0{,}088151 \\ 0{,}24325 \\ -0{,}21885 \\ -1{,}7214 \end{pmatrix} t^2; \quad C = \begin{pmatrix} 14{,}471 & -1{,}7559 & 0{,}87886 & -0{,}17531 \\ -11{,}888 & -5{,}2264 & 1{,}8535 & -0{,}74647 \end{pmatrix}.$$

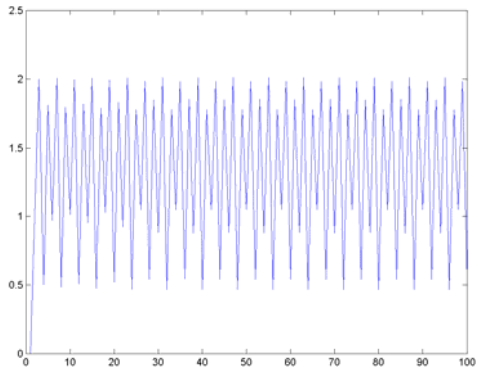 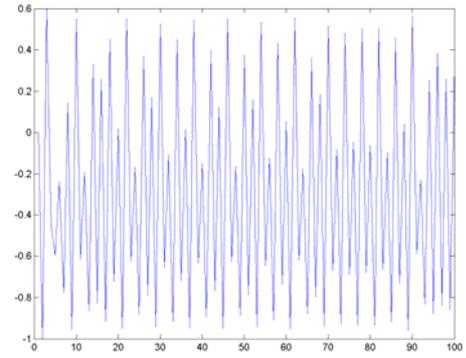

Рис. 4.10. Выходные процессы: а) $y_1(t)$; б) $y_2(t)$

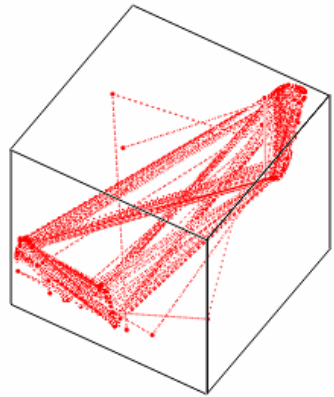

Рис. 4.11. Реконструированный аттрактор

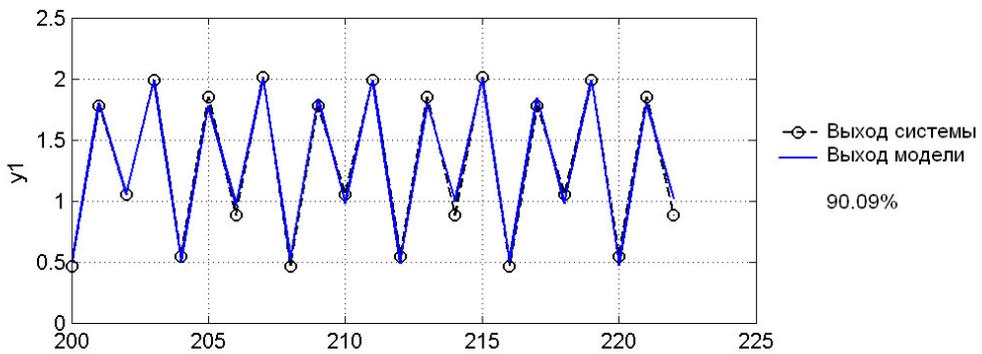

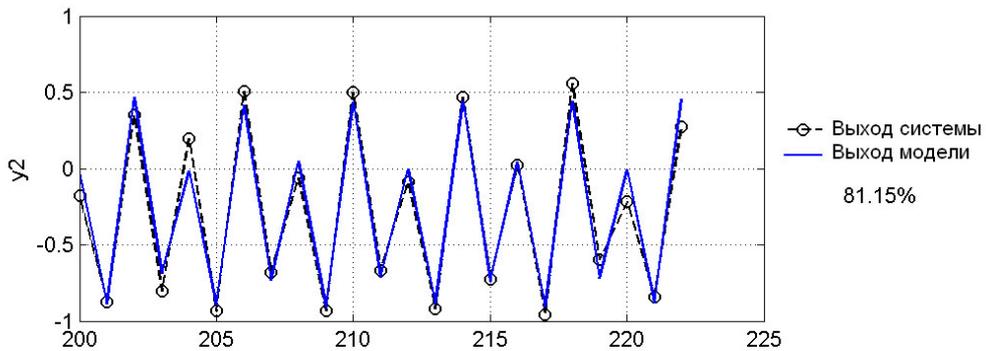

Рис. 4.12. Оценка адекватности моделирования

**Пример 4.** Одним из классических примеров хаотических систем является система Реслера:

$$\dot{x}_1 = -(x_2 + x_3),$$
$$\dot{x}_2 = x_1 + 0.2x_2,$$
$$\dot{x}_3 = 0.2 + x_3(x_1 - 5.7),$$

Simulink-модель которой приведена на рис. 4.13. Результат моделирования показан на рис. 4.14. На рис. 4.15*а* и 4.15*б* соответственно приведены аттракторы исходной и идентифицированной моделей.

Результаты восстановления аттрактора показывают, что для разработанного метода число наблюдаемых точек может быть сокращено, например для системы Реслера с 1000 до 200.

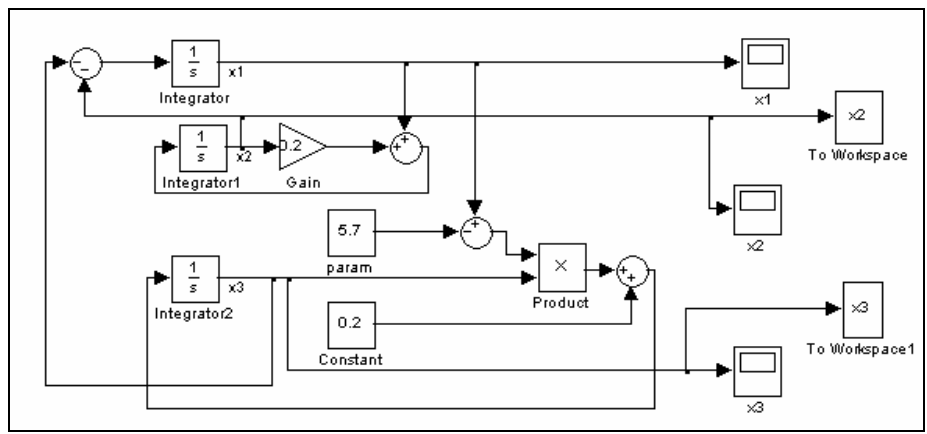

Рис. 4.13. Модель системы Реслера

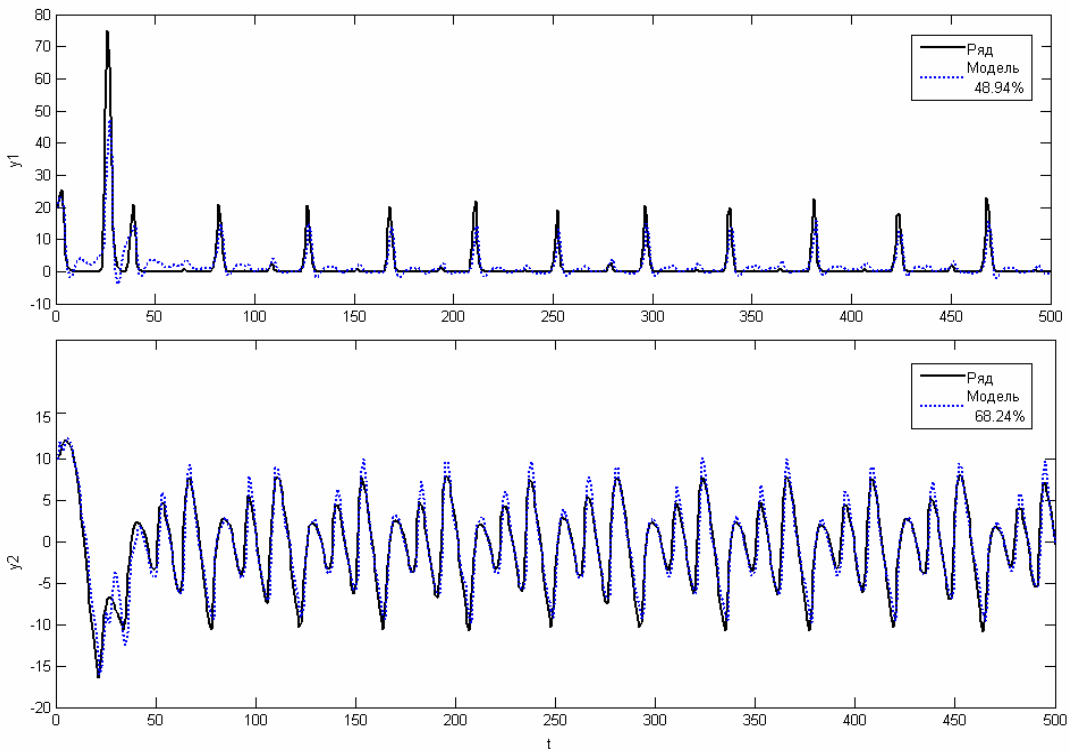

Рис. 4.14. Сравнение динамики исследуемого ряда и построенной модели

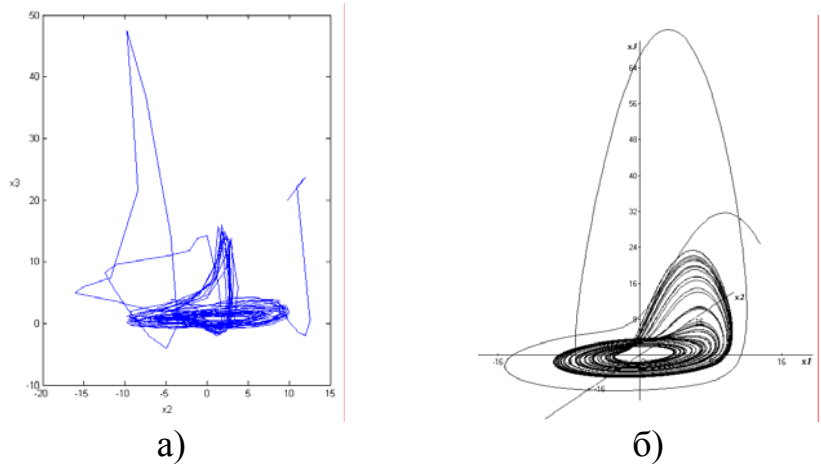

Рис. 4.15. Аттрактор системы Реслера: а) построенный
по реконструированной модели; б) построенный по исходной модели

**Пример 5.** Исходные данные получены с датчиков системы охлаждения алюминиевых сплавов: у1 — скорость охлаждения сплава, у2 — расход хладоносителя. На рис. 4.16 показан аттрактор системы. Результаты сравнения динамики смоделированного процесса с реальными данными приведены на рис. 4.17.

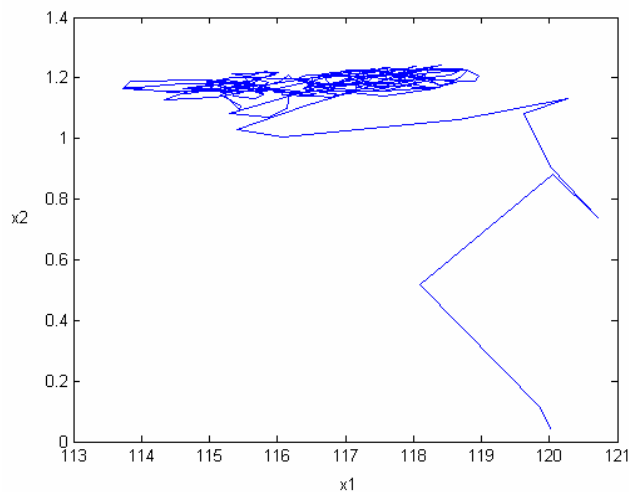

Рис. 4.16. Восстановленный аттрактор системы охлаждения

Реконструированное уравнение имеет вид (4.5) с соответствующими коэффициентами:

$$A = \begin{bmatrix} 0{,}7987 & 0{,}5871 & -0{,}1104 & -0{,}0795 & -0{,}0100 & 0{,}0100 & -0{,}0062 & 0{,}0283 & -0{,}0383 \\ -0{,}5605 & 0{,}6582 & -0{,}2665 & -0{,}3695 & 0{,}0204 & -0{,}3091 & -0{,}1216 & 0{,}0319 & 0{,}0308 \\ -0{,}1076 & 0{,}2538 & 0{,}9248 & -0{,}2668 & -0{,}0639 & -0{,}3874 & -0{,}2682 & -0{,}3668 & 0{,}6859 \\ -0{,}1254 & 0{,}2899 & -0{,}0054 & 0{,}7027 & -0{,}0487 & -0{,}7902 & 0{,}0226 & -0{,}1279 & 0{,}6137 \\ -0{,}0337 & 0{,}0967 & 0{,}2727 & -0{,}0898 & 0{,}7439 & -0{,}6723 & -0{,}3121 & -0{,}0916 & 0{,}6098 \\ -0{,}0186 & 0{,}0371 & -0{,}2327 & 0{,}2056 & 0{,}2983 & 0{,}4822 & 0{,}0862 & -0{,}3811 & -0{,}2154 \\ -0{,}0062 & 0{,}0477 & 0{,}2632 & -0{,}0479 & -0{,}0213 & -0{,}7668 & 0{,}0473 & -0{,}2913 & -0{,}1507 \\ 0{,}0110 & 0{,}0150 & 0{,}1896 & 0{,}1428 & -0{,}3417 & -0{,}0715 & -0{,}3949 & 0{,}5512 & -0{,}0141 \\ 0{,}0537 & -0{,}0289 & -0{,}5188 & 0{,}0982 & -0{,}8413 & 1{,}5644 & 0{,}7244 & -0{,}1601 & -0{,}6049 \end{bmatrix};$$

$$B = \begin{bmatrix} -3{,}9240 & 0{,}7319 \\ 1{,}7215 & -1{,}3336 \\ 2{,}0934 & -1{,}6825 \\ 2{,}4342 & -0{,}8467 \\ -0{,}2892 & -0{,}8989 \\ 2{,}7220 & 1{,}4839 \\ 0{,}4833 & 1{,}0770 \\ 1{,}0878 & -0{,}0228 \\ 2{,}8889 & 0{,}5654 \end{bmatrix} \begin{bmatrix} t^2 - 2t - 0{,}93; \\ \sin(t-10); \end{bmatrix};$$

$$C = \begin{bmatrix} 0{,}1857 & 0{,}0442 & -0{,}0082 & -0{,}0066 & -0{,}0005 & -0{,}0012 & -0{,}0010 & 0{,}0019 & -0{,}0022 \\ 0{,}6124 & 0{,}5277 & -8{,}6861 & 9{,}1780 & -1{,}1570 & -2{,}1478 & -2{,}6497 & 3{,}0497 & -0{,}3083 \end{bmatrix}.$$

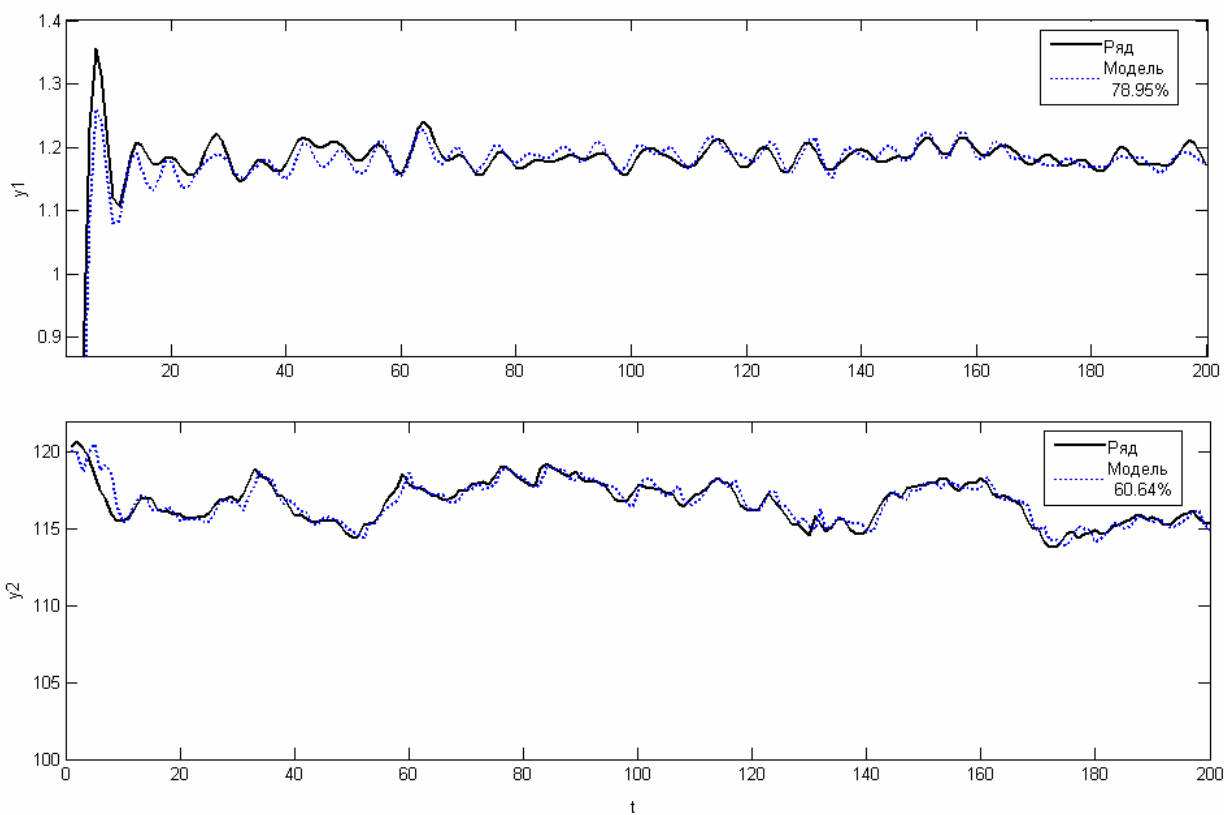

Рис. 4.17. Сравнение экспериментальных данных и построенной модели

**Пример 6.** В результате применения разработанного метода к моделирования процесса добычи нефти, являющийся исследуемых показателем энергетической безопасности с 1996 по 2002 г. (по мес.) получено: размерность пространства состояний $n = 5$, задержка — 12. Сравнение реальных данных по добычи нефти (сплошная линия) с динамическим поведением идентифицированной системы (штриховая линия) показан на рис. 4.18. Оценка адекватности модели составляет 72,5%, т. е. ошибка прогнозирования — 27.4%. Для проверки оценки построена автокорреляционная функция (рис. 4.19).

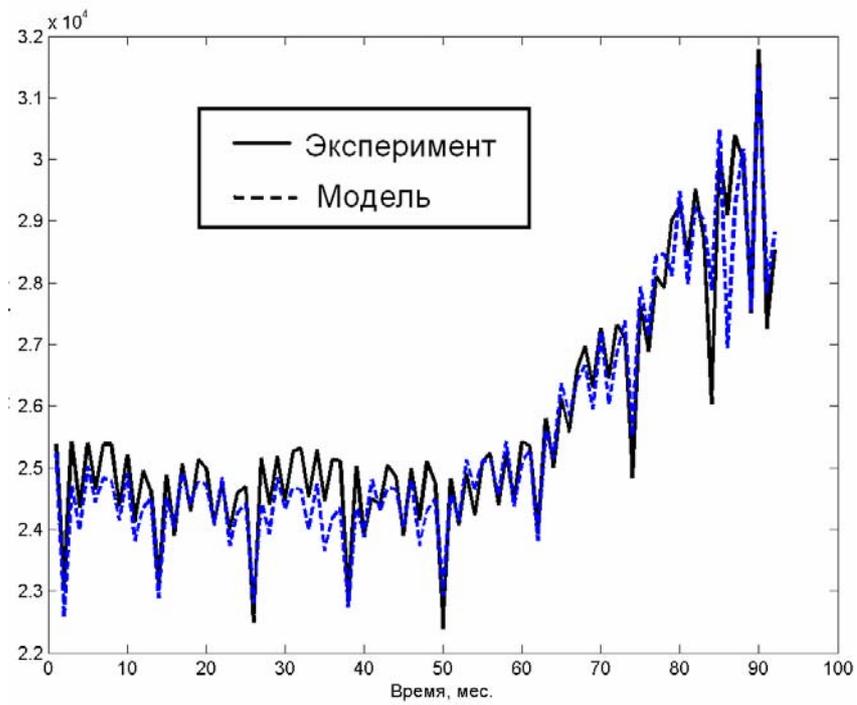

Рис. 4.18. Сравнение результатов моделирования с реальными данными

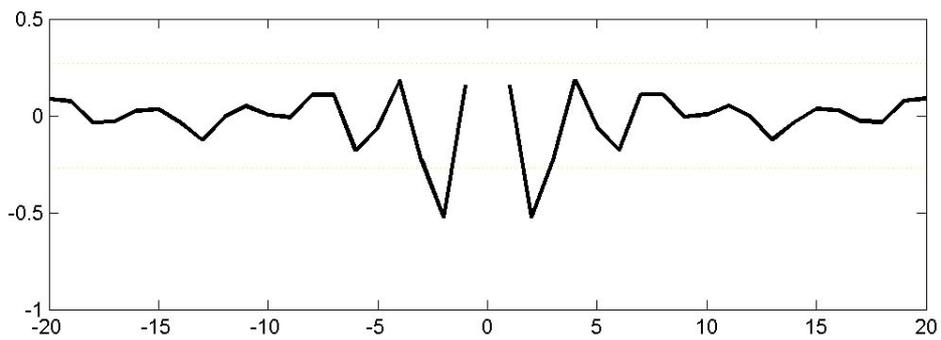

Рис. 4.19. Автокорреляционная функция

Приведенные примеры идентификации определяют достоверность предложенного метода и эффективность его применения для моделирования различных систем.

# ГЛАВА 5. ТЕХНОЛОГИЯ ПОВЫШЕНИЯ КАЧЕСТВА ФУНКЦИОНИРОВАНИЯ УПРАВЛЯЕМЫХ ТЕХНИЧЕСКИХ СИСТЕМ

На основе использования созданных инвариантных геометрических методов и моделей разработана технология, обеспечивающая повышение качества функционирования управляемых технических систем с нелинейной динамикой; приведены результаты, полученные при внедрении разработанных методов при моделировании и управлении промышленных систем.

## 5.1. Технология обеспечения повышения качества функционирования управляемых систем

Наибольшую эффективность применения разработанных в диссертации методов наблюдается при моделировании и управлении системами, в которых происходят нерегулярные и хаотические явления, связанные, например, с процессами теплообмена, потока вязких жидкостей и химическими реакциями [260–280]. Конструкции промышленных устройств, в которых протекают указанные нелинейные явления, делают, как правило, математическое моделирование задачей трудно выполнимой, реконструкция эволюционных уравнений позволяет строить модели качественного поведения на минимальном инерциальном множестве.

Проведенное компьютерное моделирование и анализ результатов внедрения показал, что построенные модели позволяют реализовать требования к качеству функционирования системы, связанные с управлением нелинейных явлений. Учет нерегулярного поведения технических систем позволил уменьшить энергетические затраты на управление, сократить время переходных процессов и обеспечить заданные параметры надежности систем.

Разработана оригинальная технология повышения качества функционирования управляемых технических систем с нелинейной динамикой. Технология включает шесть этапов.

*1. Проведение экспериментов и обработка исходных данных.* Включает использование методов, разработанных и используемых в теории нейронных сетей: методы предварительной обработки данных, методы оценки статистических характеристик и др. [283], а также описанного в четвертой главе модифицированного метода реконструкции аттрактора и методов оценки инвариантных характеристик [251].

*2. Идентификация параметров модели системы.* Используется предложенный в четвертой главе метод моделирования по экспериментальным данным систем, редуцированных на центральное многообразие [278].

*3. Формулировка требований к качеству управления.* Представляет собой формализацию критериев качества и надежности, а также применение сформулированных в пятой главе геометрических критериев локальной управляемости, наблюдаемости и критерия компактности множества достижимости [281].

*4. Проектирование системы управления с применением геометрических методов.* Применяются разработанные в пятой главе методики построения алгоритмов управления с учетом характеристик построенных моделей (аттрактора, спектра показателей Ляпунова, инфинитезимальных операторов и др.) и заданных критериев качества [283].

*5. Имитационное моделирование схемы управления и адаптация.* Заключается в настройке параметров управляющих механизмов в различных режимах функционирования [276].

*6. Внедрение разработанной САУ.* Обеспечивает анализ результатов внедрения и возможность модификации параметров модели при значительном изменении условий эксплуатации промышленной системы.

Применение в практических задачах обеспечения повышения качества систем в различных областях промышленности рекомендуется в следующих случаях.

1. В условиях функционирования существующих систем управления промышленными объектами, когда необходимо уменьшить колебания выходных параметров.

2. В условиях существующих режимов функционирования, когда необходимо улучшить качество управления: увеличить быстродействие системы, удовлетворить заданным ограничениям на управляющие сигналы и контролируемые параметры.

3. При наличии существующих моделей процессы теплообмена и потока жидкостей (т. е. те, которые исходя из своей физической природы описываются уравнениями в частных производных, допускающих группы симметрий), не удовлетворяющим заданным требованиям к качеству и надежности объекта.

4. Когда физические явления являются диссипативными (что может быть проверено просто путем численного дифференцирования экспериментальных данных).

5. При конструктивной возможности изменения числа управляющих параметров нелинейным явлением в условиях изменения режима функционирования объекта.

6. Для прогнозирования значений нелинейных диссипативных систем или систем, движение которыми ограничивается центральным многообразием.

Анализ применения разработанных методов для различных временных рядов показал, что в следующих случаях повышение качества управления не обеспечивается.

1. При наличии ограничений на изменение контролируемых параметров в режимах функционирования, в которых не проводились эксперименты.

2. При наличии значительного постоянного неизменяемого (случайного) управляющего сигнала.

3. В случаях, если нет возможности (если это требуется разработанным критерием управляемости) добавления управляющего механизма.

Однако в ряде ситуаций модели могут быть использованы в совокупности с другими методами и технологиями. Например, модели могут быть автоматически настраиваемыми с использованием механизмов переобучения.

Таким образом, в условиях указанных ограничений внедрение разработанной технологии обеспечивает повышение качества функционирования управляемых технических систем и промышленных объектов.

### *5.2. Методика применения прогнозирующих моделей*

Разработанные модели, редуцированные на центральное многообразие могут быть использованы в системах принятия решений и совместно с методами управления существующих промышленных контроллеров.

Методы и программные средства моделирования внедрены в НИИ «Энергия» Главного управления информационных систем Спецсвязи РФ в рамках информационной системы поддержки принятия решений. Исследования дали возможность построить аттракторы для моделей изменения параметров, анализ которых позволил увеличить качество прогнозирования необходимых значений.

Параметры модели имеет вид:

$$A = \begin{bmatrix} 0,57622 & 0,13284 & -0,20088 & 0,37433 & -0,023112 & 0,37285 & -0,31885 \\ -0,17116 & 0,31824 & 0,27878 & -0,43661 & -0,27536 & -0,66245 & -0,70696 \\ 0,069311 & -0,24891 & -0,15565 & -0,35583 & -0,9155 & 0,10894 & -0,24585 \\ 0,19068 & -0,29474 & -0,62437 & -0,58184 & 0,30318 & -0,29116 & -0,35317 \\ 0,17889 & 0,020402 & 0,43483 & -0,66442 & 0,1957 & 0,50556 & 0,11334 \\ -0,3242 & 0,79269 & -0,57987 & 0,099279 & 0,066329 & 0,45846 & 0,68769 \\ 0,049435 & 0,36525 & 0,070016 & -0,04069 & 0,12744 & 0,078338 & -0,63774 \end{bmatrix},$$

$$\Psi = 10^{-3}\left(9.4t^7 + t^4 + 0.059\sin(t)\right) \times$$

$$\times (0,0385; -0,1320; -0,4010; -0,5072; -0,2073; 0,0841; -0,0460)^T,$$

$$C = 10^3 \cdot [3,5234 \quad -0,0210 \quad 2,1493 \quad -2,3071 \quad -0,0303 \quad -3,1276 \quad 1,9016].$$

На рис. 5.1 приведено сравнение динамики построенных моделей с реально принятыми значениями.

Для систем принятий решений самостоятельное значение имеют вид аттракторов, предоставляющие важную информацию о поведении системы. На рис. 5.2*а* приведена динамика одного из показателей, характеризующего энергетическую безопасность, на рис. 5.2*б* — пример прогноза в области стационарности нелинейной модели, на рис. 5.2*в* — восстановленный аттрактор.

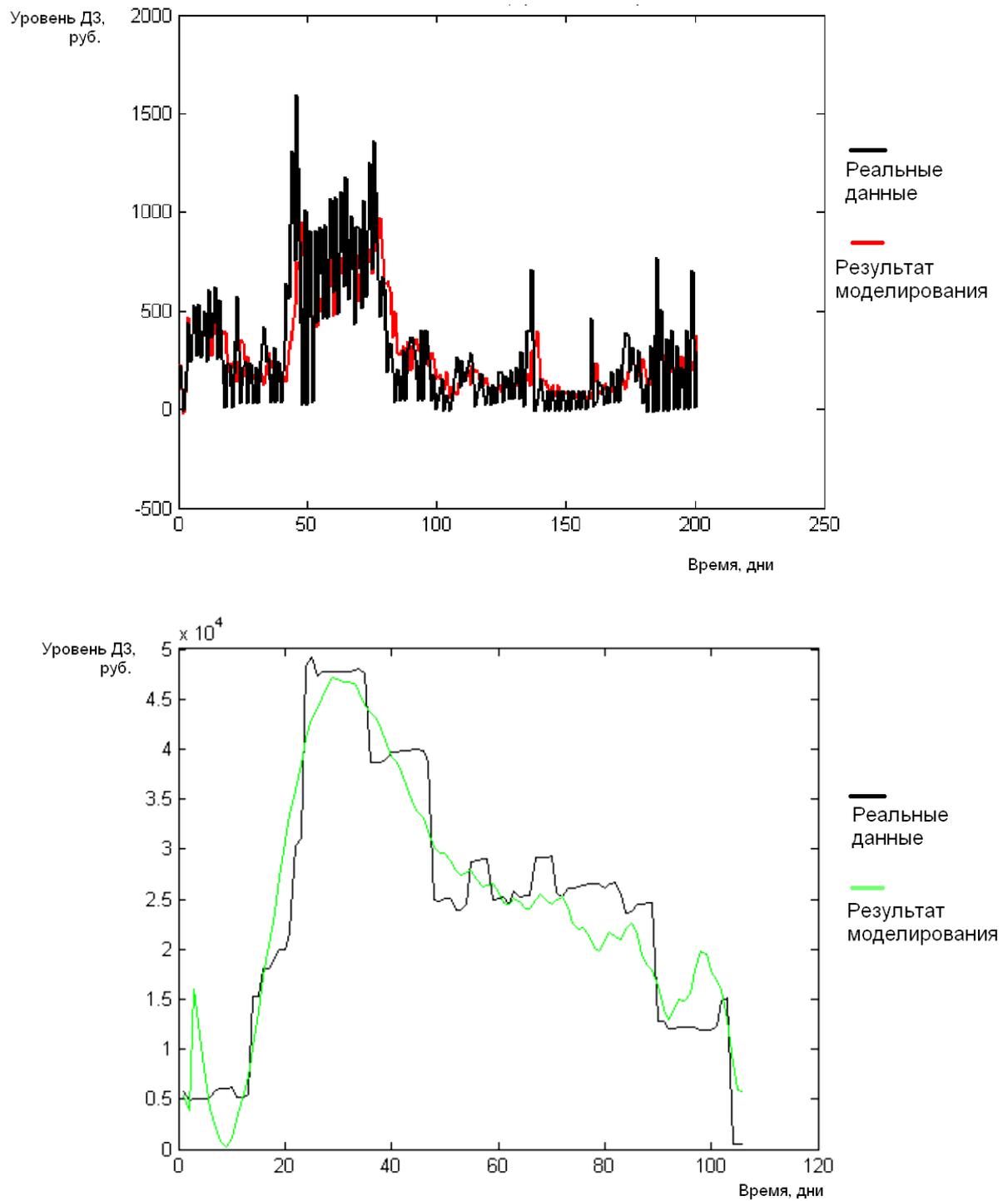

Рис. 5.1. Сравнение динамика реальных значений показателей с моделями

Построенная модель, также как и классические модели (регрессионные, разложение в ряды Фурье и Тейлора) для подобных временных рядов не позволяет строить прогноз более чем на 2–3 шага, вид фазового портрета говорит о фазовом переходе состояний системы, в котором система может находиться какое-то время (как это бы при $t < 50$), либо продолжит переход на другую окрестность. Окончательное решение остается за лицом, принимающее решение, в распоряжении которого имеется экспертный анализ остальных показателей.

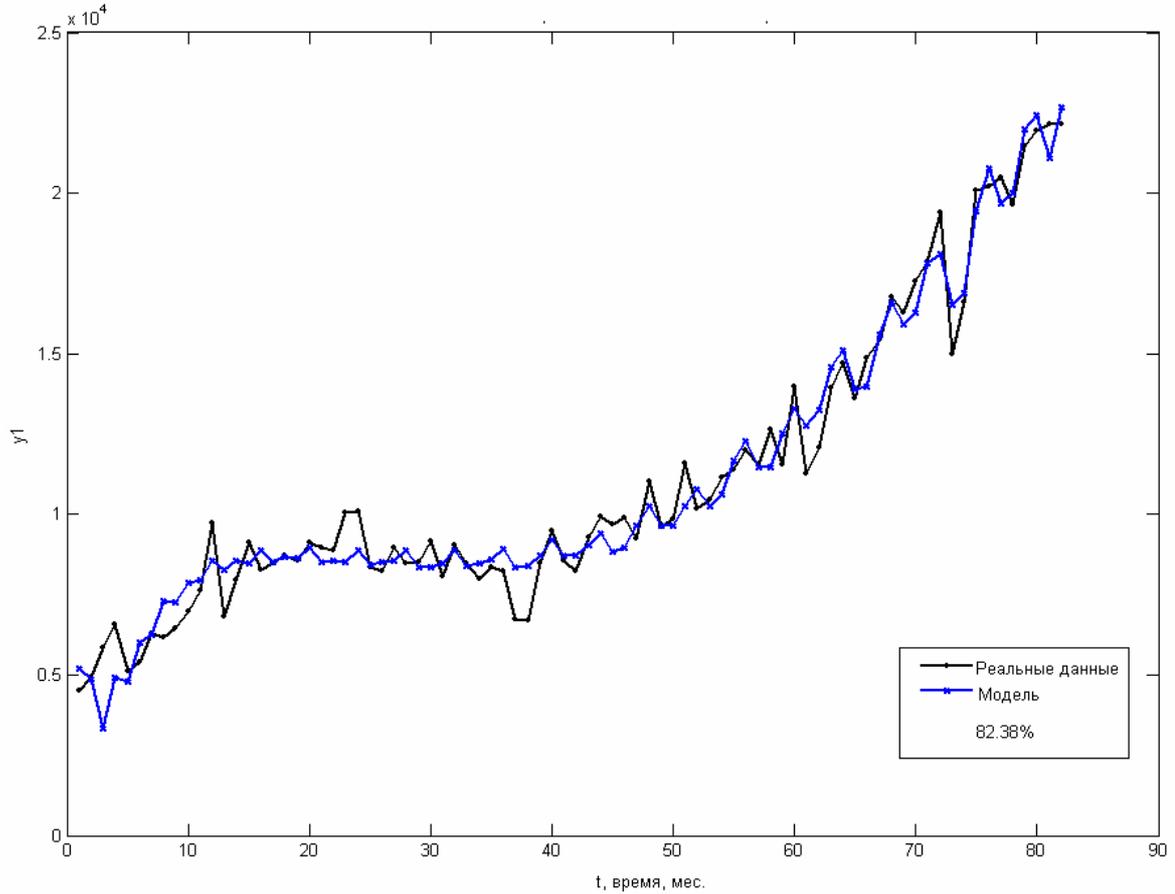

а)

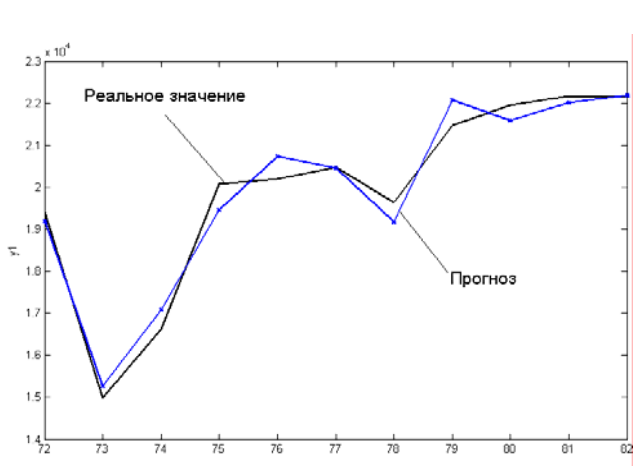

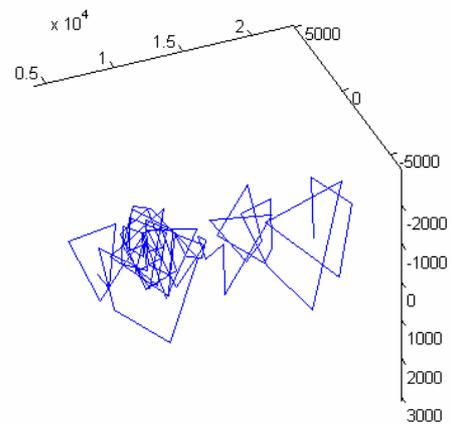

б) в)

Рис. 5.2. Результаты моделирования

На рис. 5.3 представлена функциональная схема системы принятия решений. Цель исследования конкретных объектов определяет выбор модели и набора параметров. Генератор модели на основе информации, хранящейся во БД, формирует задачи оптимизации и идентификации, записывая их в специальном формате. По существу генератор создает целое семейство оптимизационных задач. Выходной файл генератора моделей передается на вход блока, реализующего алгоритмы анализа и принятия решений, который после решения выдает информацию также в специальном виде. В базе знаний накапливается информация об исследованных задачах (правила агрегирования, поведение объектов в динамике, значения оптимальных решений и параметров и т. д.) для последующего использования и построения новых моделей на основе полученных знаний.

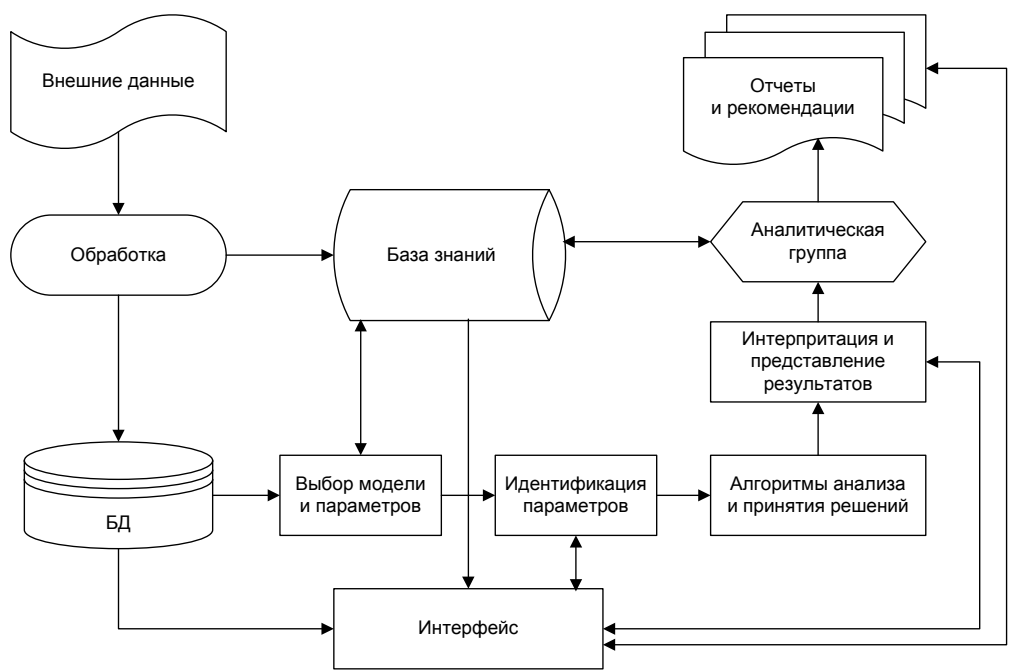

Рис. 5.3. Функциональная схема информационной системы поддержки принятия решений

Результаты внедрения позволили улучшить качество принимаемых решений в области энергетической безопасности.

Аналогично может быть использованы методы моделирования в условиях, когда на выбор метода управления накладываются ограничения, выдвигаемые типом используемых промышленных контроллеров [258].

В качестве промышленного контроллера использовались контроллеры Modicon, в которых на аппаратном уровне существует команда ПИД-регулирования. В команде PID2 реализуется схема управления, цель которого – поддержка нулевого значения отклонения E, вычисляемого как разность измеренного параметра $\xi$ и заданного значения s (рис. 5.4). При этом реализуется алгоритм управления с обратной связью, аналогичный традиционным пневматическим и аналоговым электронным контроллерам, с использованием фильтра высокочастотных параметров и источников шума.

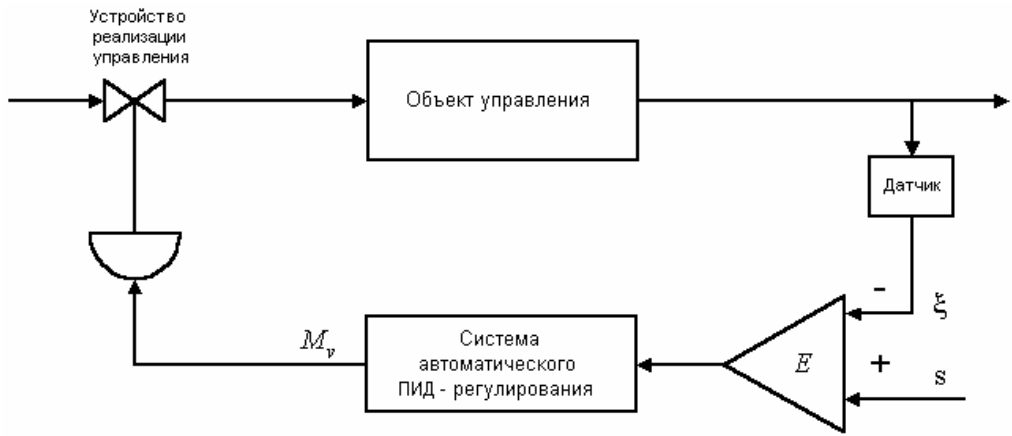

Рис. 5.4. Схема ПИД-регулирования: $\xi$ – измеряемое значение; s – заданное значение параметра; $M_v$ – управляющее воздействие; E – отклонение $\xi$ от s

При пропорциональном регулировании управление получается как произведение отклонения E на константу $K_p$ со смещением:

$$M_v = K_p E + \text{смещение}$$

Пропорционально-интегральное управление устраняет смещение, интегрируя $E$ по времени:

$$M_v = K_p \left( E + K_i \int_0^t E dt \right),$$

где $K_i$ — константа интегральной составляющей регулятора.

Алгоритм ПИД-управления имеет вид:

$$M_v = K_p \left( E + K_i \int_0^t E dt + K_d \frac{d\xi}{dt} \right),$$

где $K_d$ – постоянная дифференциальной составляющей регулятора.

Для решения задачи моделирования и построения системы автоматического регулирования сформирована методика, состоящая из четырех шагов.

*1. Идентификация системы.* По экспериментальным данным в переходном режиме и в режиме нормального функционирования происходит построение системы.

*2. Проектирование системы управления.* Выбирается схема управления промышленной системой и, используя описанные выше средства MATLAB, рассчитываются параметры ПИД–регулятора. К классически применяемым методам можно отнести метод корневого годографа и метод синтеза систем регулирования по желаемым ЛАХ (логарифмическим амплитудным характеристикам), которые реализованы в Control System Toolbox в виде отдельного приложения SISO-tool с пользовательским интерфейсом.

*3. Имитационное моделирование схемы управления.* Осуществляется для тестирования и оптимизации параметров. Для наглядности рекомендуется использовать Simulink. Типовая схема приведена на рис. 5.5. В подсистеме «Объект» находится собственно идентифицированная модель промышленного объекта.

*4. Настройка параметров регулирования.* Для выполнения расчета параметров регулирования при заданных ограничениях и критериях качества этапа используется пакет Nonlinear Control Design, в котором задаются ограничения на вид переходных процессов (рис. 5.6).

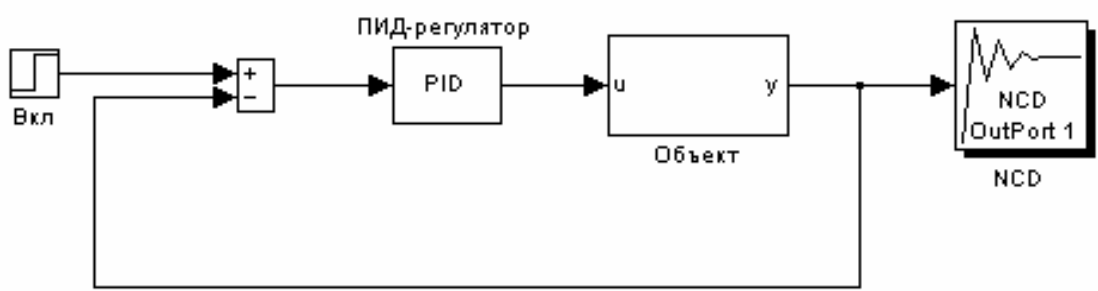

Рис. 5.5. Модель схемы управления в Simulink

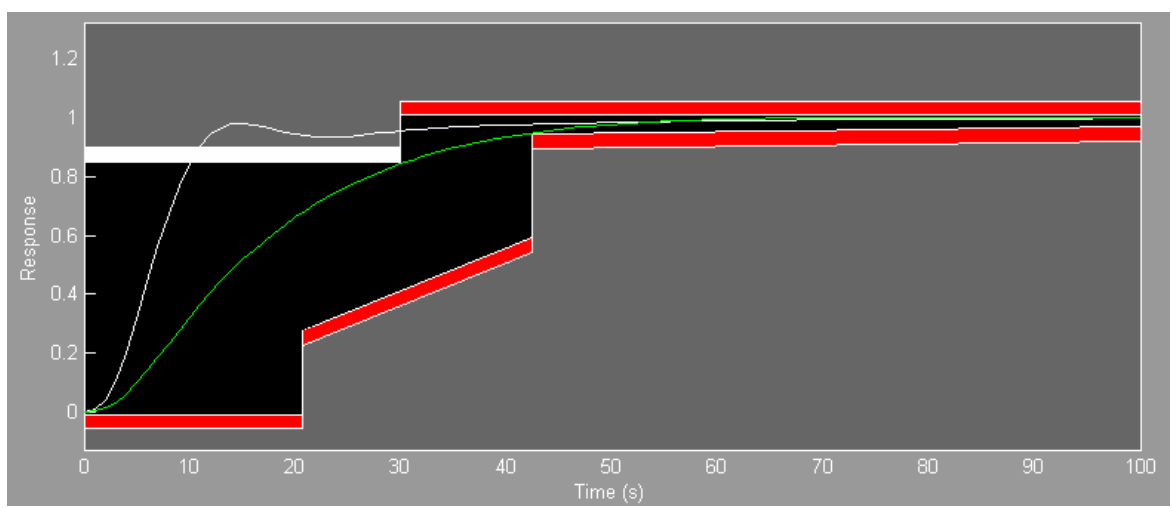

Рис. 5.6. Настройка параметров регулирования по заданному виду переходного процесса в Nonlinear Control Design

Для обоих рассмотренных случаев разработана типовая схема взаимодействия программных средств, которая представлена на рис. 5.7.

В программной реализации используются следующие программные средства:
– среда визуального моделирования Simulink, предназначена для создания базы знаний моделей и занесения их с M-file;

- System Identification Toolbox — подпрограмма идентификации динамических моделей;
- Control System Toolbox — исследование и моделирования динамических систем.

Применение этой методики было осуществлено на предприятиях филиала Мосэнерго, кампании «Кампина», ООО «Энергостроймонтаж».

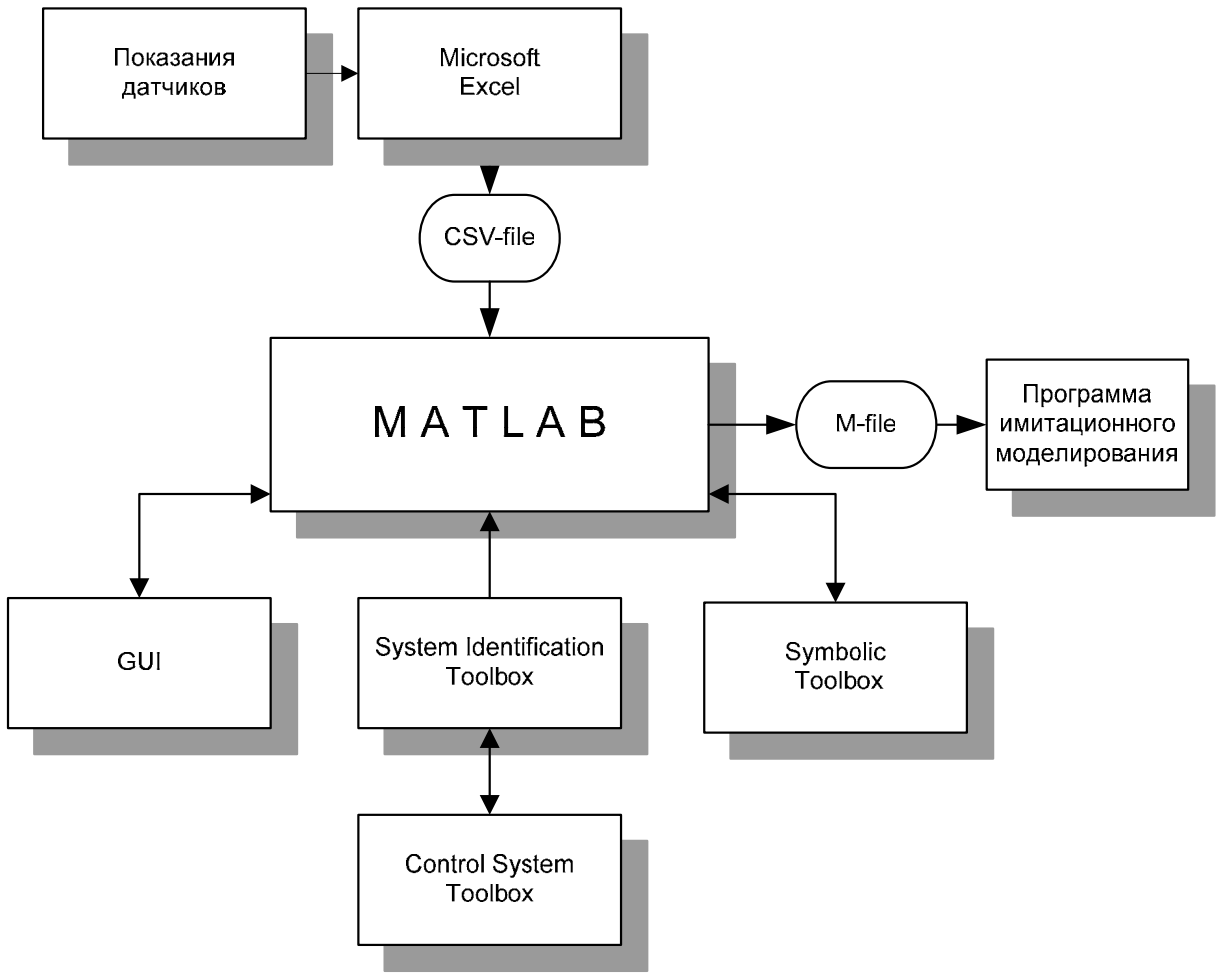

Рис. 5.7. Взаимодействие программных средств

На рис. 5.8. приведен результат сравнения результатов моделирования системы разогрева котла. На вход котла подается отфильтрованная вода, а на вход турбины уже идет так называемый острый пар. На входе и выходе снимается температура воды и острого пара соответственно. В состав входит две одинаковые цепочки *подача в котёл—разогрев—подача в турбину* пара. Включение оборудования происходит следующим образом. Последовательно осуществляется подача воды в котел до определенного уровня, разогрев горелок и, соответственно, воды в котле, затем система управления осуществляет выход на плановый расход воды. Цель моделирования — обеспечение безопасного режима функционирования.

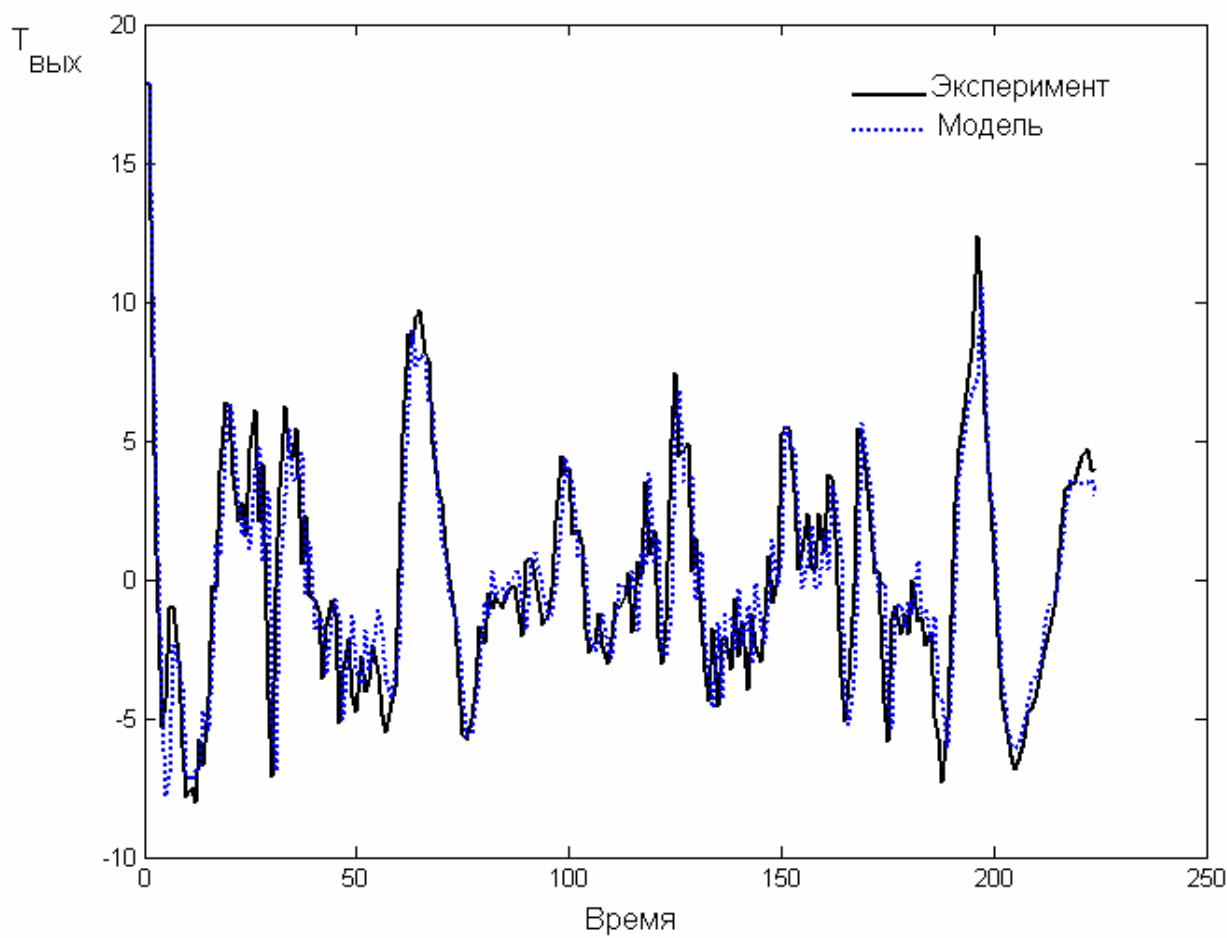

Рис. 5.8. Функционирование системы запуска котла на ТЭЦ

В результате внедрения система автоматического управления обеспечивает заданным требованиям.

### 5.3. Управление системой теплообмена с вязкой средой

В качестве объекта исследования использована система теплообмена, осуществляющая процесс нагрева потока вязкой жидкости в компании «Марс». Исходными данными являются только показания датчиков в переходных и нормальных режимах функционирования.

Система состоит из бесконтактного теплообменника, на вход которого в качестве теплоносителя подается пар. Измеряемыми параметрами являются: давление пара (теплоносителя) и выходная температура нагретого продукта. Воздействие на процесс осуществляется управляемым клапаном. Необходимо регулировать значения давления и температуры выходного продукта при заданных ограничениях на скорость роста давления пара и максимально допустимое значение температуры продукта. В системе для повышения качества управления был дополнительный клапан и связанный с ним контур регулирования, что, однако не позволяло обеспечивать заданные показатели качества.

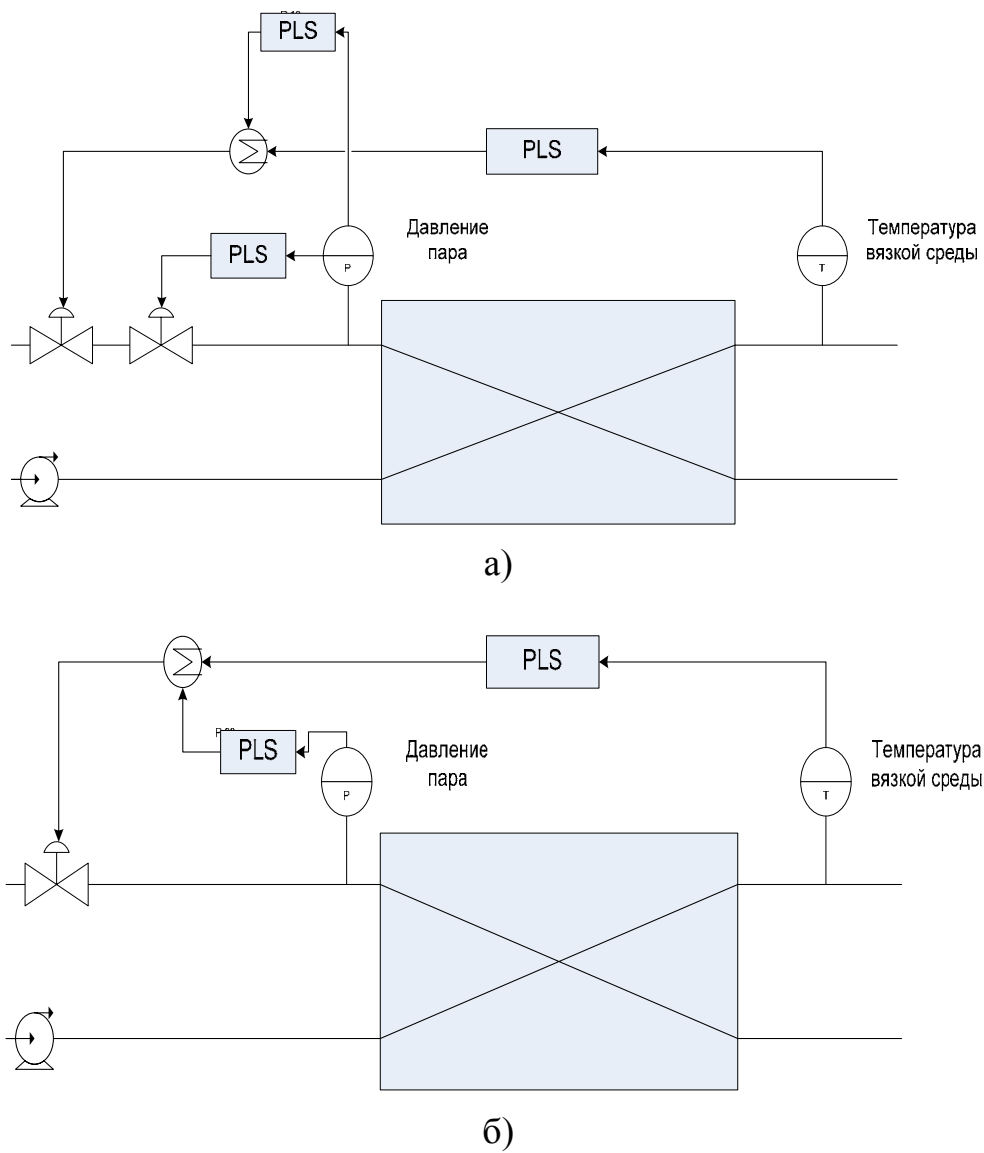

Рис. 5.9 Схема регулирования теплообменником:
а) исходная; б) после применения созданной технологии

Сложность задачи — нелинейность процесса теплообмена, и, как следствие, динамически-сложное поведение системы. Это проявляется в том, что при постоянном значении управления возникают колебания выходных параметров. На рис. 5.10 показано изменение состояния управляемого клапана (рис. 5.10*а*) давления (рис. 5.10*б*), температуры продукта (рис. 5.10*в*) во времени. Показания аналоговых датчиков промасштабированы для регистров промышленного контроллера от 0 до 4000. Значения состояния клапана изменяется в пределах 1–100% (на графике: реальное значение $\times$ 4000/100) давление пара изменяется в пределах от 1÷10 бар (на графике: реальное значение $\times$ 4000/10); температуры: 0÷150 $^0$С (на графике: реальное значение $\times$ 4000/150).

Разработана схема управления, представленная на рис. 5.9*б*. Для рассматриваемой системы реконструирован фазовый портрет (рис. 5.12); размерность пространства состояний $n = 6$.

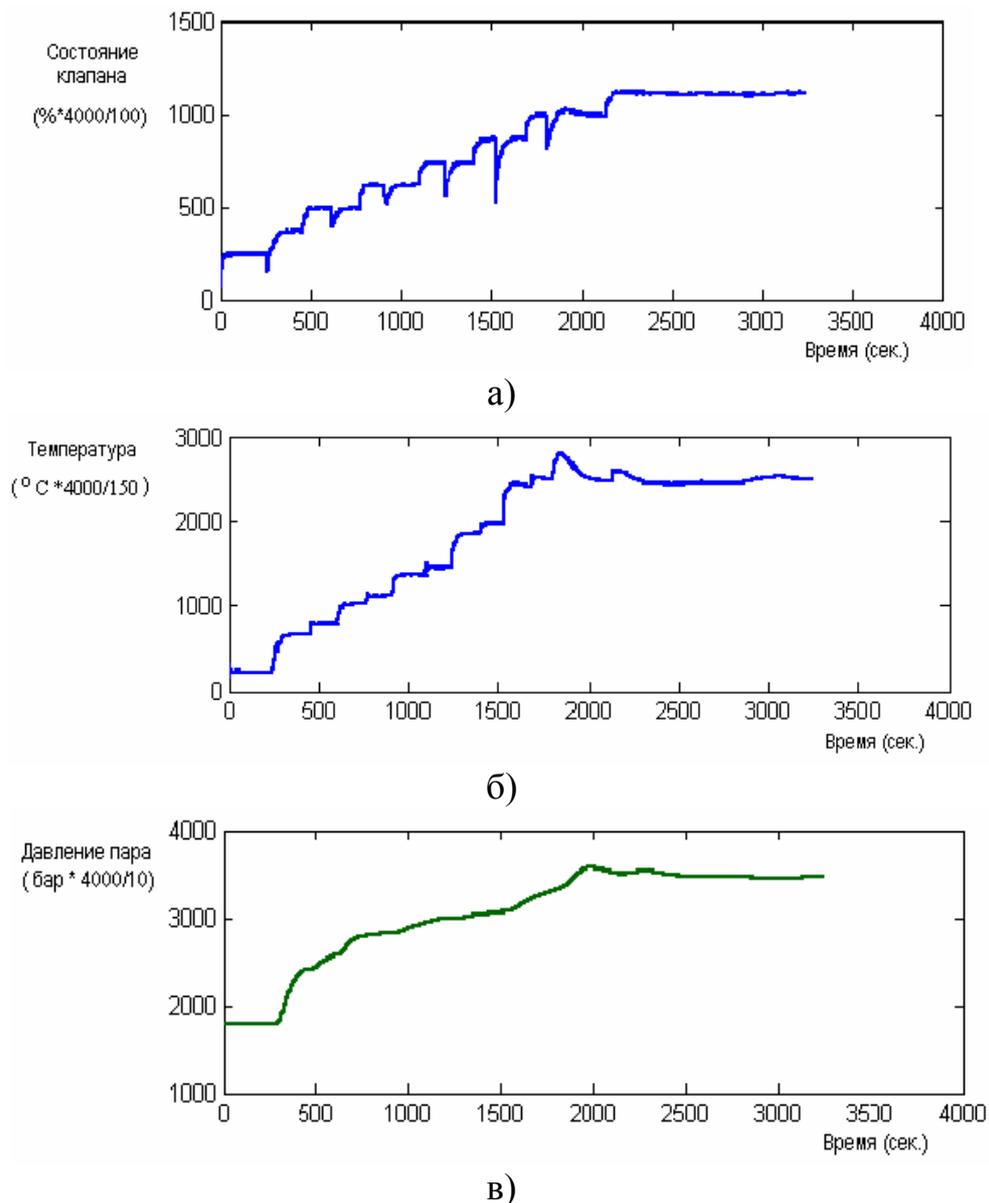

Рис. 5.10. Динамика процесса при «ручном» регулировании процесса

Применение разработанного в пятой главе метода дает уравнения с соответствующими значениями:

$$A = \begin{pmatrix} 0,998 & 0,006 & -0,009 & 0,007 & -0,001 & -0,007 \\ 0,008 & 0,913 & 0,238 & 0,142 & 0,091 & -0,215 \\ 0,006 & 0,132 & 0,091 & -0,928 & -0,459 & -0,038 \\ -8,12 \cdot 10^{-5} & 0,008 & -0,177 & 0,356 & -0,956 & 0,097 \\ -0,001 & -0,06 & 0,803 & 0,203 & -0,145 & 0,139 \\ 0,002 & -0,006 & 0,006 & -0,003 & 0,010 & 0,661 \end{pmatrix};$$

$$\Psi = \exp(10t)\left(-7,24 \cdot 10^{-4}; -0,087; 0,209; -0,074; -0,440; -0,061\right)^{T};$$

$$C = \begin{pmatrix} 129,8 & -144,2 & 18,43 & -23,21 & -3,233 & -4,911 \\ 582,1 & 4,995 & 2,230 & -1,596 & 2,587 & 4,100 \end{pmatrix}.$$

В таб. 5.1 показаны вычисленные значения инвариантных характеристик

Таблица 5.1. Значения инвариантных характеристик

| Размерность минимального вложения | Время τ | Старший показатель Ляпунова | Спектр показателей Ляпунова | | |
|---|---|---|---|---|---|
| 6 | 12 | 0,03230 | 0,02814 | 0,00217 | –0,0537 |
|   |   |   | –0,0823 | –0,0072 | –0,0034 |

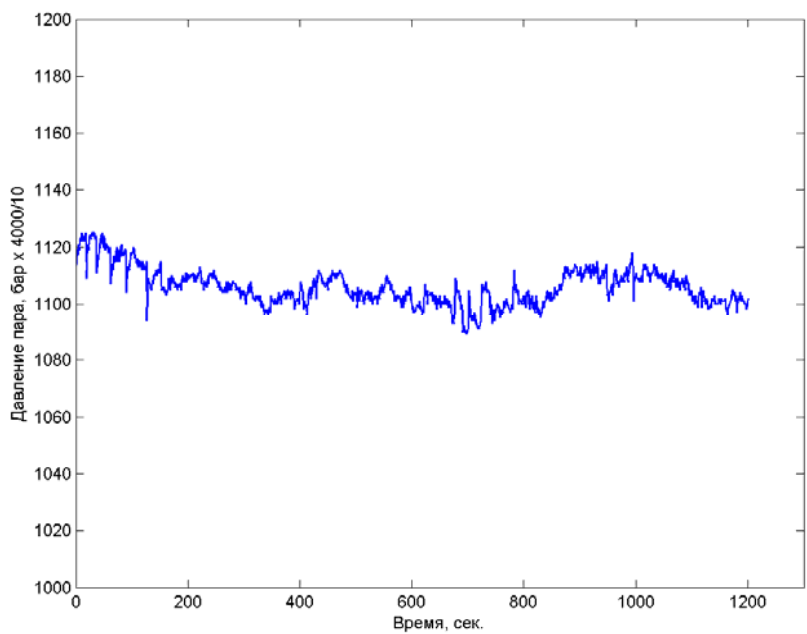

а)

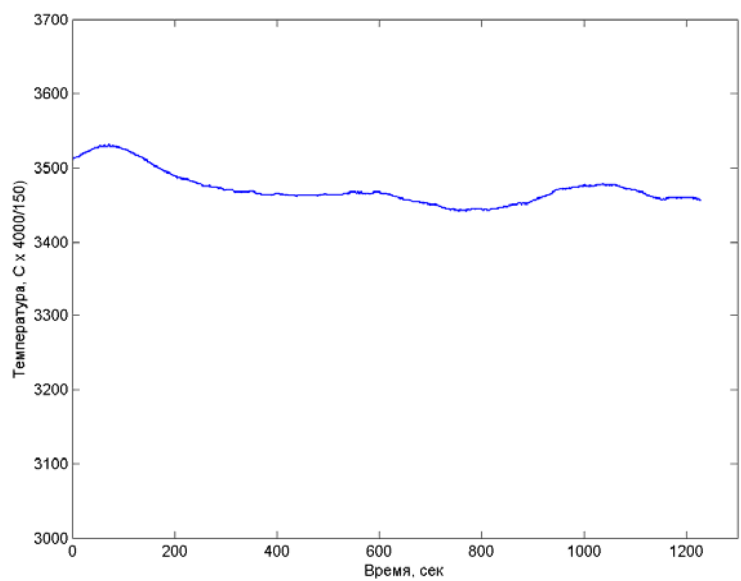

б)

Рис. 5.11. Динамическое поведение системы теплообмена в установившемся режиме

Сравнение моделируемого выходного процесса с реальными данными показано на рис. 5.13. Видно, что модель демонстрирует сложное динамическое поведение, адекватное рассматриваемому теплообменному процессу. На рис. 5.14 и 5.15 показаны графики переходных процессов во времени для решения задачи оптимального быстродействия.

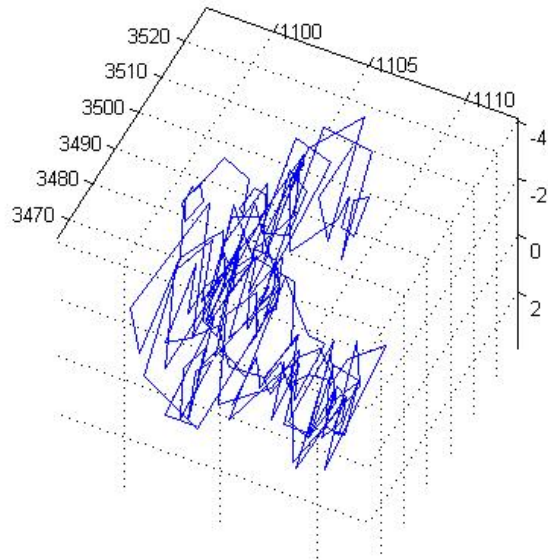

Рис. 5.12. Реконструированный фазовый портрет

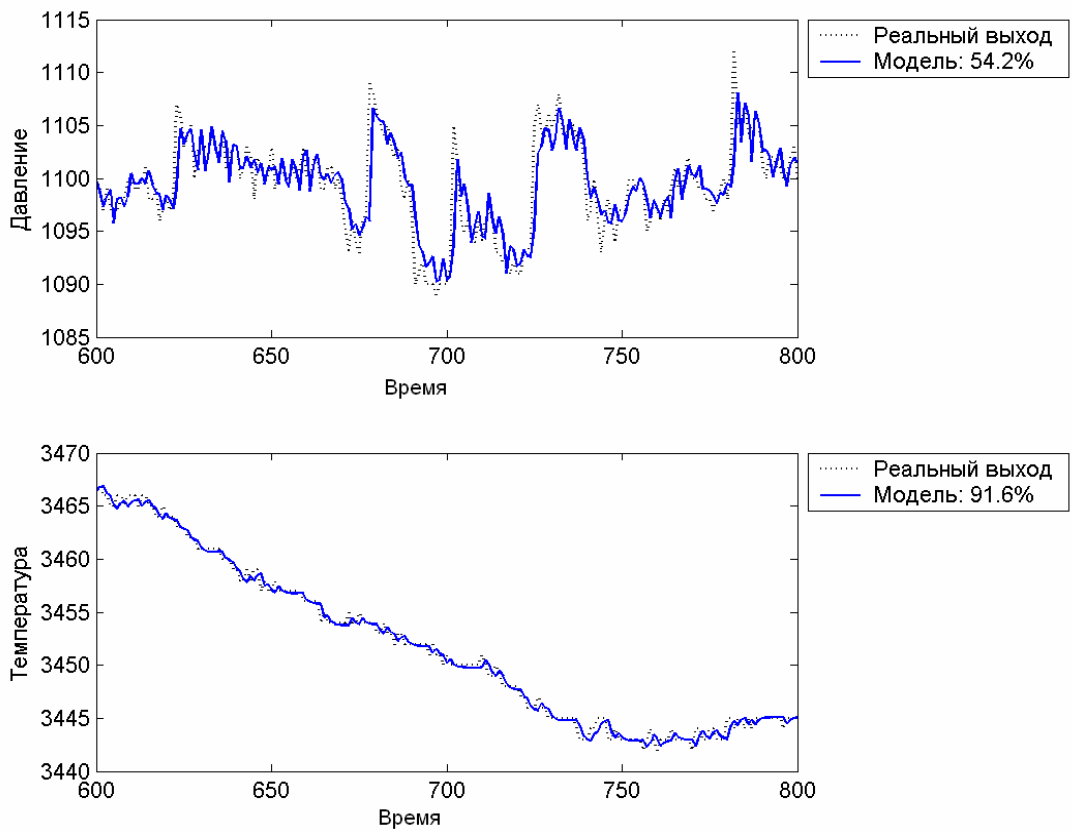

Рис. 5.13. Сравнение результатов моделирования с реальными данными

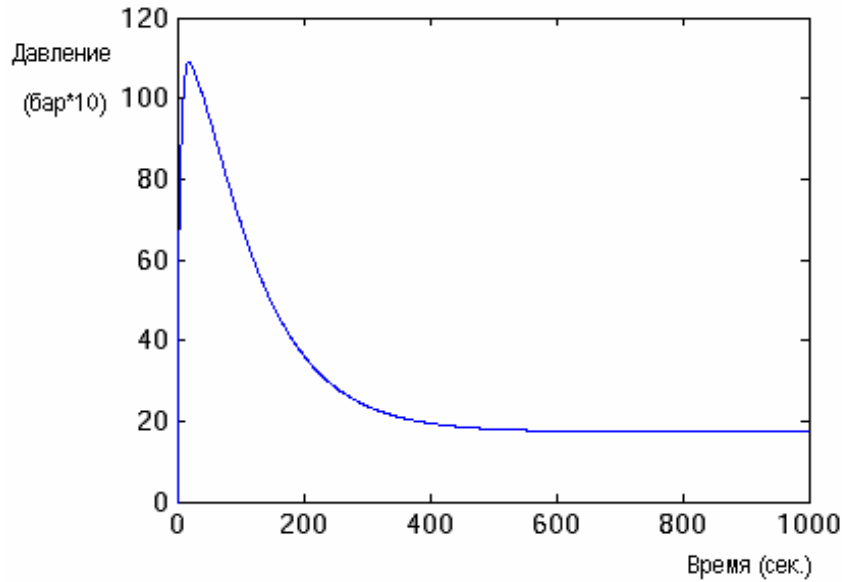

Рис. 5.14. Динамика изменения давления при оптимальном быстродействии

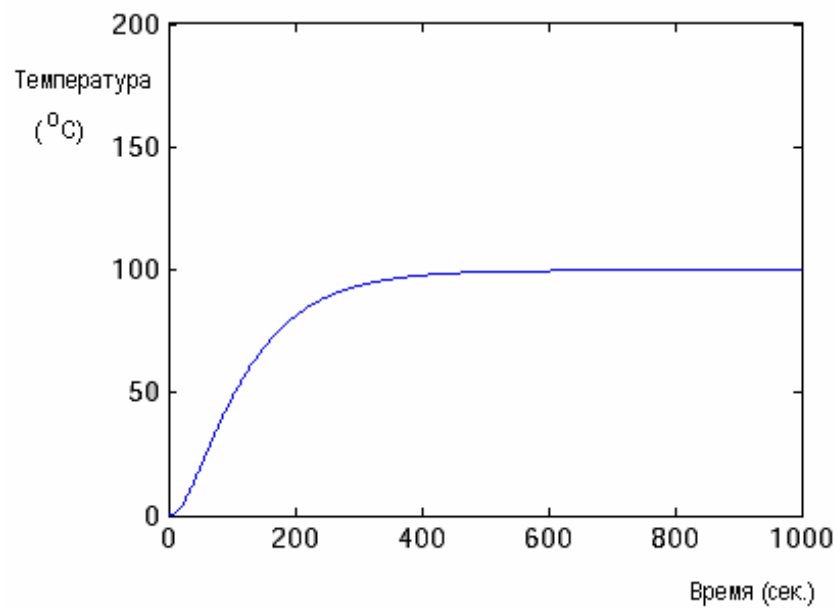

Рис. 5.15. Динамика изменения температуры при оптимальном быстродействии

Полученная модель позволила решить задачу проектирования системы регулирования теплообменника по заданным критериям качества. К системе для безопасного функционирования предъявляются следующие требования: ограничения на максимально допустимое значение температуры, минимум скорости роста давления, минимально допустимое значение быстродействия.

Применяя разработанные алгоритмы и модели, получены результаты функционирования системы с учетом заданных критериев качества, которые показаны на рис. 5.16 и 5.17.

Внедрение подтвердило полученный результат. В результате выход системы на заданный режим функционирования уменьшился с 4000 с до 2500 с (1,6 раза).

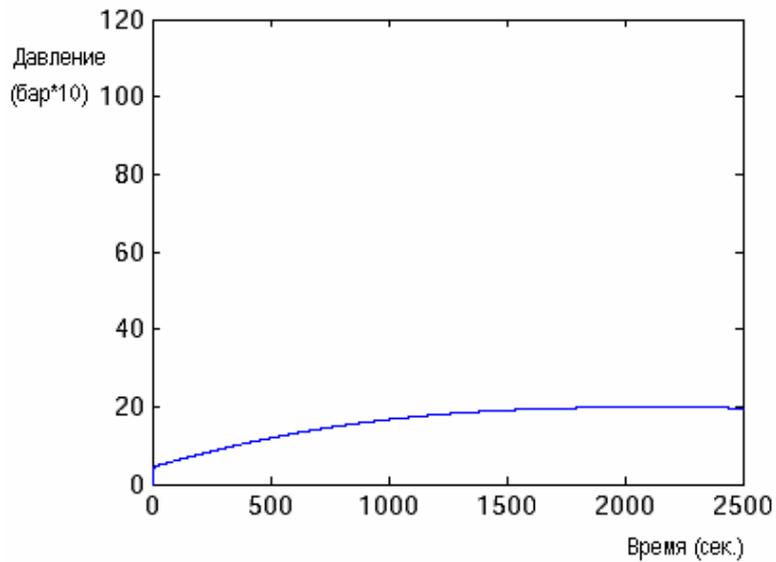

Рис. 5.16. Динамика изменения давления с учетом критериев качества

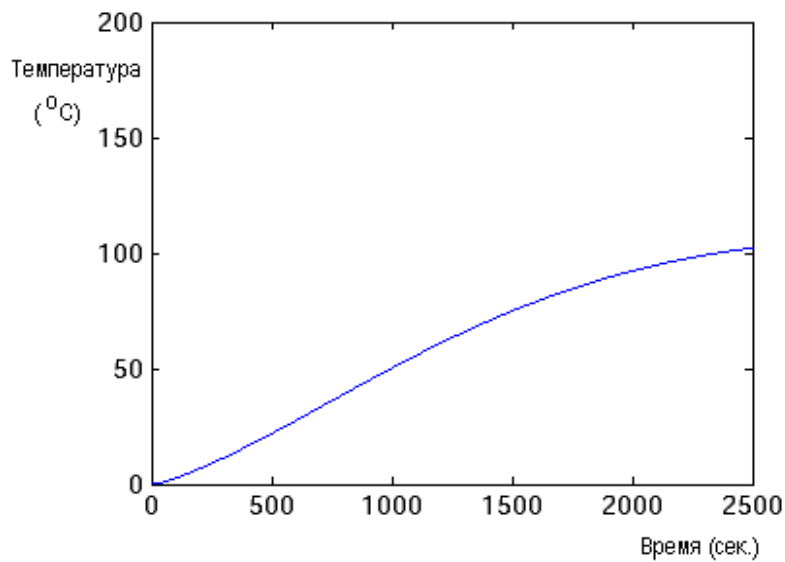

Рис. 5.17. Динамика изменения температуры с учетом критериев качества

Разработан регулятор для технологического процесса нагрева вязкой жидкости, при этом контролируемым параметром является расход вязкой среды. Работа выполнялась также для компании «Марс» по заказу подрядчика-изготовителя системы ООО «Центр передовых технологий «Базис».

Входной параметр $u(t)$ — состояние клапана (%); выходной — $y(t)$ — скорость потока вязкой жидкости (кг/ч). Результат моделирования на центральном многообразии приведен на рис. 5.18. Адекватность управляемой модели 88%.

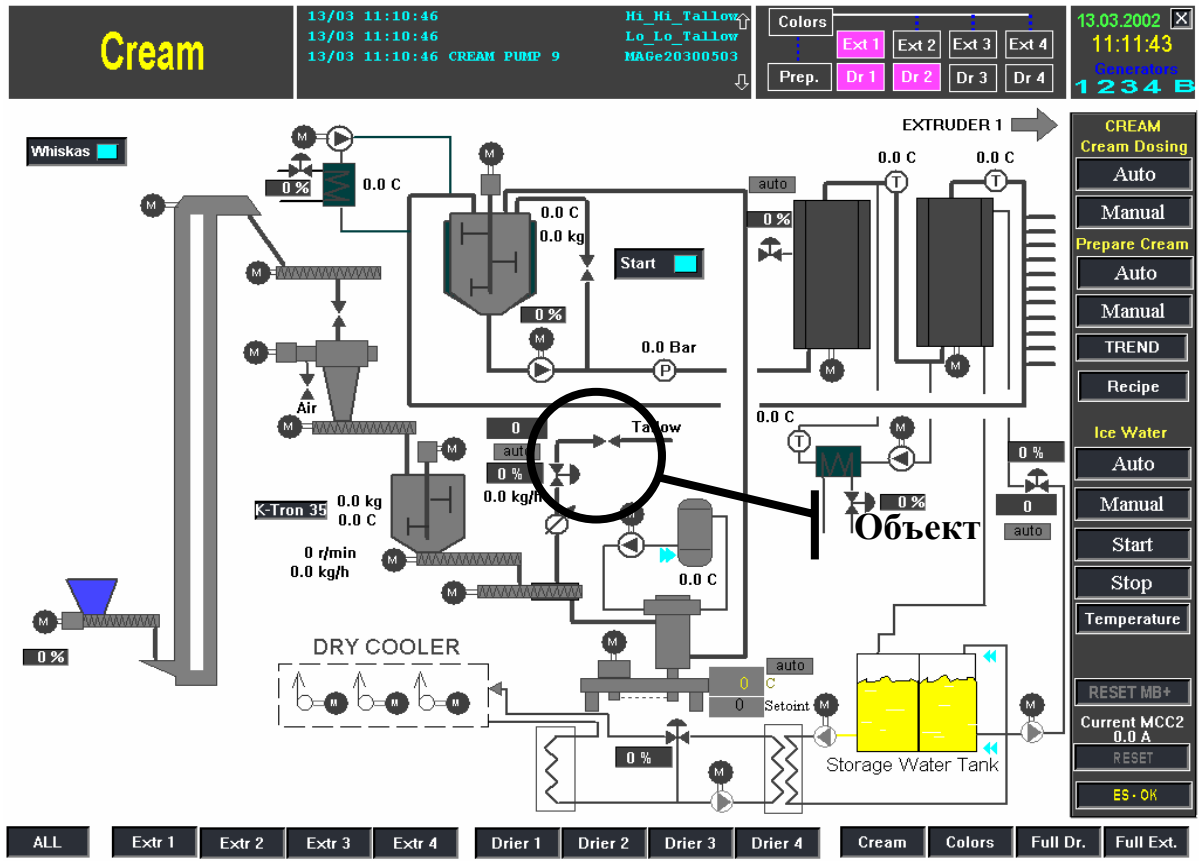

Рис. 5.18. Технологический процесс

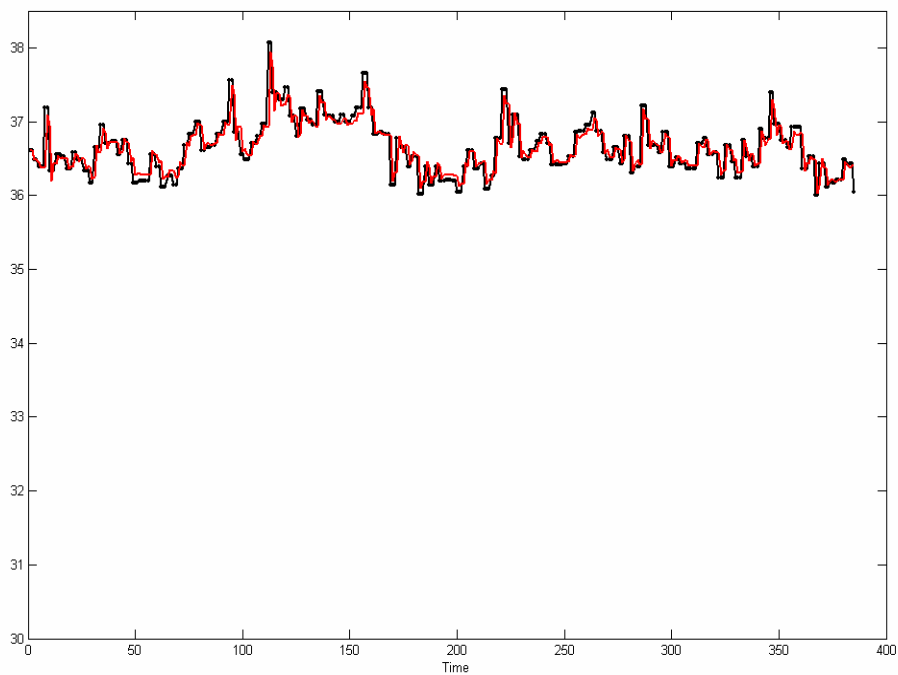

Рис. 5.19. Сравнение динамики идентифицированной модели с реальным процессом

Наличие построенной модели позволило сократить затраты, связанные с браком выходного продукта при разогреве, изменение параметров функционирования системы и обеспечить уменьшение быстродействия при запуске в 3 раза.

### 5.4. Управление процессом охлаждения алюминиевых слитков

Проведено моделирование процесса охлаждения алюминиевых слитков, изготовляемых для оборонной промышленности ОАО «Ступинская металлургическая компания». Химический состав алюминиевых сплавов и температура, при которой он готовится, зависит от марки. Металл плавится в специальной печи. После готовности расплав перекачивается в другую печь (миксер), где отбираются пробы и при необходимости добавляются различные присадки. Отливка круглых сплошных слитков из алюминиевых деформируемых сплавов в кристаллизаторы с тепловыми насадками ПН–10 допускается только после полной готовности расплава, оснастки, инструмента и рабочего места, а также соответствия режимов литья при данной технологии (согласно ТИ 410–025–02) (рис. 5.19).

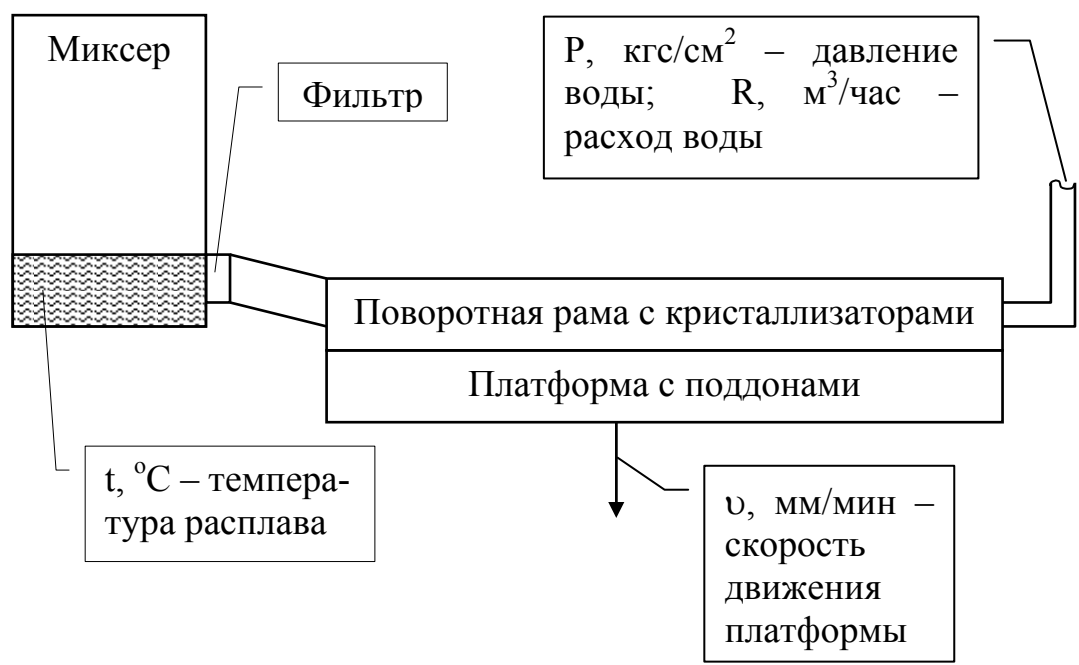

Рис. 5.19. Технология охлаждения

Рассмотрен процесс отливки круглых сплошных слитков диаметром 190 мм из сплава АД31 (6063). Температура расплава контролируется по показаниям термоэлектрического термометра, расположенного в миксере и, для литья слитков данного диаметра, составляет от 690 до 710°C включительно. Расплавленный металл через фильтр по литейному желобу поступает на поворотную раму с кристаллизаторами. В литейном желобе добавляется лигатурный пруток и отбирается жидкая проба для контроля газонасыщенности методом первого пузырька по ГОСТ 21132.0. На поворотной раме расположено 24 кристаллизатора цилиндрической формы, в каждый кристаллизатор подается вода для охлаждения металла и газо-масляная смесь для получения более качественной поверхности слитка. Металл по желобам поступает в кристаллизаторы и кристаллизуется за счет охлаждения. В этот момент платформа, приводимая в движение электродвигателем, начинает опускаться со скоростью (скорость литья) от 90 до 100 мм/мин, пока не будут отлиты слитки необходимой длины. Скорость литья задается вручную литейщиком. Для слитков диаметром 190 мм из сплава АД31 (6063) расход масла составляет от 0,12 до 0,15 л/мин включительно, давление охлаждающей воды от 0,7 до 0,9 кгс/см$^2$ включительно.

Такие параметры, как температура расплава, давление и расход воды, скорость движения платформы и длина слитка визируются электроникой литейной машины и записываются с периодичностью 3 с.

Для слитков диаметром 190 мм из сплава АД31 (6063) осуществляется охлаждение, измеряемыми параметрами системы являются давление воды (изм.— кгс/см$^2$; рис. 5.20*а*), расход охлаждающей воды (изм. — л/мин; рис. 5.20*б*). Моделируется автономный процесс при постоянном значении входного сигнала (постоянной скорости охлаждения сплава) и при определенных технологических условиях. До процесса моделирования настройка этих параметров происходила вручную. Построенная модель используется для построения дополнительного регулятора в системе.

В результате применения разработанного алгоритма — размерность пространства состояний $n = 6$, $\tau = 12$, старший показатель 0,3454950. Реконструированный фазовый портрет системы приведен на рис. 5.21. Результат идентификации показан на рис. 5.22.

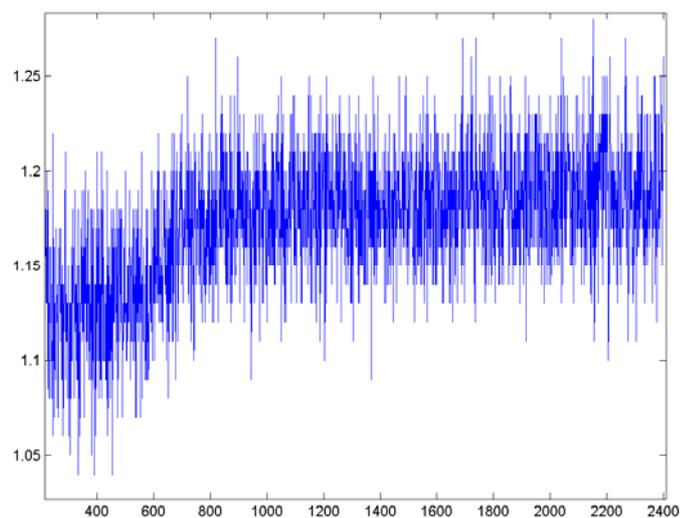

а)

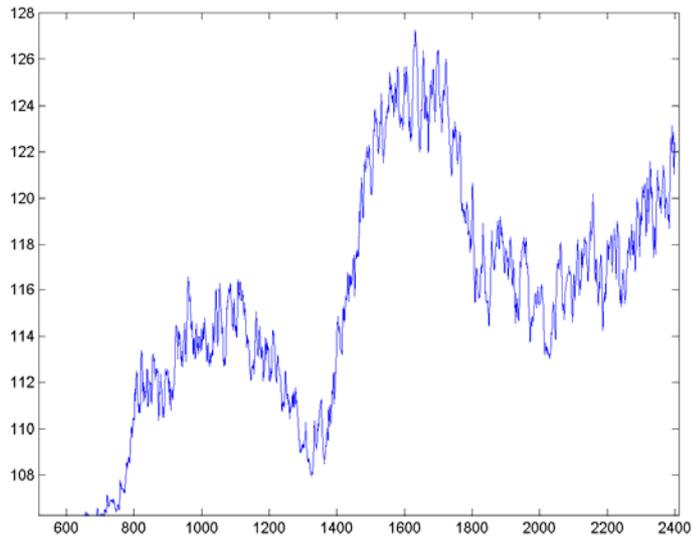

б)

Рис. 5.20. Динамика параметров системы охлаждения

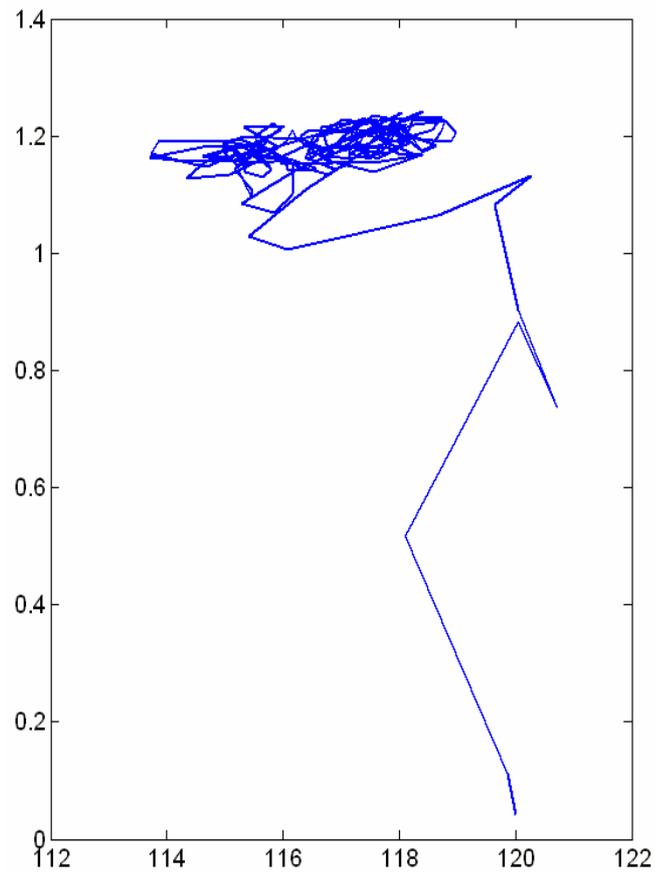

Рис. 5.21. Реконструированный аттрактор

Полученный результат позволил построить автоматический робастный регулятор в виде наблюдателя, управляющий расходом охлаждающей жидкости, графики управления (рис. 5.23).

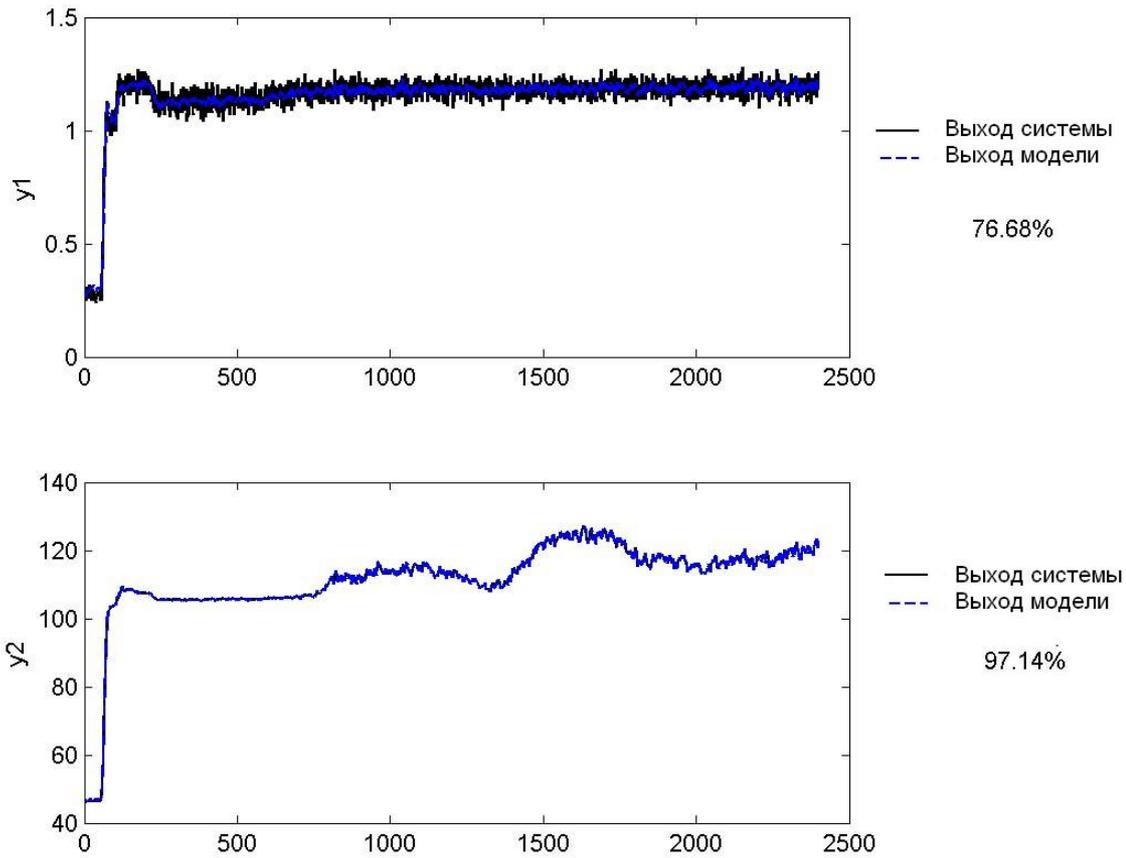

Рис. 5.22. Сравнение результатов моделирования с реальными данными

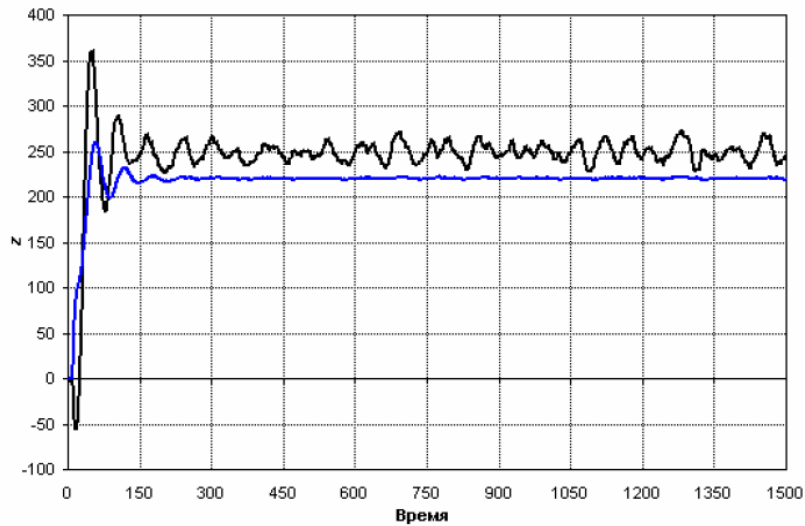

Рис. 5.23. Динамика состояний и управлений: $x_1$, $x_2$, $u_1$, $u_2$

Внедрение созданной методики позволило исключить ручную подрегулировку параметров. На основе модели построен регулятор, в результате улучшилось качество изделия, что связано с уменьшением колебаний при охлаждении.

# Заключение

На основании приведенных в работе результатов внедрения можно сделать вывод о решении практически важной задачи разработки теоретико-методологических основ повышения качества функционирования технических систем с нелинейной динамикой. Внедрение результатов позволило значительно увеличить надежность и технико-экономические показатели систем, имеющих важное значение для различных отраслей промышленности.

Работа над развитием разработанного геометрического метода не закончена. В монографии приведены два приложения, являющихся продолжением предложенного подхода.

Автор благодарит д. т. н., проф. М. В. Ульянова, д. ф.-м. н., проф. В. А. Антонца за поддержку, внимание и проявленный интерес к результатам.

# Список использованных источников

# ПРИЛОЖЕНИЯ

*Борисов Ю. Ю.*

## *Методика построения нейросетевых прогнозирующих моделей на основе анализа свойств аттракторов*

**Введение**

Задачи прогнозирования и построения математических моделей различных процессов и явлений имеет первостепенное значение во многих областях науки и жизнедеятельности человека.

В большинстве технических, экономических и социальных системах возникают процессы, являющиеся результатом взаимодействия множества составляющих, что не позволяет строить адекватные математические модели исходя только из априорных знаний. Вместе с тем часто имеется потребность строить не модели явлений, а эволюционные модели изменений динамики конкретного процесса, являющегося наблюдаемым параметром сложной системы. Особый интерес представляют эволюционные модели, дающие качественные прогнозирующие значения моделируемого процесса. Такие модели могут быть использованы в системах принятия решений, управлении, прогнозирования и оценки качества сложных систем. Таким образом, прогнозирующие модели строятся на основании наблюдения за процессами с учетом имеющейся информации и предположений о структуре и классе системы.

Значительное время единственным доступным теоретическим и практическим средством прогнозирования временных рядов являлись статистические методы. Однако фундаментальные ограничения статистических подходов, исходящие из сложности проверки предположений о вероятностных характеристиках и стохастических закономерностях исследуемой реализации приводят к невозможности объяснения многих явлений, и, как следствие, устранению неточных прогнозов.

В последнее время все большее развитие получает теория нелинейных динамических систем, в рамках которой разработаны методики, позволяющие по наблюдаемой скалярной реализации восстанавливать аттрактор, качественно эквивалентный исследуемой детерминированной системе [1]. В тоже время, аппарат нейронных сетей является мощным практическим инструментом аппроксимации функций и используется во многих работах для оценки оператора эволюции исследуемых динамических систем. Однако для построения нейросетевых моделей не достаточно широко применяются современные методики нелинейной динамики.

Исследование восстановленного аттрактора позволит выбрать его наиболее информативные характеристики, минимизируя при этом размерность входного вектора и структуру нейросетевой модели [2–3].

Совместное использование методов нелинейной динамики и аппарата нейронных сетей позволит разработать методику построения прогнозирующих моделей реальных технических, экономических систем.

Таким образом, разработка и исследование методов построения нейросетевых прогнозирующих моделей на основе исследования реконструированных по наблюдаемым данным аттракторов является актуальной.

Статья посвящена разработке методики построения нейросетевых прогнозирующих моделей на основе исследования и обработки аттракторов, реконструированных по наблюдаемым данным методом Паккарда-Такенса.

## 1. Постановка задачи

Пусть задан временной ряд, являющийся результатом функционирования структурно-сложной наблюдаемой системы, обладающий автоколебательной регулярной или хаотической динамикой для которого может быть восстановлен аттрактор.

Пусть в результате эксперимента получен скалярный временной ряд, являющийся функцией состояния неизвестной динамической системы
$$x(t+1) = f(x(t)),$$
$$y(t) = \Phi(x(t)).$$

Согласно теореме Такенса и условию Мане, фазовый потрет, восстановленный в виде
$$Z(t) = (y(t), y(t-\tau), ..., y(t-(m-1)\tau)) = (z_1(t), z_2(t), ..., z_m(t)), \qquad (1.1)$$
топологически эквивалентен аттрактору исходной динамической системы, обеспечивая глобальное отображение состояние $Z(t)$ в $x(t)$. Здесь $x(t)$ — вектор состояния исходной динамической системы; $y(t)$ — наблюдаемый скалярный временной ряд; $z_i(t)$ — $i$-ая компонента вектора, определяющего точку на восстановленном аттракторе; $\tau$ — параметр задержки (время пересечения траекториями сечения Пуанкаре); $m$ — размерность реконструкции, определяемая из условия $m \geq n^h + 1$; $n^h$ — хаусдорфова размерность.

Предложено представить модель оператора эволюции в виде вектор-функции
$$Z(t+1) = \psi(\tilde{Z}(t)),$$
отображающей характеристики аттрактора в области $\tilde{Z}$ на координаты вектора состояния в следующий момент времени, что обеспечивает возможность учета локального, глобального и синтетического прогноза при реконструкции уравнений состояния исследуемых систем.

Проблема параметрической идентификации нейросетевой модели с заданной структурой $S$ определяется как поиск параметров $W_S$, минимизирующих величину ошибки
$$E = \sum_t \left( Z(t+1) - \psi(W_S, S, \tilde{Z}(t)) \right)^2,$$
где $\psi$ — нейросетевая оценка оператора эволюции; $S$ — структура нейросетевой модели (количество слоев, число нейронов в каждом слое, вид функций активации); $W_S$ — набор матриц весовых коэффициентов; $\tilde{Z}(t)$ — область на восстановленном аттракторе системы.

## 2. Построение прогнозирующих моделей

В работе исследуются авторегрессионные модели вида $Z(t+1) = \psi(\tilde{Z})$, где $\tilde{Z}$ — некоторое подмножество точек на восстановленном аттракторе; $\psi$ — преобразование, использующее предыдущие состояния для вычисления последующего состояния.

Показано что, различные методы прогнозирования различаются лишь способом формирования окрестности $\tilde{Z}$ и методом оценки функции $\psi$.

Будем рассматривать окрестность $\tilde{Z}$ как матрицу, центральный элемент которой является текущей точкой $Z(t)$ восстановленного аттрактора системы, а центральный столбец состоит из упорядоченных по возрастанию расстояния от точки $Z(t)$ ее ближайших соседей — $Z_1(t_1), Z_2(t_2),...,Z_{2r}(t_{2r})$. Строки матрицы $\tilde{Z}$ соответствуют непрерывным участкам траекторий, центрами которых являются ближайшие соседи точки $Z(t)$. Таким образом, матрица $\tilde{Z}$ имеет вид:

$$\tilde{Z} = \begin{pmatrix} Z_{2r-1}(t_{2r-1}+k\Delta t) & ... & Z_{2r-1}(t_{2r-1}+\Delta t) & Z_{2r-1}(t_{2r-1}) & Z_{2r-1}(t_{2r-1}-\Delta t) & ... & Z_{2r-1}(t_{2r-1}-k\Delta t) \\ ... & ... & ... & ... & ... & ... & ... \\ Z_1(t_1+k\Delta t) & ... & Z_1(t_1+\Delta t) & Z_1(t_1) & Z_1(t_1-\Delta t) & ... & Z_1(t_1-k\Delta t) \\ 0 & ... & 0 & Z(t) & Z(t-\Delta t) & ... & Z(t-k\Delta t) \\ Z_2(t_2+k\Delta t) & ... & Z_2(t_2+\Delta t) & Z_2(t_2) & Z_2(t_2-\Delta t) & ... & Z_2(t_2-k\Delta t) \\ ... & ... & ... & ... & ... & ... & ... \\ Z_{2r}(t_{2r}+k\Delta t) & ... & Z_{2r}(t_{2r}+\Delta t) & Z_{2r}(t_{2r}) & Z_{2r}(t_{2r}-\Delta t) & ... & Z_{2r}(t_{2r}-k\Delta t) \end{pmatrix}.$$

Нулевые элементы в центральной строке матрицы $\tilde{Z}$ объясняются отсутствием данных в точках $Z(\tilde{t})$, где $\tilde{t} > t$.

Для глобальных моделей матрица $\tilde{Z}$ имеет вид:

$$\tilde{Z} = \begin{pmatrix} 0 & ... & 0 & 0 & 0 & ... & 0 \\ ... & ... & ... & ... & ... & ... & ... \\ 0 & ... & 0 & 0 & 0 & ... & 0 \\ 0 & ... & 0 & Z(t) & Z(t-\Delta t) & ... & Z(t-k\Delta t) \\ 0 & ... & 0 & 0 & 0 & ... & 0 \\ ... & ... & ... & ... & ... & ... & ... \\ 0 & ... & 0 & 0 & 0 & ... & 0 \end{pmatrix}. \qquad (2.1)$$

Для локальных моделей матрица $\tilde{Z}$ может быть записана в виде (2.2) или (2.3).

$$\tilde{Z} = \begin{pmatrix} 0 & ... & Z_{2r-1}(t_{2r-1}+\Delta t) & 0 & 0 & ... & 0 \\ ... & ... & ... & ... & ... & ... & ... \\ 0 & ... & Z_1(t_1+\Delta t) & 0 & 0 & ... & 0 \\ 0 & ... & 0 & 0 & 0 & ... & 0 \\ 0 & ... & Z_2(t_2+\Delta t) & 0 & 0 & ... & 0 \\ ... & ... & ... & ... & ... & ... & ... \\ 0 & ... & Z_{2r}(t_{2r}+\Delta t) & 0 & 0 & ... & 0 \end{pmatrix}, \qquad (2.2)$$

$$\tilde{Z} = \begin{pmatrix} 0 & ... & Z_{2r-1}(t_{2r-1}+\Delta t) & Z_{2r-1}(t_{2r-1}) & 0 & ... & 0 \\ ... & ... & ... & ... & ... & ... & ... \\ 0 & ... & Z_1(t_1+\Delta t) & Z_1(t_1) & 0 & ... & 0 \\ 0 & ... & 0 & Z(t) & 0 & ... & 0 \\ 0 & ... & Z_2(t_2+\Delta t) & Z_2(t_2) & 0 & ... & 0 \\ ... & ... & ... & ... & ... & ... & ... \\ 0 & ... & Z_{2r}(t_{2r}+\Delta t) & Z_{2r}(t_{2r}) & 0 & ... & 0 \end{pmatrix}. \qquad (2.3)$$

Матрица (2.2) определяет метод локального прогнозирования, основанный на усреднении значений ближайших соседних точек на аттракторе системы, в то время как в (2.3) используются значения изменений данных точек. Математические модели, использующие виды окрестностей отличные от (2.1), (2.2), (2.3) являются синтетическими.

Разработан алгоритм, позволивший экспериментально доказать эффективность применения синтетических моделей в задачах прогнозирования. Алгоритм можно переставить в виде шести шагов.

**Шаг 1.** Выполнить реконструкцию аттрактора системы по заданному временному ряду.

**Шаг 2.** Сформировать очередную конфигурацию матрицы $\tilde{Z}$.

Выполнить $N$ раз шаг 3 и шаг 4:

**Шаг 3.** Построить нейросетевую модель.

**Шаг 4.** Получить ошибку прогноза на тестовом множестве для построенной модели.

**Шаг 5.** Оценить качество прогноза для данной конфигурации матрицы $\tilde{Z}$ с учетом качества $N$ нейросетевых моделей.

**Шаг 6.** Выбрать $k$ конфигураций матрицы $\tilde{Z}$, для которых были получены наилучшие результаты.

Предложено преобразование $\psi$ представить как последовательное выполнение этапов предобработки и вычисления выходных характеристик имеющейся модели:

$$\psi(\tilde{Z}) = \psi_n\left(\psi_p\left(\tilde{Z}\right)\right),$$

где $\psi_p$ — функция предобработки локальной окрестности; $\psi_n$ — нейросетевая модель.

Установлено, что методики предобработки $\psi_p$ решают три следующие задачи: фильтрация, учет прогноза альтернативных моделей для текущей ситуации (положение на аттракторе) и учет ошибки прогноза идентичных по структуре моделей для похожих ситуаций. Основная идея предлагаемой комбинированной методики предобработки заключается в оценке характеристик локальной окрестности, направленных на решение каждой из выделенных задач.

Для каждой из выделенных задач введен соответствующий класс. Методы, осуществляющие фильтрацию, принадлежат классу 1 ($class1$); методы, учитывающие прогноз альтернативных моделей — классу 2 ($class2$); методы, использующие ошибки прогноза идентичных по структуре моделей — классу 3 ($class3$).

Систематизированы методы предобработки в соответствии с введенной классификацией (табл. 1.1).

Таблица 1.1
Классификация методов предобработки

| № | Описание | Класс |
|---|---|---|
| 1 | Точки вида $y(t-k_1),...,y(t-k_n)$: $m1(k_1,...,k_n)$ | $class1$ |
| 2 | Усреднение по $k_1,...,k_n$ предыдущим точкам: $m2(k_1,...,k_n)$ | $class1, class2$ |
| 3 | Усреднение по $k_1,...,k_n$ ближайшим соседям: $m3(k_1,...,k_n)$ | $class1, class2$ |
| 4 | Взвешенное усреднение по $k_1,...,k_n$ предыдущим точкам: $m4(k_1,...,k_n)$ | $class1, class2$ |
| 5 | Ошибки прогнозирования идентичных по структуре моделей: $m5(k_1,...,k_n)$ | $class3$ |

Разработана общая схема обучения и функционирования нейросетевой модели (рис. 2.1).

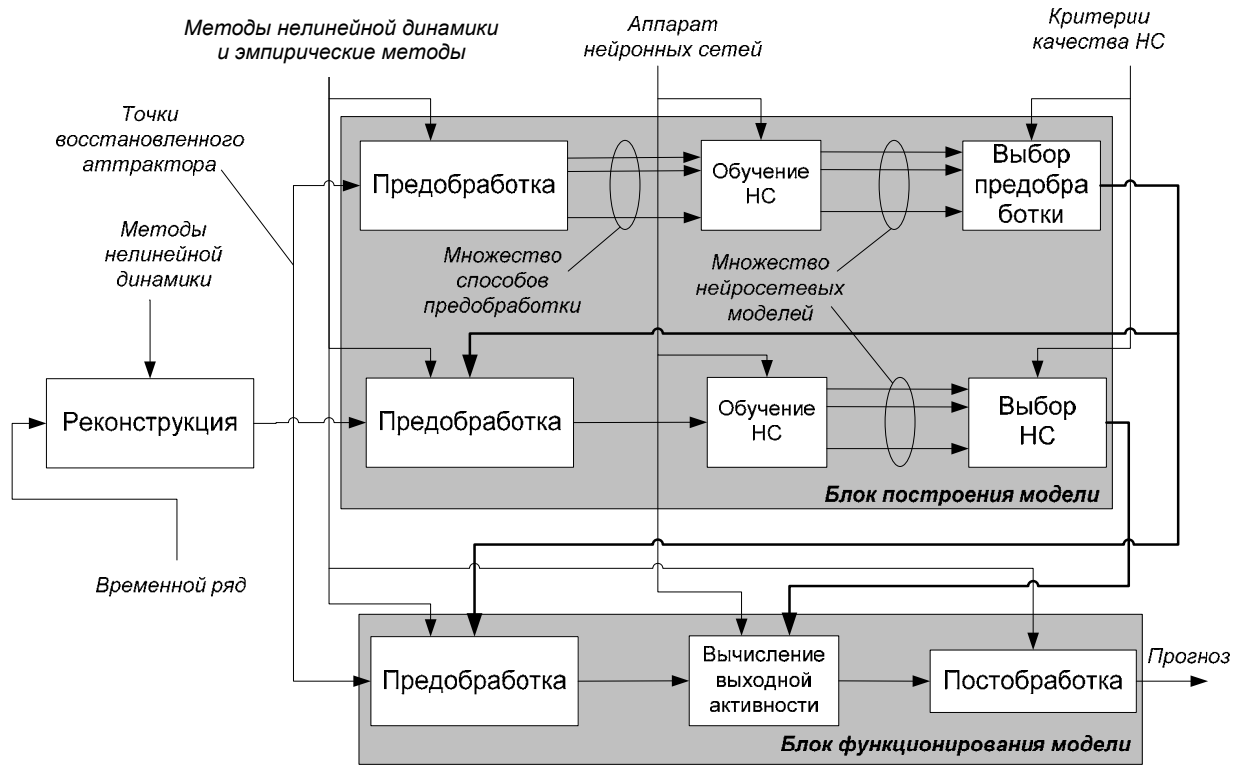

Рис. 2.1. Общая схема обучения и функционирования нейросетевой модели

Алгоритм поиска оптимальных методов предобработки является частью разработанной методики построения нейросетевых прогнозирующих моделей динамических систем, которую можно представить в виде следующей последовательности шагов.

**Шаг 1.** Восстановить аттрактор системы как последовательность точек вида $Z(t) = \left( z_1(t), z_2(t), ..., z_m(t) \right)$.

**Шаг 2**. Выполнить алгоритм поиска оптимальных методов предобработки, определив при этом минимальную размерность входного вектора $M$.

Последовательно изменяя количество скрытых нейронов $i = \overline{1, 2M+1}$, выполнить шаг 3.

**Шаг 3**. Провести $L$ раз процесс обучения нейронной сети по алгоритму Левенберга-Марквардта, используя найденный набор методов предобработки.

**Шаг 4**. Выбрать нейросетевую модель, обеспечивающую наименьшую ошибку на тестовом множестве данных.

В результате применения приведенной выше методики подбирается структура и параметры нейросетевой модели так, чтобы минимизировать ошибку обобщения для минимально возможного числа нейронов в распределительном слое.

### 3. Оценка качества прогнозирования

По выходной активности нейросетевых моделей не может быть сделан вывод о предполагаемой точности получаемого прогноза. Предложено для оценки точности прогноза нейросетевой модели использовать критерий в виде суммарной среднеквадратичной ошибки прогнозирования точек, принадлежащих ближайшим соседним траекториям аттрактора:

$$E_\psi = E_\psi(Z(t),k) = \sum_k \left(Z_k(t_k+1) - \psi(\tilde{Z}_k(t_k))\right)^2, \qquad (3.1)$$

где $E_\psi(Z(t),k)$ — суммарная ошибка прогнозирования $k$ ближайших соседей точки $Z(t)$; $Z_k(t_k+1)$ — $k$-ая ближайшая соседняя точка по отношению к точке $Z(t)$ в следующий момент времени; $\psi(\tilde{Z}_k(t_k))$ — прогноз для $k$-ой ближайшей соседней точки.

Разработана методика выбора прогнозирующей модели, в соответствии с которой, на очередном шаге прогнозирования выбирается прогноз той модели, которая минимизирует критерий (3.1). Так, прогноз на основе методики выбора модели строится в виде:

$$\hat{y} = a_1\hat{y}_1 + a_2\hat{y}_2 + ... + a_n\hat{y}_n \qquad (3.2)$$

$$a_i = \begin{cases} 1, \text{ если } E_{\psi(i)} = \min_j\left(E_{\psi(j)}\right), \overline{j=1,n}, \\ 0, \text{ иначе,} \end{cases}$$

где $\hat{y}_i$ — прогноз, построенный $i$-ой моделью; $E_{\psi(i)}$ — ошибка $i$-ой модели при прогнозировании будущих значений ближайших соседних траекторий.

Методику выбора модели для прогнозирования одномерных временных рядов можно представить в виде трех этапов.

**Этап 1.** Провести реконструкцию аттрактора.

**Этап 2.** Построить $N$ нейросетевых моделей.

**Этап 3.** Провести выбор прогноза одной из $N$ моделей с помощью функции (3.2).

Схема применения методики выбора модели для прогнозирования одномерных временных рядов показана на рис. 3.1.

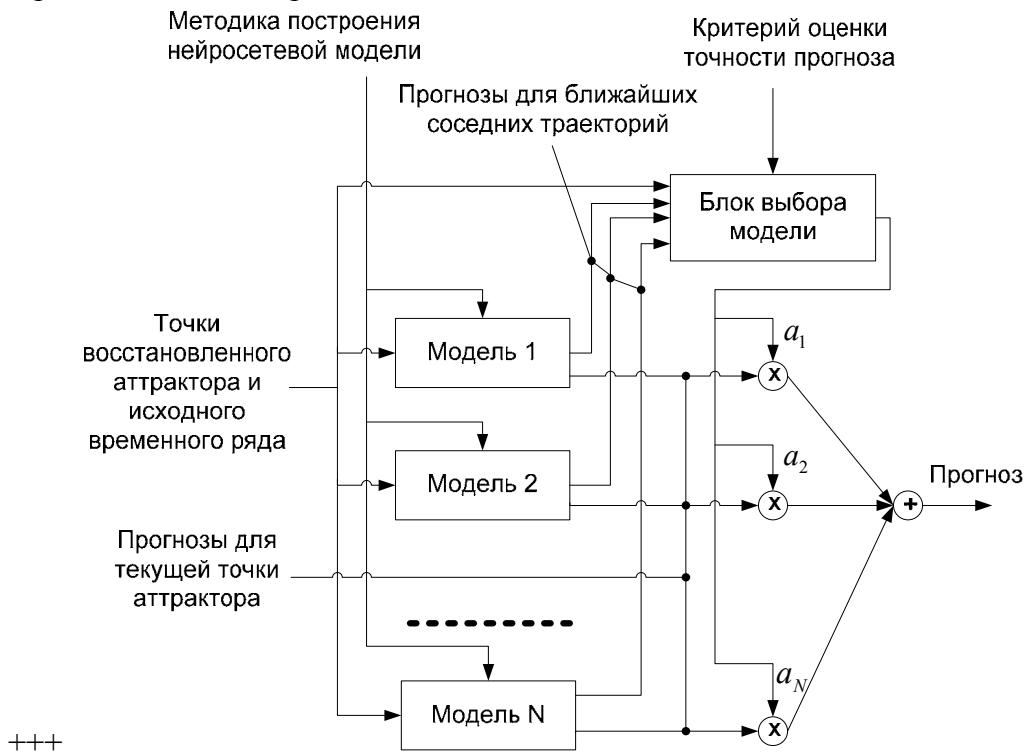

+++

Рис 3.1. Схема применения методики выбора модели на каждом шаге прогнозирования

Разработан метод оценки параметра доходности в задаче портфельного инвестирования. Так, предложено доходность $i$-го финансового инструмента вычислять в виде:

$$R_i = \begin{cases} \hat{y}, & \text{если } E_{\psi(i)} < d, \\ 0, & E_{\psi(i)} \geq d, \end{cases} \qquad (3.3)$$

где $R_i$ — доходность $i$-го финансового инструмента; $E_{\psi(i)}$ — ошибка $i$-ой модели при прогнозировании будущих значений ближайших соседних траекторий; $d$ — минимально допустимая величина суммарной среднеквадратичной ошибки прогнозирования ближайших соседних траекторий на аттракторе, при которой получаемый прогноз $\hat{y}$ считается достоверным.

Оценка параметра доходности в виде (3.3) позволяет исключить из портфеля те ценные бумаги, для которых построить достоверный прогноз с помощью имеющихся моделей не представляется возможным.

Представлена методика пошаговой реконструкции, суть которой состоит в изменении параметров реконструкции на каждом шаге прогнозирования так, чтобы точки аттрактора на интервале прогнозирования принадлежали локальной области аттрактора с наибольшей степенью устойчивости. Если (1.1) записать в виде

$$Z(t) = \Lambda\big(F(y(t))\big) = \big(f(t), f(t-\tau), ..., f(t-(m-1)\tau)\big) = \big(z_1(t), z_2(t), ..., z_m(t)\big), \qquad (3.4)$$

то параметрами реконструкции будут являться $m, \tau, F$, где $F$ — некоторое преобразование известного временного ряда.

Методика пошаговой реконструкции основывается на классификации локальных областей фазового пространства. Предложено считать локальную область фазового пространства устойчивой, если принадлежащие ей соседние точки в момент времени $t$ остаются соседними и в момент времени $t+1$. Если же на следующем временном интервале изначально соседние траектории более не являются соседними, то локальная область считается неустойчивой. Количественная характеристика локальной устойчивости $\lambda_D$ определена как максимальное расстояние в следующий момент времени между соседними точками, принадлежащими области $D$, и имеет вид:

$$\lambda_D = \frac{1}{\max\limits_{i,j} \delta\big(Z_i(t_i+1), Z_j(t_j+1)\big)}, \quad \big(Z_i(t_i), Z_j(t_j)\big) \subset D, \qquad (3.5)$$

где $D$ — локальная область аттрактора системы; $\delta\big(Z_i(t_i+1), Z_j(t_j+1)\big)$ — функция расстояния между точками $Z_i(t_i+1)$ и $Z_j(t_j+1)$.

Для эффективного применения локальных прогнозирующих моделей в области $D$ необходимо максимизировать введенный коэффициент локальной устойчивости (3.5) в данной области. Вводится соответствующий критерий качества реконструкции

$$J_1(m, \tau, F) = \lambda_D,$$

где $m, \tau, F$ — параметры реконструкции.

Установлено, что точность оценки $\lambda_D$ в (3.5) зависит от количества локальных траекторий, проходящих через область $D$. Таким образом, необходимо также увеличить концентрацию локальных траекторий в области $D$. Данное требование определяет следующий критерий качества реконструкции:

$$J_2(m, \tau, F) = |D|,$$

где $m, \tau, F$ — параметры реконструкции; $|D|$ — мощность множества $D$.

В общем случае, максимизировать критерии $J_1$ и $J_2$ одновременно не представляется возможным. Фиксируя величину локальной устойчивости (критерий $J_1$), и максимизируя точность ее оценки (критерий $J_2$), получим компромиссное решение в виде параметров реконструкции $m, \tau, F$. Формально это можно записать в виде:

$$J(m, \tau, F) = \begin{cases} J_2, J_1 \geq \lambda_d, \\ 0, J_1 < \lambda_d, \end{cases} \quad (3.6)$$

где $\lambda_d$ — заданная допустимая величина локальной устойчивости.

Для аттрактора, соответствующего найденному компромиссному решению, строится прогноз с помощью локальной модели вида:

$$Z(t+1) = \frac{1}{n} \sum_{j=1}^{n} Z(t_j + 1), \; Z(t_j) \subset D, \quad (3.7)$$

где $Z(t+1) = \left(z_1(t+1), z_2(t+1), ..., z_m(t+1)\right)$ — прогнозируемая точка аттрактора системы; $Z(t_j + 1) = \left(z_1(t_j + 1), z_2(t_j + 1), ..., z_m(t_j + 1)\right)$ — ближайшие соседи прогнозируемой точки на аттракторе.

Для визуализации структурной устойчивости прогнозирующей модели предложен следующий вид оператора $F$:

$$z_1(t) = \frac{sign(\hat{y}(t+1))}{E_\psi}, \quad (3.8)$$
$$z_2(t) = sign(y(t) - y(t-1)),$$

где $E_\psi$ — ошибка модели при прогнозировании будущих значений ближайших соседних траекторий, $\hat{y}(t+1)$ — прогноз модели для момента времени $t+1$.

Представлена разработанная методика пошаговой реконструкции для прогнозирования временных рядов в виде следующей последовательности шагов.

**Шаг 1**. Применяя к исходному временному ряду $y(t)$ преобразования $F$, извлечь $n$ признаков вида $Z_1(t), Z_2(t), ..., Z_n(t)$, $t = \overline{1, T}$.

Для $C_n^m$ возможных комбинаций признаков $Z_1(t), Z_2(t), ..., Z_m(t)$ и значений параметра задержки $\tau = \overline{1, \tau_{max}}$ выполнить шаги 2–3.

**Шаг 2**. Выполнить реконструкцию аттрактора в $m$-мерном пространстве

**Шаг 3**. Для точки аттрактора $Z(t)$ найти ближайшие соседние точки и рассчитать для них композитный критерий $J(m, \tau, F)$ (3.6).

**Шаг 4**. Восстановить аттрактор с параметрами, максимизирующими $J(m, \tau, F)$.

**Шаг 5**. Построить прогноз для точки $Z(t+1)$ с помощью локальной модели (3.7).

Таким образом, на основе анализа ближайших соседних траекторий восстановленного аттрактора получены критерии оценки качества получаемого прогноза и разработана методика выбора модели для прогнозирования временных рядов.

## 4. Пример моделирования

Рассмотрим задачу прогнозирования поведения структурно-сложных систем на примере американского фондового индекса *S&P 500*. На рис. 4.1а показан исходный вид индекса *S&P 500*. Исключив тренд, получаем стационарную составляющую данного индекса (рис. 4.1b).

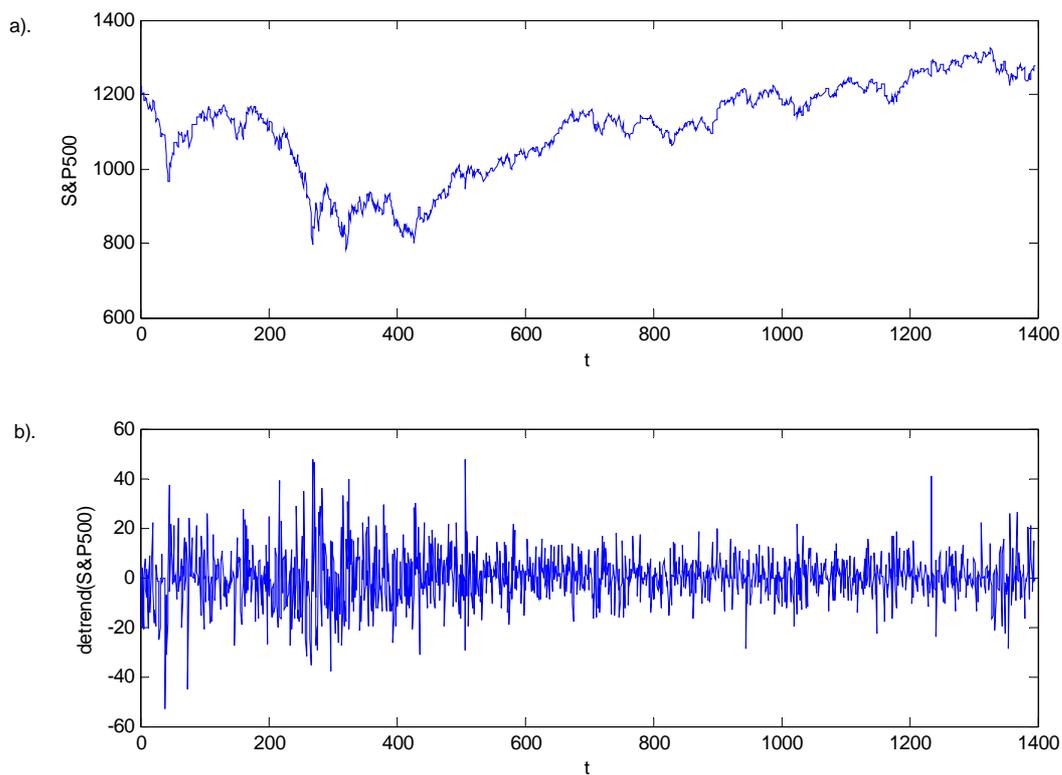

Рис. 4.1. Исходный вид индекса S&P500 (a) и индекс S&P500 за вычетом тренда (b)

Аттракторы, восстановленные по исходной реализации и стационарной составляющей индекса, представлены на рис. 4.2 и рис. 4.3.

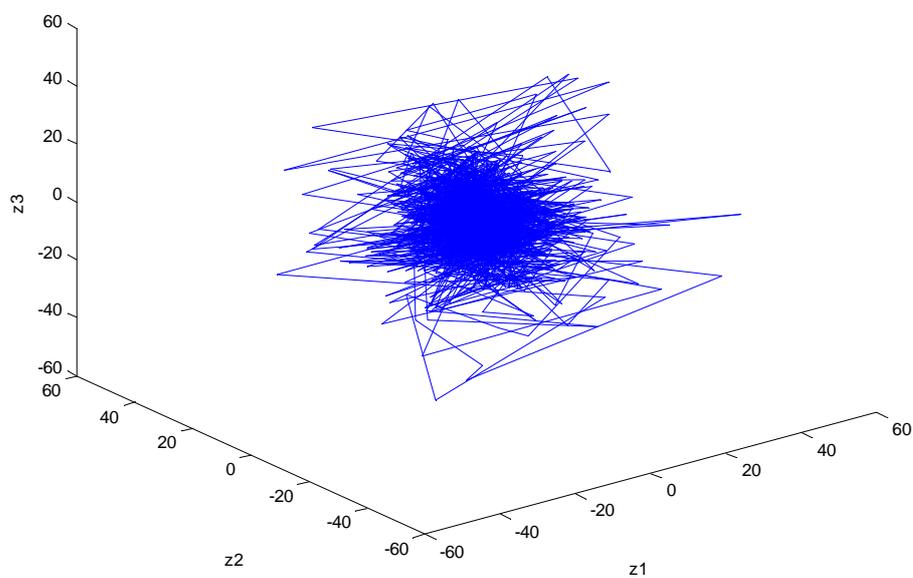

Рис. 4.2. Аттрактор, восстановленный по индексу S&P500

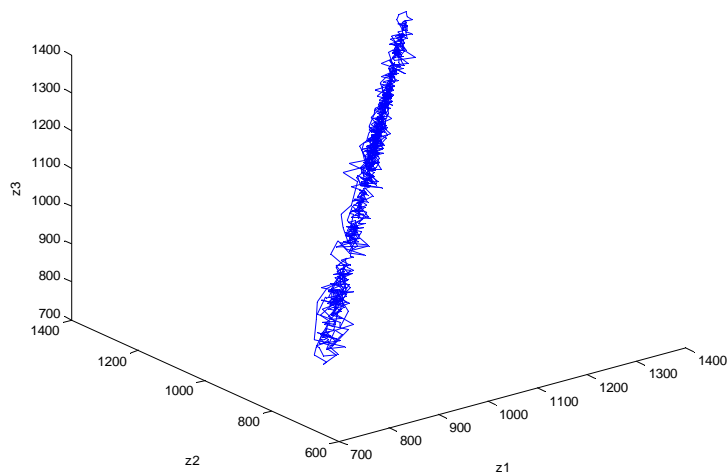

Рис.4.3. Аттрактор, восстановленный по стационарной составляющей индекса S&P500

На практике интерес представляет не столько абсолютное значение индекса, сколько направление его движения через установленный интервал прогнозирования.

Рассмотрим применение разработанной методики построения нейросетевой модели для прогнозирования знака дневного изменения стационарной составляющей индекса *S&P500*.

На рис. 4.4 и рис. 4.5 показаны примеры работы полученной прогнозирующей модели на обучающем и тестовом множестве данных.

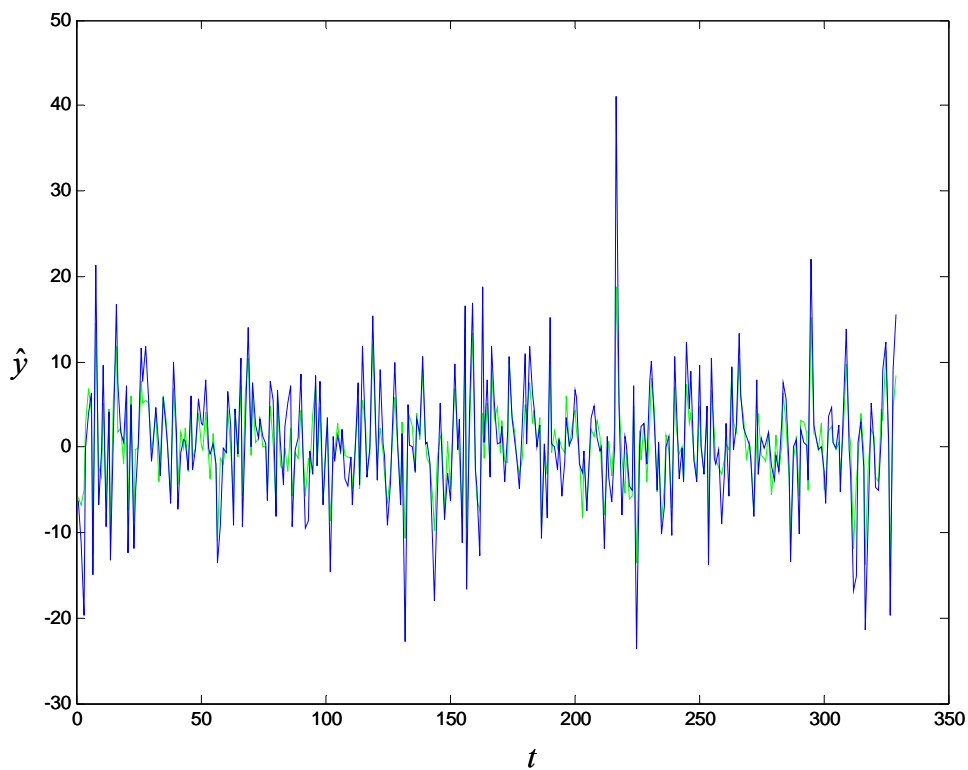

Рис. 4.4. Прогнозирование нейросетевой моделью обучающего множества данных

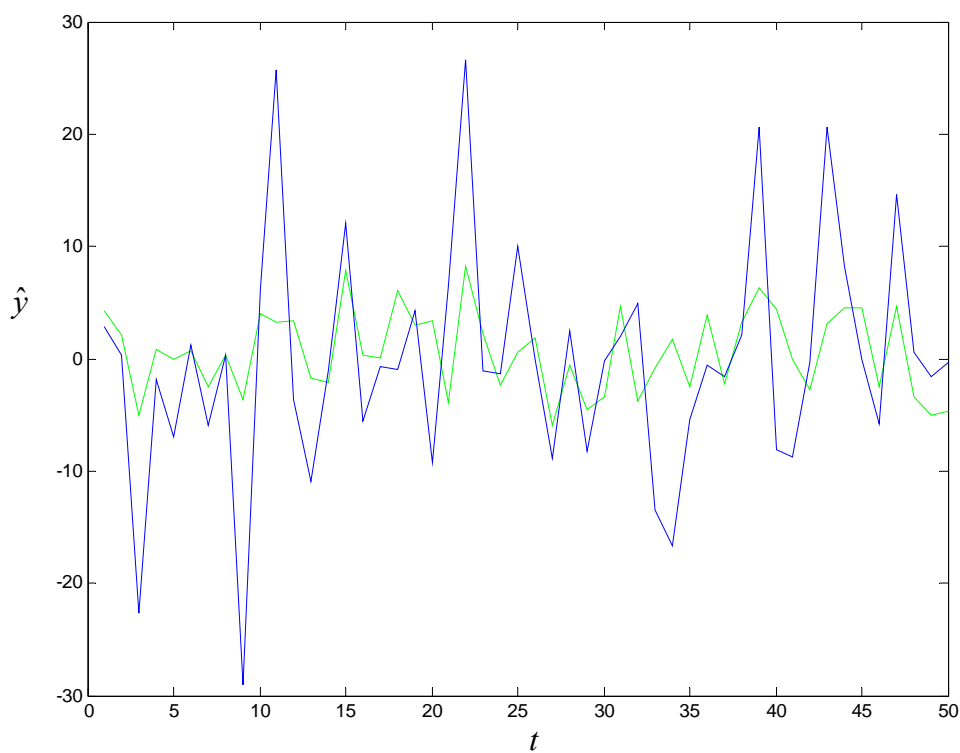

Рис. 4.5. Прогнозирование нейросетевой моделью тестового множества данных

Методы предобработки, для которых получена наименьшая ошибка на тестовом множестве данных, отображены в табл. 4.1. Структура наилучшей прогнозирующей модели показана в табл. 4.2.

Таблица 4.1
Методы предобработки, обеспечившие наименьшую ошибку прогнозирования на тестовом множестве данных

| № | Методы предобработки | Процент верно определенного направления на тестовом множестве, % |
|---|---|---|
| 1 | m1(0,1,3),m2(15,13,10,6),m3(1,2),m5(1,2,3) | 68 |
| 2 | m1(0,3), m2(15,13,7,6,2), m3(1,2),m5(1,2,3) | 64 |
| 3 | m1(0,1,2,3),m2(15,13,2),m3(1,2),m5(1,2,3) | 62 |

Таблица 4.2
Структура наилучшей прогнозирующей модели

| Размерность входного вектора | Число нейронов в скрытом слое | Число нейронов в выходном слое | Вид функции активации | Суммарное количество связей |
|---|---|---|---|---|
| 12 | 24 | 1 | Сигмоидальная и линейная функция | 312 |

Исключение из интервала прогнозирования точек, для которых неверно прогнозируется направление хотя бы одной из четырех ближайших соседних траекторий, увеличивает адекватность наилучшей прогнозирующей модели до 79%. Таким образом, разработана методика построения прогнозирующих моделей на основе анализа аттракторов нелинейных динамических систем, реконструированных на основании экспериментальных данных.

*Литература*

1. *Никульчев Е. В.* Геометрический метод реконструкции систем по экспериментальным данным // Письма в ЖТФ.— 2007.— Т. 33.— Вып. 6.— С. 83–89.
2. *Борисов Ю. Ю.* Построение прогнозирующих моделей динамических систем на основе исследования окрестностей реконструированных аттракторов // Автоматизация и современные технологии.— 2007.— №2.— С. 32–37.
3. *Борисов Ю. Ю.* Метод пошаговой реконструкции для построения локальных прогнозирующих моделей хаотических временных рядов. // Известия вузов. Проблемы полиграфии и издательского дела.— 2007.— № 2.— С. 51–56.

*Козлов О. В.*

## Выявление симметрий реконструированных фазовых траекторий динамических систем

Для реконструкции динамических моделей нелинейных систем по экспериментальным данным в работе [1] предложен геометрический метод, основанный на выделении локальных областей фазовых траекторий, близких к периодическим, и построение конечнопараметрических преобразований, переводящих одну область в другую, т. е. построение группы симметрий фазовых траекторий, которая характеризуется преобразованием графиков.

Доклад посвящен разработке методики, позволяющей выявлять практические значимые симметрии (аффинные преобразования и поворот) при построении динамических моделей по группам симметрий реконструированных аттракторов.

Пусть имеется траектория $X(m)$ дискретной динамической системы (контур), состоящая из $m$ действительных точек в $n$-мерном фазовом пространстве, которую можно представить в виде:

$$X(m) = \begin{bmatrix} x_{1,1} & x_{2,1} & \cdots & x_{n,1} \\ x_{1,2} & x_{2,2} & \cdots & x_{n,2} \\ \vdots & & & \\ x_{1,m} & x_{2,m} & \cdots & x_{n,m} \end{bmatrix}.$$

Разработана методика, результатом которой является получение нормализованных характеристик траектории и показателей для наиболее практически важных симметрий (переноса, масштабирования и поворота) в условиях слабого нарушения симметрии. При формировании методики решения многомерной задачи выявления преобразований симметрий контура был модифицирован подход, основанный на применении дискретного преобразования Фурье к двумерной задаче [2].

Применим дискретное преобразование Фурье (ДПФ) для каждой точки рассматриваемой локальной области фазовой траектории в виде:

$$s_{d,k} = \sum_{p=1}^{m} x_{d,p} e^{\frac{-2\pi i}{m}(k-1)(p-1)}, \; 1 \leq d \leq n. \tag{1}$$

Результатом преобразования будет спектр следующей структуры:
$$S_d = \begin{bmatrix} s_{d,0} & s_{d,1} & \cdots & s_{d,m} \end{bmatrix}^{\mathrm{T}}, 1 \le d \le n.$$

Пары $(S_1\ S_{n-1}), (S_2\ S_{n-2}), \ldots, \left(S_{(n-1)/2}\ S_{n/2}\right)$ являются комплексно сопряженными числами. Заметим, что они также определяют характеристики контура, но в другом пространстве. Переход от спектра к контуру осуществляется с помощью обратного дискретного преобразования Фурье (ОДПФ) вида:

$$x_{d,k} = \frac{1}{m} \sum_{p=1}^{m} s_{d,p} e^{\frac{2\pi i}{m}(k-1)(p-1)}, 1 \le d \le n. \qquad (2)$$

Рассмотрим получение симметрий переноса, масштабирования и поворота контура относительно начала координат. Первый элемент спектра определяет положение центра контура, так как является усредненной суммой координат всех точек контура. Собственно, первый элемент спектра и является показателем симметрии переноса контура:

$$K_{\text{пер}} = \begin{bmatrix} s_{1,0} & s_{2,0} & \cdots & s_{n,0} \end{bmatrix}. \qquad (3)$$

Воспользуемся следующим свойством преобразования Фурье: поворот или масштабирование спектра приводит к повороту или масштабированию контура на тот же угол или коэффициент. Пары комплексно сопряженных элементов спектра $(S_1\ S_{n-1}), (S_2\ S_{n-2})$ и т. д., после ОДПФ (2) представляют собой эллипсы (если не меняя размера спектра просто приравнять к нулю все остальные элементы), в частности пара $(S_1\ S_{m-1})$ — наибольший эллипс, аппроксимирующий контур. Коэффициент масштабирования удобно рассматривать как отношение длины вектора второго элемента к единичному вектору, такой же направленности:

$$K_{\text{мшт}} = \sqrt{|s_{1,1}|^2 + |s_{2,1}|^2 + \cdots + |s_{m,1}|^2}. \qquad (4)$$

Показатель поворота контура удобно определять в виде набора углов вектора второго элемента спектра к координатным осям. Положение контура определяется с помощью $n-1$ углов. Угол поворота контура к оси измерения k определяется следующим образом:

$$K_{\text{пов},k} = \begin{cases} \arctan\left(s_{k,2}/s_{k+1,2}\right), s_{k,2} > 0, s_{k+1,2} > 0, \\ -\arctan\left(s_{k+1,2}/s_{k,2}\right) + \pi/2, s_{k,2} \le 0, s_{k+1,2} > 0, \\ \arctan\left(s_{k,2}/s_{k+1,2}\right) + \pi, s_{k,2} \le 0, s_{k+1,2} \le 0, \\ -\arctan\left(s_{k+1,2}/s_{k,2}\right) + 3\pi/2, s_{k,2} > 0, s_{k+1,2} \le 0, \end{cases} \qquad (5)$$

где $1 \le k \le n-1$.

В условиях точного соблюдения симметрии для определения симметрий двух контуров $A$ и $B$ достаточно применить к ним вышеописанную процедуру и получить показатели симметрий из разности полученных показателей относительно начала координат.

$$\begin{aligned} K_{\text{пер},AB} &= K_{\text{пер},A} - K_{\text{пер},B}, \\ K_{\text{мшт},AB} &= K_{\text{мшт},A} - K_{\text{мшт},B}, \\ K_{\text{пов},AB} &= K_{\text{пов},A} - K_{\text{пов},B}. \end{aligned} \qquad (6)$$

Однако при исследовании реальных систем имеют место нарушения симметрии или шум. При этом возникает необходимость в оценке близости контуров друг к другу. Оценка близости должна осуществляться независимо от переноса, сдвига и масштабирования, этого можно достичь путем сравнения спектров, нормализованным по этим преобразованиям. После нормализации из спектра $S$ контура будет получен спектр $S_{норм}$, инвариантный относительно переноса, масштаба и поворота исходного контура.

1. *Нормализация по сдвигу.* Необходимо перенести центр контура в начало координат, что достигается приравниванием к нулю первого элемента спектра:

$$s_{1,0} = 0, s_{2,0} = 0, \cdots, s_{n,0} = 0. \quad (7)$$

2. *Нормализация по растяжению/сжатию.* Необходимо смасштабировать спектр так, чтобы стал единичным вектор второго элемента, определяющего наибольший эллипс, аппроксимирующий контур. При этом все элементы спектра умножаются на коэффициент масштабирования:

$$S = S \cdot K_{мшт}. \quad (8)$$

3. *Нормализация по повороту.* Необходимо повернуть спектр так, чтобы вектор второго элемента, определяющего наибольший эллипс, аппроксимирующий контур, совпал с положительным направлением оси OX. Это достигается с помощью $n-1$ двухмерных поворотов спектра. Матрица поворота спектра $M_k$ в плоскости измерений $(k, k+1)$ выглядит следующим образом:

$$M_k = \begin{bmatrix} 1 & 0 & & \cdots & & 0 \\ 0 & \ddots & & & & \\ & & \cos(-K_{пов,k}) & \sin(-K_{пов,k}) & & \\ \vdots & & -\sin(-K_{пов,k}) & \cos(-K_{пов,k}) & & \vdots \\ & & & & \ddots & 0 \\ 0 & & \cdots & & 0 & 1 \end{bmatrix}. \quad (9)$$

Необходимо для каждого $1 \le k \le n-1$ составить матрицу $M_k$ и перемножить ее со спектром $S$:

$$S = S \cdot M_k. \quad (10)$$

После всех проведённых манипуляций спектр $S$ инвариантен относительно переноса, масштабирования и поворота исходного контура:

$$S_{норм} = S. \quad (11)$$

Для сравнения двух нормализованных спектров контуров $A$ и $B$ необходимо ввести критерий близости, например, сумму скалярных произведений векторов элементов спектров с дисконтированием по важности элемента:

$$K_{близ,AB} = \sum_{k=1}^{\frac{m}{2}+1} \frac{\vec{S}_{A,k} \cdot \vec{S}_{B,k}}{k}. \quad (12)$$

Разработанная методика выявления основных симметрий и показателя схожести двух контуров $A$ и $B$ имеет следующий вид:
1. Заданы контуры $A$ и $B$.
2. Применяем к ним преобразование Фурье (1).
3. Получаем их спектры $S_A$ и $S_B$.
4. Получаем показатели переноса $K_{пер,A}$ и $K_{пер,B}$ по (3).
5. Получаем коэффициенты $K_{мшт,A}$ и $K_{мшт,B}$ масштабирования (4).
6. Получаем углы поворота спектров относительно осей $K_{пов,A,k}$ и $K_{пов,B,k}$ по формуле (5).

7. Получаем основные симметрии контуров $K_{\text{пер},AB}$, $K_{\text{мшт},AB}$, $K_{\text{пов},AB}$ в соответствии с (6).
8. Осуществляем нормализацию спектров $S_A$ и $S_B$ по переносу (7).
9. Осуществляем нормализацию спектров $S_A$ и $S_B$ по масштабированию (8).
10. Последовательно осуществляем поворот по каждой последовательной паре измерений контура, рассчитывая матрицы $M_{A,k}$ и $M_{B,k}$ (9) и умножая на них спектры (10).
11. Получаем нормализованные спектры $S_{A,\text{норм}}$ и $S_{B,\text{норм}}$ (11).
12. Применяя критерий (12), получаем показатель близости контуров $K_{\text{близ},AB}$.

Ниже приведен пример реализации в среде MATLAB нормализации только одного трёхмерного ($n = 3$) контура (шаги методики 2–11) и визуализации этапов этого процесса (рис. 1).

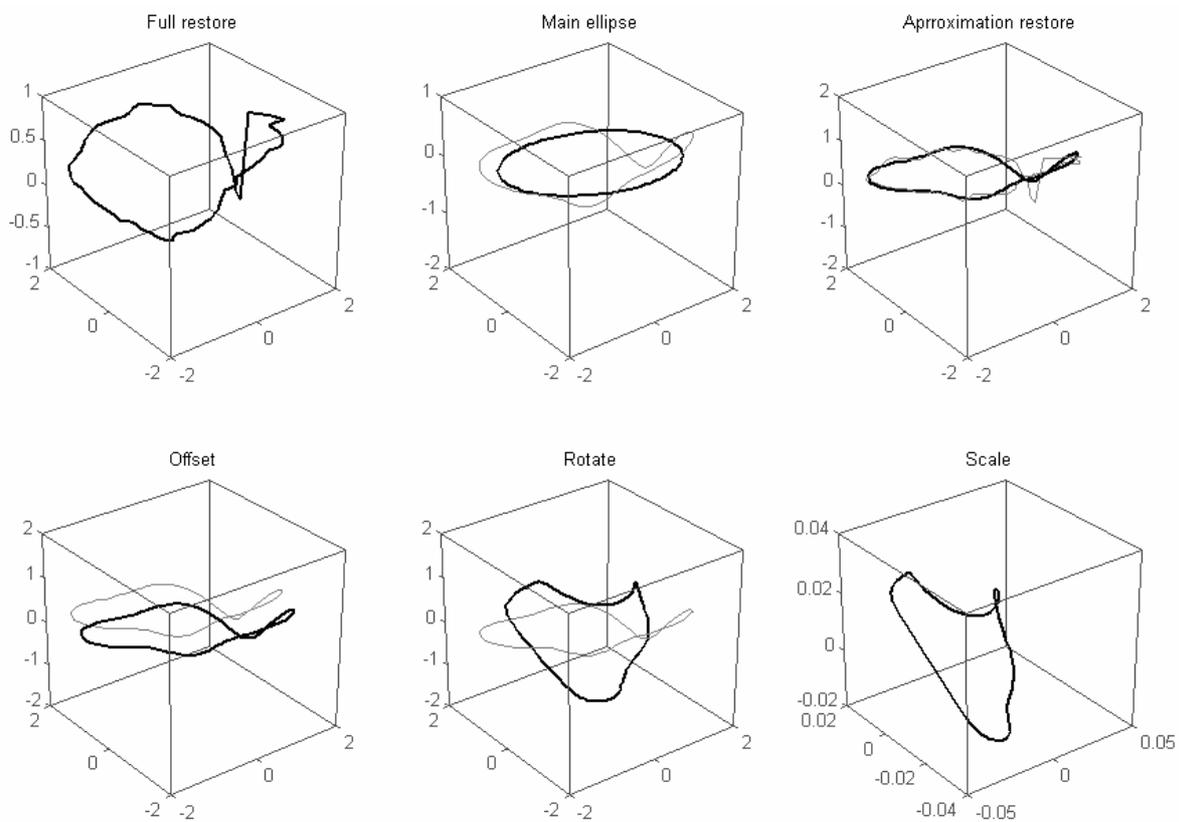

Рис. 1. Результат работы алгоритма для демонстрационного контура, состоящего из 100 точек и параметром сглаживания $K = 5$

Основная функция, принимающая на вход трёхмерный контур и параметр сглаживания, определяющий количество занчимых пар коэффициентов Фурье при восстановлении контура из спектра приведена ниже.

```
%Curve - трёхмерный контур, матрица Mx3
%K - параметр сглаживания
function PreProc3D(Curve, K)

%Получение кривой в комплексной плоскости
M = size(Curve, 1);
```

```matlab
%Применение быстрого одномерного преобразования Фурье
FCurve = fft(Curve, [], 1);

%Восстановление кривой по всем членам Фурье
RCurveAll = ifft(FCurve, [], 1);

%Восстановление кривой по К-парам членов Фурье
FCuttedCurve = FCurve;
FCuttedCurve(K+2:M-K, :) = 0;
RCCurve = ifft(FCuttedCurve, [], 1);

%Нормализация по перемещению и восстановление кривой К-парам членов Фурье
FCMCurve = FCuttedCurve;
FCMCurve(1, :) = 0;
RCMCurve = ifft(FCMCurve, [], 1);

%Нормализация по повороту и восстановление кривой К-парам членов Фурье
N = FCMCurve(2,:);
AngleXY = AngleToAxis([N(1) N(2)]);

%Матрица поворота вокруг Z
XYRotateMatrix = [ cos(AngleXY)     sin(AngleXY)    0;
                   -sin(AngleXY)    cos(AngleXY)    0;
                   0                0               1   ];

%Поворот вокруг Z
FCMRCurve = FCMCurve * XYRotateMatrix;
N = N * XYRotateMatrix;
AngleYZ = AngleToAxis([N(2) N(3)]);

%Матрица поворота вокруг X
YZRotateMatrix = [ 1    0               0;
                   0    cos(AngleYZ)    sin(AngleYZ);
                   0    -sin(AngleYZ)   cos(AngleYZ)];

%Поворот вокруг X
FCMRCurve = FCMRCurve * YZRotateMatrix;
RCMRCurve = ifft(FCMRCurve, [], 1);

%Нормализация по масштабу и восстановление кривой по по К-парам членов Фурье
Scale = sqrt(FCMRCurve(2,1)^2 + FCMRCurve(2,2)^2 + FCMRCurve(2,3)^2);
FCMRSCurve = FCMRCurve / real(Scale);
RCMRSCurve = ifft(FCMRSCurve, [], 1);

%Эллипс - главная компонента, образуемая при восстановлении контура по
%первой паре членов Фурье
FDCurve = FCurve;
F1 = FDCurve(1,:);
F2 = FDCurve(2,:);
FM = FDCurve(M,:);
FDCurve(:,:) = 0;
FDCurve(1,:) = F1;
FDCurve(2,:) = F2;
FDCurve(M,:) = FM;
FDCurve(1, :) = 0;
RDCurve = ifft(FDCurve, [], 1);

%РЕЗУЛЬТАТЫ:
%FCurve - полный спектр исходного контура
%RCurveAll - контур восстановленный по полному спектру
```

```
%FCuttedCurve - урезанный спектр состоящий только их K-пар членов Фурьё
%RCCurve - контур восстановленный из урезанного спектра
%FCMCurve - спектр, нормализованный по перемещению
%RCMCurve - контур, нормализованный по перемещению
%FCMRCurve - спектр, далее нормализованный по повороту
%RCMRCurve - контур, нормализованный по перемещению и повороту
%FCMRSCurve - спектр, далее нормализованный по масштабированию
%RCMRSCurve - контур, нормализованный по перемещению, повороту и
масштабированию
%FDCurve - спектр, состоящий только из нулевого члена и первой пары членов
Фурье
%RDCurve - эллипс, главная компонента исходного контура

%Вывод всех результатов
Cols = 2;
Rows = 3;
subplot(Cols, Rows, 1);
RCurveAll = cat(1, RCurveAll, RCurveAll(1,:));
plot3(RCurveAll(:,1), RCurveAll(:,2), RCurveAll(:,3));
title('Full restore')
set(gca, 'Projection', 'perspective');

subplot(Cols, Rows, 2);
RDCurve = cat(1, RDCurve, RDCurve(1,:));
RDCurve = real(RDCurve);
plot3(RCMCurve(:,1), RCMCurve(:,2), RCMCurve(:,3), RDCurve(:,1),
RDCurve(:,2), RDCurve(:,3));
title('Main ellipse')
set(gca, 'Projection', 'perspective');

subplot(Cols, Rows, 3);
RCCurve = cat(1, RCCurve, RCCurve(1,:));
RCCurve = real(RCCurve);
plot3(RCurveAll(:,1), RCurveAll(:,2), RCurveAll(:,3), RCCurve(:,1),
RCCurve(:,2), RCCurve(:,3));
title('Aprroximation restore')
set(gca, 'Projection', 'perspective');

subplot(Cols, Rows, 4);
RCMCurve = cat(1, RCMCurve, RCMCurve(1,:));
RCMCurve = real(RCMCurve);
plot3(RCCurve(:,1), RCCurve(:,2), RCCurve(:,3), RCMCurve(:,1),
RCMCurve(:,2), RCMCurve(:,3));
title('Offset')
set(gca, 'Projection', 'perspective');

subplot(Cols, Rows, 5);
RCMRCurve = cat(1, RCMRCurve, RCMRCurve(1,:));
RCMRCurve = real(RCMRCurve);
plot3(RCMCurve(:,1), RCMCurve(:,2), RCMCurve(:,3), RCMRCurve(:,1),
RCMRCurve(:,2), RCMRCurve(:,3));
title('Rotate')
set(gca, 'Projection', 'perspective');

subplot(Cols, Rows, 6);
RCMRSCurve = cat(1, RCMRSCurve, RCMRSCurve(1,:));
RCMRSCurve = real(RCMRSCurve);
plot3(RCMRSCurve(:,1), RCMRSCurve(:,2), RCMRSCurve(:,3));
title('Scale')
set(gca, 'Projection', 'perspective');
```

В основной функции используется вспомогательная функция, возвращающая угол между радиус-вектором точки и координатной осью:

```
%Point – точка
%Angle – угол к оси
function Angle = AngleToAxis(Point)
Angle = 0;
X = real(Point(1));
Y = real(Point(2));
if (X > 0 && Y > 0)
   Angle = -atan(X\Y);
end
if (X <= 0 && Y > 0)
   Angle = atan(Y\X) - pi/2;
end
if (X <= 0 && Y <= 0)
    Angle = -atan(X\Y) - pi;
end
if (X > 0 && Y <= 0)
    Angle = atan(Y\X) - 3*pi/2;
end
```

Таким образом, результаты тестирования реализации подтверждает эффективность решения выявления практически значимых симметрий (аффинные преобразования и поворот) в предложенной методике.

### *Литература*